\pdfoutput=1
\documentclass[cernpreprint, atlasdraft=false, LANGEDIT=true, UKenglish, texmf, orcidlogo]{atlasdoc}

\usepackage[backend=biber,subfigure=true]{atlaspackage}
\usepackage{atlasbiblatex}

\usepackage{atlasphysics}
\graphicspath{{logos/}{figures/}}

\usepackage{ANA-TOPQ-2023-44-PAPER-defs}
\usepackage{esdiff}
\usepackage{multirow}
\AtlasTitle{Measurement of differential $t$-channel single top (anti)quark production cross-sections at 13 TeV with the ATLAS detector}
\AtlasAbstract{%
The production of single top quarks and top antiquarks via the $t$-channel exchange of a virtual $W$~boson is measured in proton--proton collisions at a centre-of-mass energy of 13 TeV at the Large Hadron Collider.
The full Run 2 data sample recorded with the ATLAS detector in the years 2015--2018 is used, corresponding to an integrated luminosity of \SI{140}{\per\fb}.
The absolute and normalised production cross-sections are measured differentially as a function of the transverse momentum and absolute rapidity of the top quark and top antiquark.
In addition, the ratio of top quark to top antiquark production cross-sections is measured.
The measured distributions are compared with next-to-leading-order quantum chromodynamics predictions obtained with different combinations of matrix-element generators, parton-shower programs and proton parton distribution functions, as well as to next-to-next-to-leading-order calculations.
Overall, good agreement is observed between the measurements and the theoretical predictions.
For most measured distributions, the sensitivity to differences between the predictions is limited by the systematic uncertainties in the measurement.
The measured differential distributions are also interpreted in an effective field theory approach to constrain the Wilson-Coefficient $C_{Qq}^{3,1}$ associated with a four-quark operator.
The interpretation accounts for the effect of the selection efficiency, which is altered significantly by non-zero contributions from $C_{Qq}^{3,1}$.
}

\AtlasRefCode{TOPQ-2023-44}

\PreprintIdNumber{CERN-EP-2025-286}

\arXivId{2601.04938}

\HEPDataRecord{167734}

\AtlasJournal{JHEP}
\AtlasJournalRef{JHEP 05 (2026) 174}
\AtlasDOI{10.1007/JHEP05(2026)174}

\hypersetup{pdftitle={ATLAS document},pdfauthor={The ATLAS Collaboration}}

\begin{document}

\maketitle

\tableofcontents

\newpage
\section{Introduction}

The production of single top quarks was first observed in 2009 in proton--antiproton collisions at the Tevatron~\cite{Abazov_2009,Aaltonen_2009}
and was since then used to obtain valuable insights on the properties of the top quark~\cite{Giammanco_2018}.
At the Large Hadron Collider, the main production channel of single top quarks is the \tchan exchange of a virtual \wboson.
The total production cross-section of single top quarks and top antiquarks via the \tchan process (\tqtbarq production) was measured by both
the ATLAS and CMS collaborations at centre-of-mass energies of \SI{5.02}{\TeV}~\cite{TOPQ-2022-22}, \SI{7}{\TeV}~\cite{TOPQ-2012-21,CMS-TOP-11-021}, \SI{8}{\TeV}~\cite{TOPQ-2015-05,CMS-TOP-12-038} and \SI{13}{\TeV}~\cite{TOPQ-2019-02,CMS-TOP-17-011}.
At \SI{13}{\TeV}, differential cross-sections were measured by CMS~\cite{CMS-TOP-17-023} using a partial Run 2 data sample with an integrated luminosity of \SI{35.9}{\per\fb}.

\begin{figure}[h!]
\begin{center}
\subfigure[]{
\includegraphics[width=0.43\textwidth]{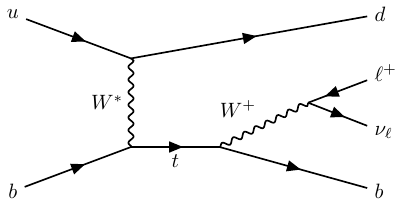}
\label{fig:Feynman_top_quark}
}
\subfigure[]{
\includegraphics[width=0.43\textwidth]{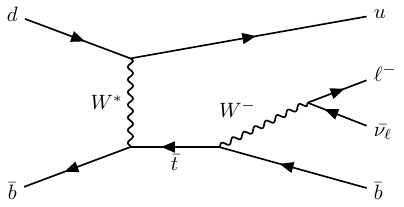}
\label{fig:Feynman_top_antiquark}
}

\caption{\label{fig:feyn_signal}Example Feynman diagrams of
\protect\subref{fig:Feynman_top_quark} single top quark
and \protect\subref{fig:Feynman_top_antiquark} single top antiquark production
via the \tchan exchange of a virtual \wboson at LO in perturbation theory with the leptonic decay channel of the top (anti)quark.}
\end{center}
\end{figure}

Example Feynman diagrams for \tchan production of single top quarks and top antiquarks at leading-order (LO) in perturbation theory are depicted in \Fig{\ref{fig:feyn_signal}}.
A light quark from one of the colliding protons interacts with a \bquark from the other proton by exchanging a virtual \wboson.
Since the integral of the valence \uquark density of the proton is twice the integral of the valence \dquark density, the $tq$ production cross-section is expected to be higher than the $\bar{t}q$ production cross-section.
Subleading diagrams containing only sea quarks in the initial-state contribute to the production cross-section in a charge-symmetric way.

In this paper, the ATLAS measurements of differential \tqtbarq production cross-sections are presented using the full Run 2 data sample, recorded with the ATLAS detector between 2015 and 2018 at $\sqrt{s} = \SI{13}{\TeV}$, corresponding to an integrated luminosity of \SI{140}{\per\fb}.
Absolute and normalised differential cross-sections are measured as functions of the transverse momentum \pTt or \pTtbar and the absolute value of the rapidity \Rapt or \Raptbar of the top quark and top antiquark.
The cross-sections for \tqtbarq production are measured separately, along with their ratio.

The measurement targets events with leptonically decaying \wbosons.
Signal events are required to contain exactly one isolated electron or muon and large missing transverse momentum \met.
Based on the LO event signature of \tqtbarq production, exactly two jets are required and exactly one of these jets must be identified as originating from a \bquark.
The purity of the event selection in signal events is enhanced by dedicated kinematic requirements and the use of a neural network (NN), trained to separate signal and background events.
The NN was trained for the ATLAS measurement of the total \tqtbarq production cross-sections at $\sqrt{s} = \SI{13}{\TeV}$~\cite{TOPQ-2019-02}, which serves as the basis for the differential analysis reported in this paper.
Consequently, the event selection, simulated event samples, and systematic uncertainties used here are almost identical and are therefore described only briefly.
Further details can be found in the paper~\cite{TOPQ-2019-02}.

Two signal regions are defined based on the sign of the lepton charge to measure the \tqtbarq production cross-sections separately.
The measured cross-sections are corrected for geometric acceptance, detector resolution, and trigger and reconstruction efficiencies.
The corrections are made by unfolding the measured distributions to parton level, using D'Agostini's iterative approach~\cite{DAgostini:1994fjx}.
The measured differential cross-sections are compared with predictions obtained at next-to-leading-order (NLO) in quantum chromodynamics (QCD) with different combinations of matrix-element event generators and parton-shower programs and with different parton distribution function (PDF) sets, and to next-to-next-to-leading-order (NNLO) predictions calculated with MCFM~\cite{Campbell:2019dru}.
Measurements of the differential cross-sections and in particular the ratio of the differential $tq/\bar{t}q$ cross-sections as a function of \Raptortbar can help to constrain the PDFs of the proton~\cite{PDFfitssingletop}.
The \tqtbarq process offers the possibility to constrain the $u/d$-quark ratio and, as future prospect, the $b$-quark density.

The differential cross-sections as a function of \pTt or \pTtbar are also of interest for effective field theory (EFT) interpretations~\cite{EFTtopquarksector}.
In the standard-model effective field theory (SMEFT)~\cite{Degrande:2020evl}, at leading-order in QCD only three dimension-6 operators
contribute to \tqtbarq production through interference with the SM: \Oqq, \OphiQ, and \OtW.
A non-zero contribution from the four-quark EFT operator \Oqq affects both the \tqtbarq production cross-sections and the kinematic properties of the events, thereby altering the acceptance.
In contrast, the \OphiQ operator only alters the inclusive production cross-section and \OtW modifies the kinematics of the top-quark decay~\cite{deBeurs:2018pvs}.
The EFT interpretation aims to constrain the contribution of \Oqq and is based on the measured \pTt and \pTtbar differential distributions at parton level.
The dependence of the acceptance on the \Oqq contribution is studied and a significant impact on the obtained constraints is found.
Therefore, the interpretation takes this acceptance effect into account.
The obtained constraints on an \Oqq contribution improve the limits set by the EFT interpretation of the total \tqtbarq cross-section measurement~\cite{TOPQ-2019-02}.

A brief introduction of the ATLAS detector and the data samples is given in \Sect{\ref{sec:atlas}} and the samples of simulated events are described in \Sect{\ref{sec:samples}}.
The object definitions and event selection are given in \Sect{\ref{sec:selection}}. \Sect{\ref{sec:background}} provides a description of the estimate of the background rate.
The unfolding procedure and needed inputs are defined in \Sect{\ref{sec:unfolding}} and \Sect{\ref{sec:inputs}}.
The uncertainties are described in \Sect{\ref{sec:uncertainties}}.
The unfolded results and an interpretation of the results in the context of EFT are presented in \Sect{\ref{sec:results}}.


%
\FloatBarrier

\section{ATLAS detector and data sample}
\label{sec:atlas}

\newcommand{\AtlasCoordFootnote}{%
ATLAS uses a right-handed coordinate system with its origin at the nominal interaction point (IP)
in the centre of the detector and the \(z\)-axis along the beam pipe.
The \(x\)-axis points from the IP to the centre of the LHC ring,
and the \(y\)-axis points upwards.
Polar coordinates \((r,\phi)\) are used in the transverse plane,
\(\phi\) being the azimuthal angle around the \(z\)-axis.
The pseudorapidity is defined in terms of the polar angle \(\theta\) as \(\eta = -\ln \tan(\theta/2)\) and is equal to the rapidity
$ y = \frac{1}{2} \ln \left( \frac{E + p_z}{E - p_z} \right) $ in the relativistic limit.
Angular distance is measured in units of \(\Delta R \equiv \sqrt{(\Delta y)^{2} + (\Delta\phi)^{2}}\).}

The ATLAS experiment~\cite{PERF-2007-01} at the LHC is a multipurpose particle detector with a forward--backward symmetric cylindrical geometry and a near \(4\pi\) coverage in solid angle.\footnote{\AtlasCoordFootnote}
It consists of an inner tracking detector surrounded by a thin superconducting solenoid
providing a \qty{2}{\tesla} axial magnetic field, electromagnetic and hadronic calorimeters, and a muon spectrometer.
The inner tracking detector covers the pseudorapidity range \(|\eta| < 2.5\).
It consists of silicon pixel, silicon microstrip, and transition radiation tracking detectors.
Lead/liquid-argon (LAr) sampling calorimeters provide electromagnetic (EM) energy measurements
with high granularity within the region \(|\eta|< 3.2\).
A steel/scintillator-tile hadronic calorimeter covers the central pseudorapidity range (\(|\eta| < 1.7\)).
The endcap and forward regions are instrumented with LAr calorimeters
for EM and hadronic energy measurements up to \(|\eta| = 4.9\).
The muon spectrometer surrounds the calorimeters and is based on
three large superconducting air-core toroidal magnets with eight coils each.
The field integral of the toroids ranges between \num{2.0} and \qty{6.0}{\tesla\metre}
across most of the detector.
The muon spectrometer includes a system of precision tracking chambers up to \(|\eta| = 2.7\) and fast detectors for triggering up to \(|\eta| = 2.4\).
The luminosity is measured mainly by the LUCID--2~\cite{LUCID2} detector which is located close to the beampipe.
A two-level trigger system was used to select events~\cite{TRIG-2016-01}.
The first-level trigger is implemented in hardware and used a subset of the detector information
to accept events at a rate close to \qty{100}{\kHz}.
This is followed by a software-based trigger that
reduced the accepted rate of complete events to \qty{1.25}{\kHz} on average
depending on the data-taking conditions.
A software suite~\cite{SOFT-2022-02} is used in data simulation, in the reconstruction
and analysis of real and simulated data, in detector operations, and in the trigger and data acquisition
systems of the experiment.

The analysis uses proton--proton ($pp$) collision data recorded in the years 2015 to 2018 \cite{DAPR-2018-01} at a centre-of-mass energy of 13$\,\TeV$ and corresponds
to an integrated luminosity of $140\,\ifb$ with a relative uncertainty of $0.83\%$~\cite{DAPR-2021-01}.
Events were selected online during data taking by single-electron or single-muon triggers~\cite{TRIG-2018-05,TRIG-2018-01}.
Multiple triggers with different thresholds in transverse momentum (\pT) were combined in a logical OR to increase the selection efficiency.
The lowest-threshold triggers utilised isolation requirements for reducing the trigger rate.


\section{Samples of simulated events}
\label{sec:samples}

Sets of simulated events from signal and background processes were produced with
MC-based event generator programs to model the physics processes.
After event generation, the response of the ATLAS detector was simulated using the \GEANT toolkit~\cite{Agostinelli:2002hh} with a full detector model~\cite{SOFT-2010-01} or a fast simulation~\cite{ATL-PHYS-PUB-2010-013,AFII} which employed a parameterisation of the calorimeter response.

Samples of simulated events involving top quarks were generated using the \POWHEGBOX[v2]~\cite{Nason:2004rx,Frixione:2007nw,Frixione:2007vw,Alioli:2009je,Alioli:2010xd,Re:2010bp,Frederix:2012dh} matrix element (ME) generator calculating in QCD NLO with the top-quark mass set to \(m_t=172.5\,\GeV\).
The production of single top-quarks in the \(t\)-channel was simulated with the
\NNPDF[3.0nlo\_nf4] PDF set, which implements the four-flavour scheme (4FS) preferred for this process~\cite{Frederix:2012dh}.
The initial gluon splitting into the $b\bar{b}$ pair is included in the ME calculation and can be carried out with massive $b$-quarks, leading to a more reliable modelling of the kinematics of the ``spectator'' $b$-quark.
All other top-quark-related processes used the \NNPDF[3.0nlo] PDF set~\cite{Ball:2014uwa} with the five-flavour scheme (5FS).
Parton showers, hadronisation, and the underlying event were modelled using \PYTHIA[8.230]~\cite{Sjostrand:2007gs}
with the A14~\cite{ATL-PHYS-PUB-2014-021} set of tuned parameters and the \NNPDF[2.3lo] PDF set~\cite{Ball:2012cx}. When generating $tW$ events, the diagram-removal scheme~\cite{Frixione:2008yi} was employed to treat the interference with \ttbar\ production~\cite{ATL-PHYS-PUB-2016-020}.
The decays of bottom and charm hadrons were simulated using the \EVTGEN[1.6.0] program~\cite{Lange:2001uf}.

The production of $W$ and $Z$~bosons in association with jets, including heavy-flavour jets, was simulated with the \SHERPA[2.2.11] generator~\cite{Bothmann:2019yzt} using NLO matrix elements for up to two partons and LO matrix elements for up to five partons. The default \SHERPA parton shower~\cite{Schumann:2007mg} and the
cluster hadronisation model~\cite{Winter:2003tt} were used. The generation employed the dedicated set of tuned parameters developed by the \SHERPA authors and the \NNPDF[3.0nlo] PDF set.
The $W$+jets and $Z$+jets samples were normalised to NNLO QCD and NLO electroweak
predictions~\cite{Anastasiou:2003ds} of the total cross-sections, obtained with the \textsc{FEWZ} package~\cite{Gavin:2010az,QUACKENBUSH2013209}.

Samples of on-shell diboson production ($WW$, $WZ$, and $ZZ$, abbreviated as $VV$ in the following) were simulated with the same \SHERPA configuration as described above for \Wjets and \Zjets production.
The matrix elements considered contain all diagrams with four electroweak vertices and were calculated at
NLO in QCD for up to one additional parton and at LO for up to three additional parton emissions.
The diboson event samples were normalised to the total cross-sections provided by \SHERPA.

Multijet events with fake or non-prompt leptons are modelled using different strategies depending on the lepton type.
Non-prompt muons are estimated from collision data while fake electrons or non-prompt electrons from hadron decays are described with simulated dijet samples.
The dijet event sample was simulated using Pythia 8.186 with LO matrix elements using the NNPDF2.3lo PDF set and additional radiation (jets) is produced through the parton shower using the A14 set of tuned parameters.

For interpreting the measurement in the framework of EFT, samples of \(t\)-channel and \(s\)-channel production were generated
with \MGNLO[2.7.3] using the SMEFTatNLO-NLO UFO model~\cite{Degrande:2020evl} with the 5FS
and the \NNPDF[3.0nlo] PDF set. The operator \(\Oqq\) is considered, which introduces
a four-quark contact interaction. Separate samples of simulated events were generated for each
\(\Cqq\in\left\{-0.6,\,-0.2,\,0.0,\,0.4,\,1.0\right\}\,\TeV^{-2}\)
for single top-quark and top-antiquark production.
The generated events were showered with \PYTHIA[8.244] using the A14 set of tuned parameters and
the \NNPDF[2.3lo] PDF set.
In these samples, the top-quark (top-antiquark) is assumed to decay into \(W^+b\) (\(W^-\bar{b}\)) with a branching ratio of \SI{100}{\%}.

For understanding the simulation of single-top-quark events, several parameters can be varied and compared to measured differential distributions.
These include PDFs, \alphas, and the calculation scheme (i.e.\ 4FS vs.\ 5FS). Among these, the variations of the PDFs have the largest impact.

Fixed order differential calculations of the top-quark and top-antiquark kinematics are available up to NNLO in pertubation theory within the MCFM framework~\cite{Campbell:2020fhf,Campbell:2021qgd} and made in the 5FS\@.
The dynamic double deep inelastic scattering scale is used as suggested in ~\cite{Campbell:2021qgd}.
In this scale choice, the momentum transfer along the $W$~boson $Q^{2}$ is used as the scale for the light-quark-line corrections $\mu^{2}=Q^{2}$ and for the heavy-quark-line corrections, $\mu^{2}=Q^{2}+m_{t}^{2}$ with the top quark mass $m_{t}$ set to \SI{172.5}{\GeV}.
The uncertainty in the calculations comprises the uncertainty in the renormalisation and factorisation scale choice and the uncertainty in the value of \alphas.
The signal samples for \tqtbarq production are normalised to the reference cross-sections calculated with MCFM at NNLO~\cite{ATL-PHYS-PUB-2025-035}.
Within the parton level ME generators interfaced to parton shower programs, calculations are available up to NLO with implementations in the \POWHEGBOX[v2] (ST\_tch\_4f) and \MGaMC~\cite{Alwall:2014hca} frameworks, in both the 4FS and 5FS, and interfaced to either the \PYTHIA[8] or the \HERWIG[7]~\cite{Bahr:2008pv,Bellm:2015jjp} parton shower generator programs.
The \POWHEGBOX implementation is the same as used for the event generation of the nominal signal sample.
An overview of the considered fixed order calculations and ME event generators is given in \Tab{\ref{tab:generators}}.

PDFs are provided by several groups worldwide, e.g. {\textsc ABM}~\cite{Alekhin:2017kpj},
CTEQ~\cite{Hou:2019qau}, {\textsc MSHT}~\cite{Bailey:2020ooq}, {\textsc NNPDF}~\cite{Ball:2014uwa,NNPDF:2017mvq,NNPDF:2021njg},
and {\textsc HERAPDF}~\cite{H1:2015ubc}.
The ATLAS Collaboration also provides its own PDF sets~\cite{STDM-2012-20,STDM-2020-32}.
The PDFs from each group can differ in, for example, the data used, the value of \(\alpha_s\), the treatments of the heavy-quark masses and the implementation of higher-order QCD corrections.
Predictions for the \tqtbarq production with different PDF sets are obtained using the \POWHEGBOX[v2] implementation.
An overview of the PDFs used for the comparison is given in \Tab{\ref{tab:PDFs}}, where available the 4FS variant is used.
All PDF sets are employed at NLO precision to match that of the ME calculation, except for ATLASPDF21, which is available only at NNLO precision.

\begin{table}[ht]
\caption{\label{tab:generators} Summary of fixed order calculations and ME event generators used for the production of top-quark and top-antiquark kinematic-distributions of the \tqtbarq production.}
\begin{center}
\begin{tabular}{llll}
\hline
\hline
Program & Precision & FS &  Note\\
\hline
\hline
MCFM  & LO / NLO / NNLO & 5FS & FO calculation\\
\MGaMC & NLO & 4FS / 5FS & Event generation \\
\POWHEG ST\_tch\_4f & NLO & 4FS & Event generation \\
\hline
\end{tabular}
\end{center}
\end{table}

\begin{table}[ht]
\caption{\label{tab:PDFs} Summary of PDF sets employed for the production of top-quark and top-antiquark kinematic-distributions of the \tqtbarq production using the \POWHEG ST\_tch\_4f generator. }
\begin{center}
\begin{tabular}{llll}
\hline
\hline
PDF & LHAPDF ID~\cite{Buckley_2015} & Precision &  FS  \\
\hline
\hline
\NNPDF[3.0]  & 260400--260500 & NLO & 4FS  \\
\NNPDF[4.0] & 334700 & NLO & 4FS \\
{\textsc MSHT20}     & 27700 & NLO & 4FS  \\
\CT[18]  & 14400 & NLO & 5FS \\
\PDFforLHC[21] & 93500--93541 & NLO & 4FS \\
ATLASPDF21  & 65700--65753 & NNLO & 5FS \\
ABMP16  & 42930--42961 & NLO & 4FS \\
\hline
\end{tabular}
\end{center}
\end{table}


\FloatBarrier

\section{Object reconstruction and event selection}
\label{sec:selection}

The final state of the \tqtbarq signal processes comprises a charged lepton, a neutrino, a \bquark
and a light quark (see \Fig{\ref{fig:feyn_signal}}) and is reconstructed by identifying corresponding objects
measured in the detector.
The presence of a high-$\pt$ neutrino is indicated by large missing transverse momentum.
In the context of this analysis, ``charged lepton'' refers to electrons and muons only.

\subsection{Object definitions}

Electron candidates are reconstructed from Inner Detector (ID) tracks matched to EM calorimeter energy clusters~\cite{EGAM-2018-01},
where for the pseudorapidity of the clusters $\eta_{\text{cluster}}$, $|\eta_{\text{cluster}}|<2.47$ is required.
The detector transition region $1.37<|\eta_{\text{cluster}}|<1.52$ is excluded.
Electron candidates are required to have $\pt>10\,\GeV$.
A likelihood-based discriminant is constructed to reject non-prompt electrons~\cite{EGAM-2018-01}.
Muon candidates are reconstructed by combining ID and muon spectrometer tracks~\cite{MUON-2018-03}. All tracks must be in the range of $|\eta|<2.5$.
Leptons must be consistent with the primary vertex via impact-parameter requirements, and are further subject to isolation criteria optimised
to reject non-prompt and fake candidates~\cite{EGAM-2018-01,MUON-2018-03}.

Jets are reconstructed from particle-flow objects~\cite{PERF-2015-09} with the anti-$k_t$ algorithm~\cite{Cacciari:2008gp,Fastjet} using a radius parameter of $0.4$.
The jet energy is calibrated by applying several simulation-based corrections and techniques
correcting for differences between simulation and data~\cite{JETM-2018-05}.
The jets must satisfy $\pt>\SI{30}{\GeV}$ and $|\eta|<4.5$.
Pile-up suppression is performed via the jet-vertex tagger (JVT)~\cite{PERF-2014-03} and forward JVT (fJVT)~\cite{PERF-2016-06} discriminants.
Jets containing \(b\)-hadrons are identified (\(b\)-tagged) with the \texttt{DL1r} algorithm,
which uses a deep feed-forward neural network with several \(b\)-tagging algorithms as inputs~\cite{FTAG-2019-07}.
As requirement on the \texttt{DL1r} discriminant, the 60\% efficiency working point is used.
Since the \texttt{DL1r} algorithm uses measurements from the ID,
the identification of \(b\)-jets is limited to the region with \mbox{\(\abseta < 2.5\)}.

An overlap-removal procedure is applied to prevent double-counting of objects satisfying multiple reconstruction criteria.
Reconstructed objects are removed in the following order:
electrons sharing an ID track with a muon, jets within \(\Delta R = 0.2\) of an electron,
electrons within \(\Delta R = 0.4\) of a remaining jet, jets within \(\Delta R = 0.2\) of a muon if they have less than three associated tracks with $\pT > \SI{0.5}{\GeV}$,
and muons within \(\Delta R = 0.4\) of a remaining jet.

The missing transverse momentum \vecpmiss is reconstructed as the negative vector sum of the $\pt$ of
the reconstructed leptons and jets, and ID tracks that point to the primary vertex
but are not associated with a reconstructed object~\cite{PERF-2016-07}.
The magnitude of \vecpmiss is denoted by \met.

The neutrino momentum is required for the reconstruction of the top (anti)quark.
The \(x\)- and \(y\)-components of the neutrino momentum are approximated
by the components of \vecpmiss.
The \(z\)-component, \(p_z(\nu)\), is determined by constraining the mass of
the reconstructed \(W\)~boson to match the measured world average $m(W)=\SI{80.4}{\gev}$~\cite{PhysRevD.110.030001}.
If the resulting quadratic equation has two real solutions, the one with the smallest \(|p_z(\nu)|\) is chosen.
In the case of complex solutions, a kinematic fit is performed such that the imaginary part vanishes and
the distance between the transverse components of the neutrino momentum and \vecpmiss is minimised~\cite{Chwalek:2010vka}.
The \(W\)~boson is formed by adding the four-vectors of the reconstructed neutrino and
the charged lepton and the top (anti)quark is reconstructed by adding the four-vector of the \(W\)~boson and the \(b\)-jet.

\subsection{Event selection}
The event selection imposes requirements based on the expected event signature of \tqtbarq production.
Candidate events must contain exactly one charged lepton with $p_\text{T} >28\,\GeV$ matched to the trigger object.
Two separate signal regions are defined according to the sign of the lepton charge,
referred to as $\ell^+$ SR and $\ell^-$ SR in the following.

Exactly two jets with $p_\text{T} > 30\,\GeV$ are required, of which exactly one must be $b$-tagged.
The $b$-tagged jet is explicitly required to have $|\eta| < 2.5$.
For events with forward jets with \(2.3<|\eta|\), the requirement \(\pT(\text{jet}) > \SI{35}{\gev}\) is imposed
leading to an improved modelling of the \(|\eta|\) distribution of untagged jets in the given regime.
To reduce the multijet background, requirements are imposed on the missing transverse energy and the transverse $W$ mass defined as
$\mtw = \sqrt{2 \pT(\ell) \MET \left(1-\cos \Delta \phi \left(\vecpmiss, \ell \right) \right)}$.
Events are required to have $\met > 30\,\GeV$ and $\mtw > \SI{50}{\gev}$.
To further suppress the multijet background and to remove poorly reconstructed leptons
with low \(\pt\), an additional requirement is applied based on the azimuthal angle $\phi$
between the charged lepton and the leading jet ($j_1$), i.e.\ the jet with the largest \(\pt\).
The imposed requirement is $\pT\left(\ell\right) > \SI{40}{\gev} \cdot |\Delta \phi\left(j_1, \ell \right)|/\pi$
since the charged lepton and the leading jet are typically emitted back-to-back in multijet events.
An additional selection criterion is imposed on the invariant mass of the charged lepton and the $b$-tagged jet,
$m(\ell b)< \SI{160}{\gev}$, to exclude the off-shell region for top-quark decays which is not modelled well.

To reduce the contribution from background processes further, an additional selection criterion is introduced using the NN
trained for the inclusive analysis to separate background and signal events~\cite{TOPQ-2019-02}.
The NN is implemented using the NeuroBayes package~\cite{feindt-2004,Feindt:2006pm} and trained on kinematic variables formed from reconstructed objects and global event observables.
Seventeen input variables are used which describe angular distances and differences between the transverse momentum of reconstructed objects,
the invariant mass and absolute rapidity of reconstructed objects and the sum of the transverse momenta of the reconstructed jets, charged lepton and neutrino.
The modelling of the input variables was studied in \Refn{\cite{TOPQ-2019-02}} and agrees with the data within uncertainty.
The simulated nominal signal samples and all MC simulated background processes are used for the training.

The NN assigns an output-score \Dnn$\in[0,1]$ to each event.
Signal events are assigned a high \Dnn score and background events are assigned low scores.
The requirement of a high \Dnn score reduces the contribution from background processes significantly, but it also reduces the signal acceptance.
A very low contribution from background events is required to keep the associated rate uncertainties under control, but a high number of data events is needed to keep the statistical uncertainties low.
The \Dnn requirement is therefore optimised to balance these effects and the requirement $\Dnn>0.93$ is chosen.
Introducing the \Dnn requirement enhances the signal-to-background ratio in the $\ell^+$~SR from 0.25 to 6.1 and in the $\ell^-$~SR from 0.17 to 3.8.
The impact of the \Dnn requirement on the systematic uncertainties related to the modelling of the signal process was found to be negligible.

%
%

%
%
%
%
%
%
%
%
%
%
%
%
%
%
%
%
%
%
%


%
\FloatBarrier

\section{Background estimation}
\label{sec:background}

The background contribution from processes with one prompt lepton is estimated from MC simulations and includes \ttbar production, single-top production via the $s$-channel and the $tW$ process,
diboson production and the production of single $W$ and $Z$~bosons in association with jets.
The production processes of a $W$~boson in association with jets are referred to as $W+b\bar{b}$ and $W+c(\bar{c})$ in the following.
While $W+c$ production gives a significant contribution, the $W+b$ process is heavily suppressed.

The multijet background is estimated by using a combination of simulated dijet events and data-driven techniques.
Events with misidentified jets are modelled from simulated dijet events by re-labelling jets that deposit most of their energy in the EM calorimeter as electrons.
Events with non-prompt muons are modelled from dedicated collision events which are enriched in non-prompt muons by applying the nominal selection with inverted or modified muon isolation requirements.
Both of these methods are detailed in \Refn{\cite{ATLAS-CONF-2014-058}}.

The normalisation of each background process is extracted from a dedicated simultaneous binned profile maximum-likelihood fit using the same methodology as~\Refn{\cite{TOPQ-2019-02}}.
The likelihood function is defined as the product of Poisson probability terms across all bins. The fitted event yields in each bin depend on nuisance parameters that account for systematic uncertainties.
All nuisance parameters, except those representing MC statistical uncertainties, are constrained by Gaussian terms in the likelihood function.
Control regions for the multijet background and the complete \Dnn distributions
excluding $\Dnn > 0.93$, i.e. the signal region for the differential measurement, are used in the fit.
No additional control regions are included for backgrounds with a prompt lepton.
The complete set of experimental and background-related uncertainties are included via nuisance parameters.
The event yields of the multijet backgrounds are left floating while the yields of all other backgrounds are constrained to their predictions within the associated uncertainties.
For the background processes involving top quarks, the relative uncertainties in the theoretical calculations of the cross-sections of $\pm 4\%$ for \ttbar and $\pm 3.7\%$ for \tWschan are assigned~\cite{TOPQ-2019-02}.
An uncertainty of $\pm 40\%$ is assigned to the expected rate of $W+b\bar{b}$, covering differences seen in previous measurements~\cite{TOPQ-2018-06} between the \SHERPA prediction and ATLAS collision data.
The uncertainties in the $W+c(\bar{c})$ and the $VV,Z$+jets backgrounds are estimated as $\pm 20\%$ based on the theoretical uncertainties in the processes and uncertainties in the associated heavy flavour production.
Separate nuisance parameters are used for the cross-section uncertainties
of $W+b\bar{b}$ production and $W+c(\bar{c})$ production in the regions with positive and negative lepton charge.
The post-fit agreement of the expected and observed distributions is shown in \Fig{\ref{fig:postFit}}.
The difference between prediction and observation in \Fig{\ref{fig:SRppost}} can be attributed to the signal modelling uncertainties being not included.
No significant pulls of nuisance parameters are observed and a consistent inclusive \tqortbarq cross-section is obtained.

\begin{figure}[htbp]
\centering
\subfigure[]{
\includegraphics[width=0.45\textwidth]{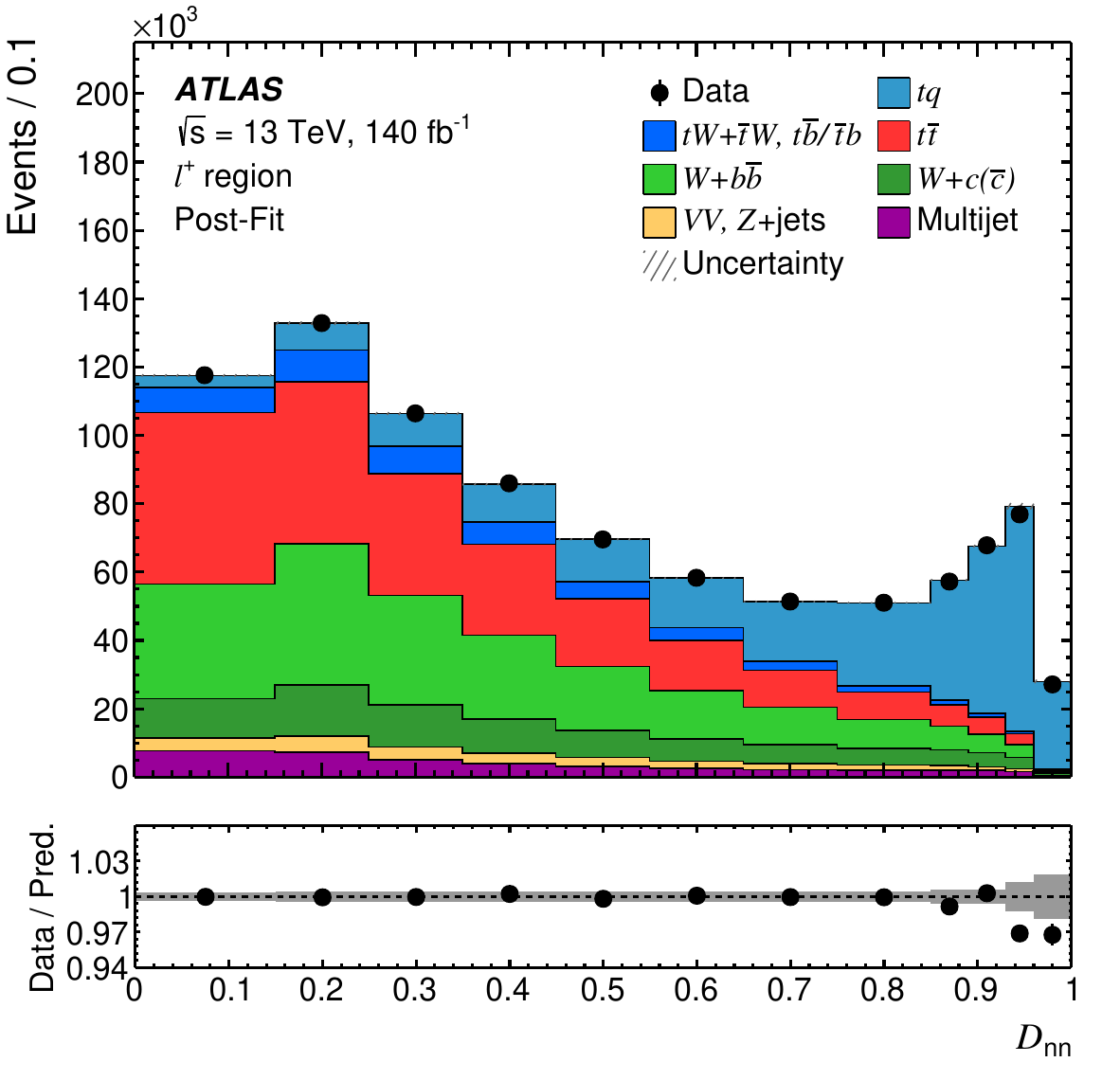}
\label{fig:SRppost}
}
\subfigure[]{
\includegraphics[width=0.45\textwidth]{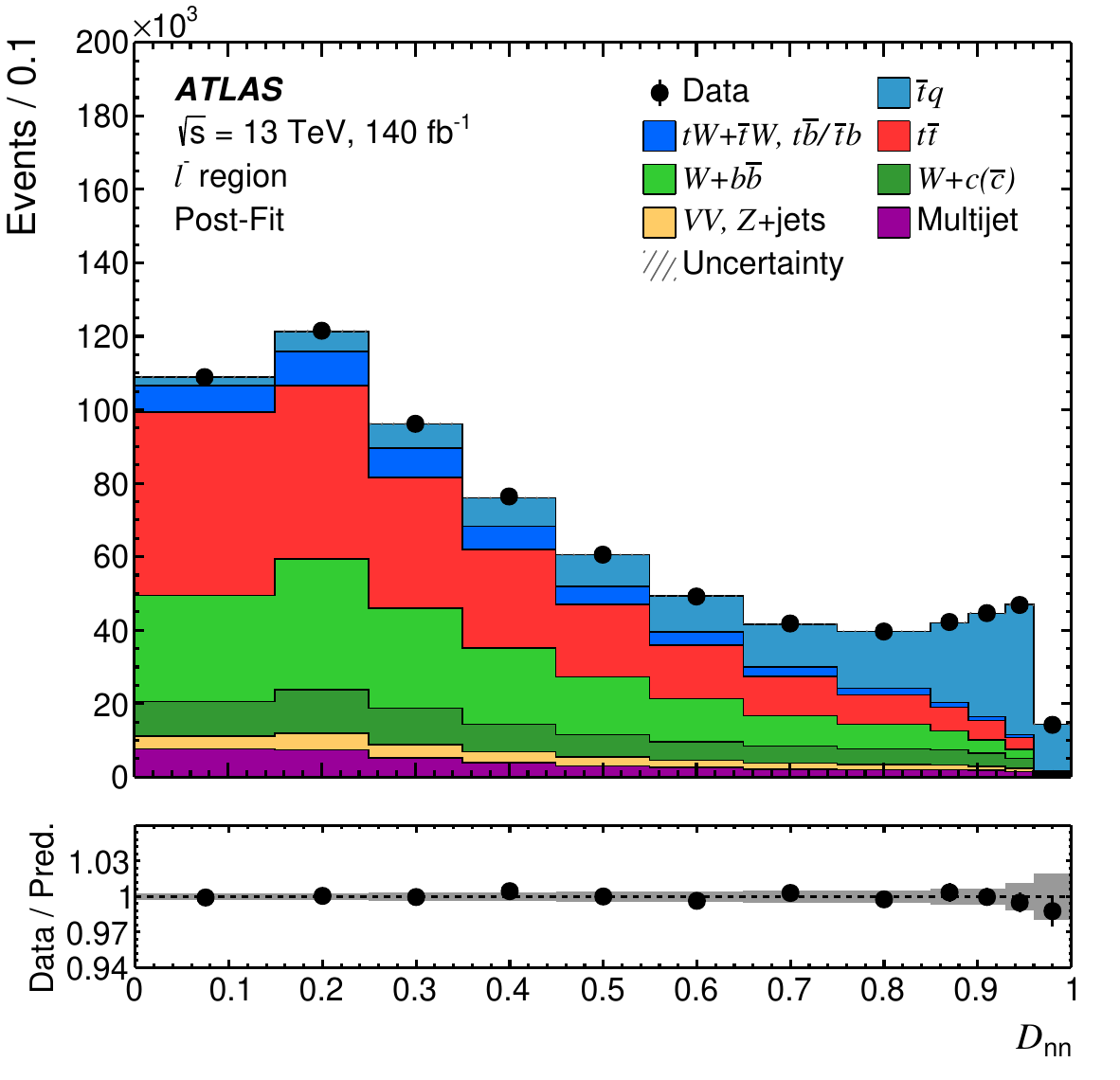}
\label{fig:SRnpost}
}
\caption{ Post-fit agreement between data and the expected distributions in events containing a
\protect\subref{fig:SRppost} positively charged lepton and \protect\subref{fig:SRnpost} negatively charged lepton.
The experimental, background-related and MC statistical uncertainties are incorporated in the uncertainty bands.
The region $\Dnn > 0.93$ is excluded from the fit.
}
\label{fig:postFit}
\end{figure}

Per-process scale factors are obtained from the post-fit over pre-fit event yield ratio and the relative uncertainty in the post-fit yield is taken as the uncertainty in the expected rate of the respective process.
The expected event yields in the SRs per process after the scale factors are applied to the background processes are presented in \Tab{\ref{tab:Yields}}.
Comparisons between the observed and expected distributions in both signal regions incorporating all uncertainties are shown in \Fig{\ref{fig:DataMCcomp}}.
Here, $\ell^\pm\nu b$ is used to refer to the reconstructed top (anti)quark.

\begin{figure}[htbp]
\centering
\subfigure[]{
\includegraphics[width=0.45\textwidth]{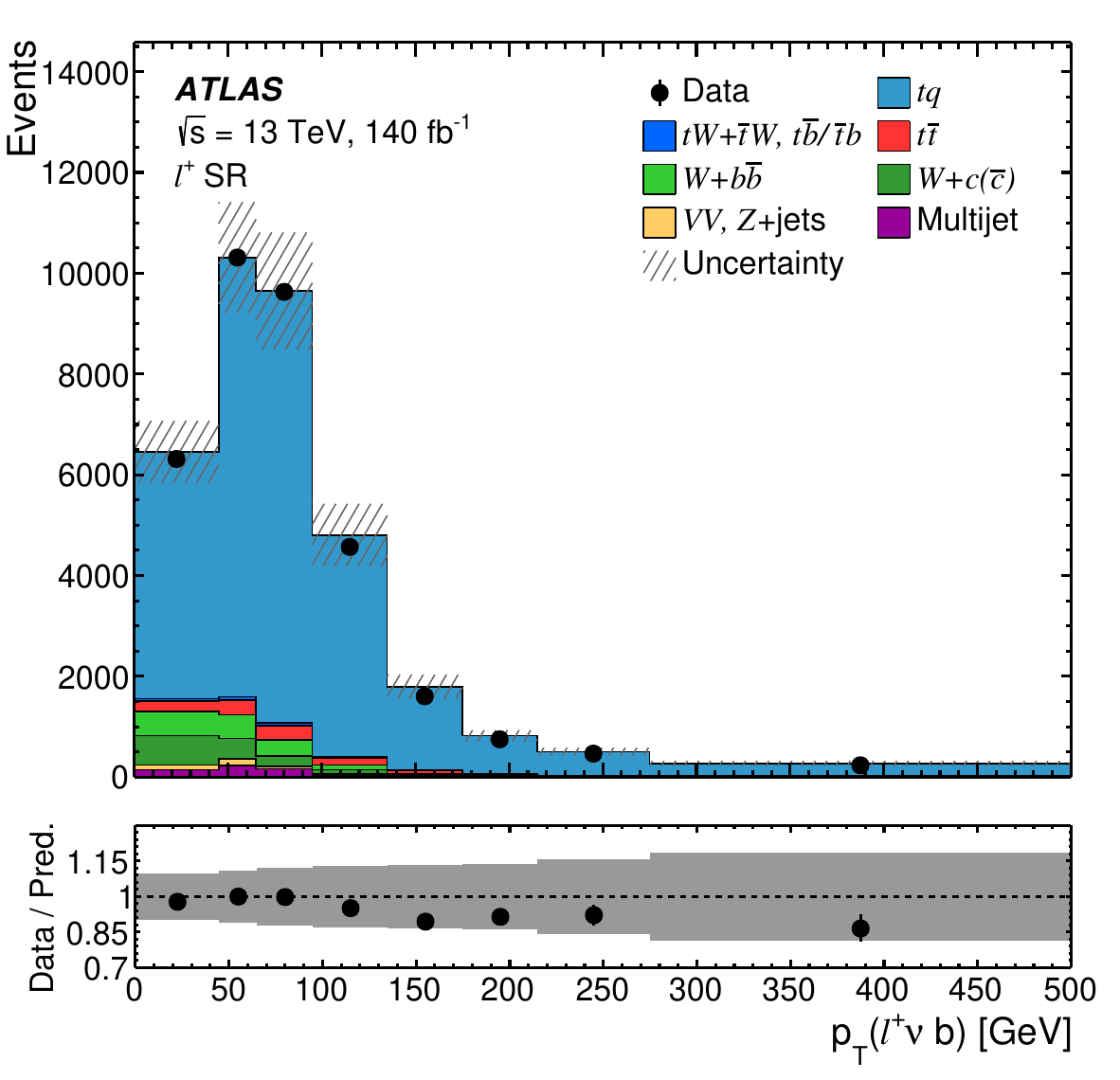}
\label{fig:DataMCcomppTSRp}
}
\subfigure[]{
\includegraphics[width=0.45\textwidth]{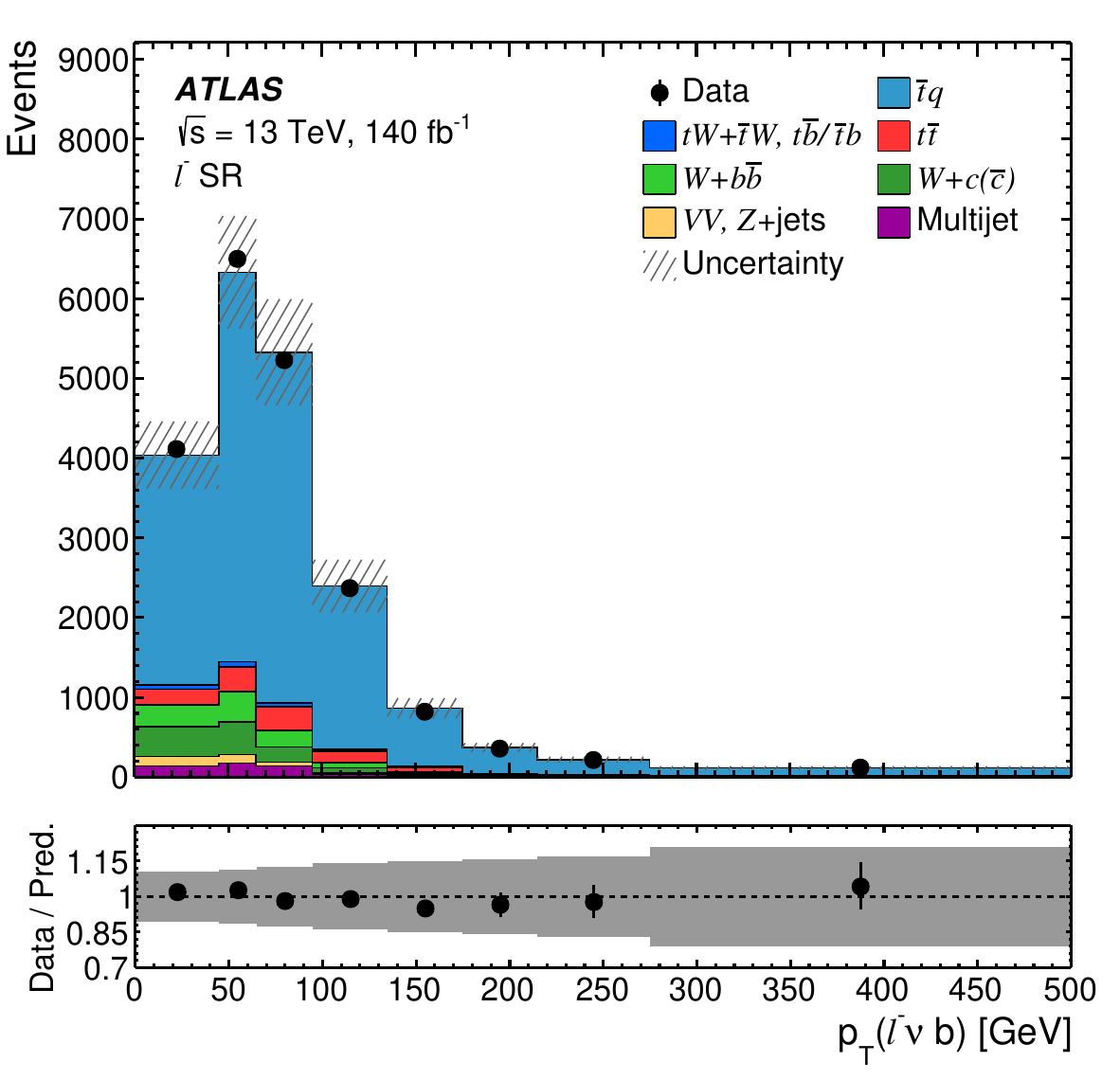}
\label{fig:DataMCcomppTSRn}
} \\
\subfigure[]{
\includegraphics[width=0.45\textwidth]{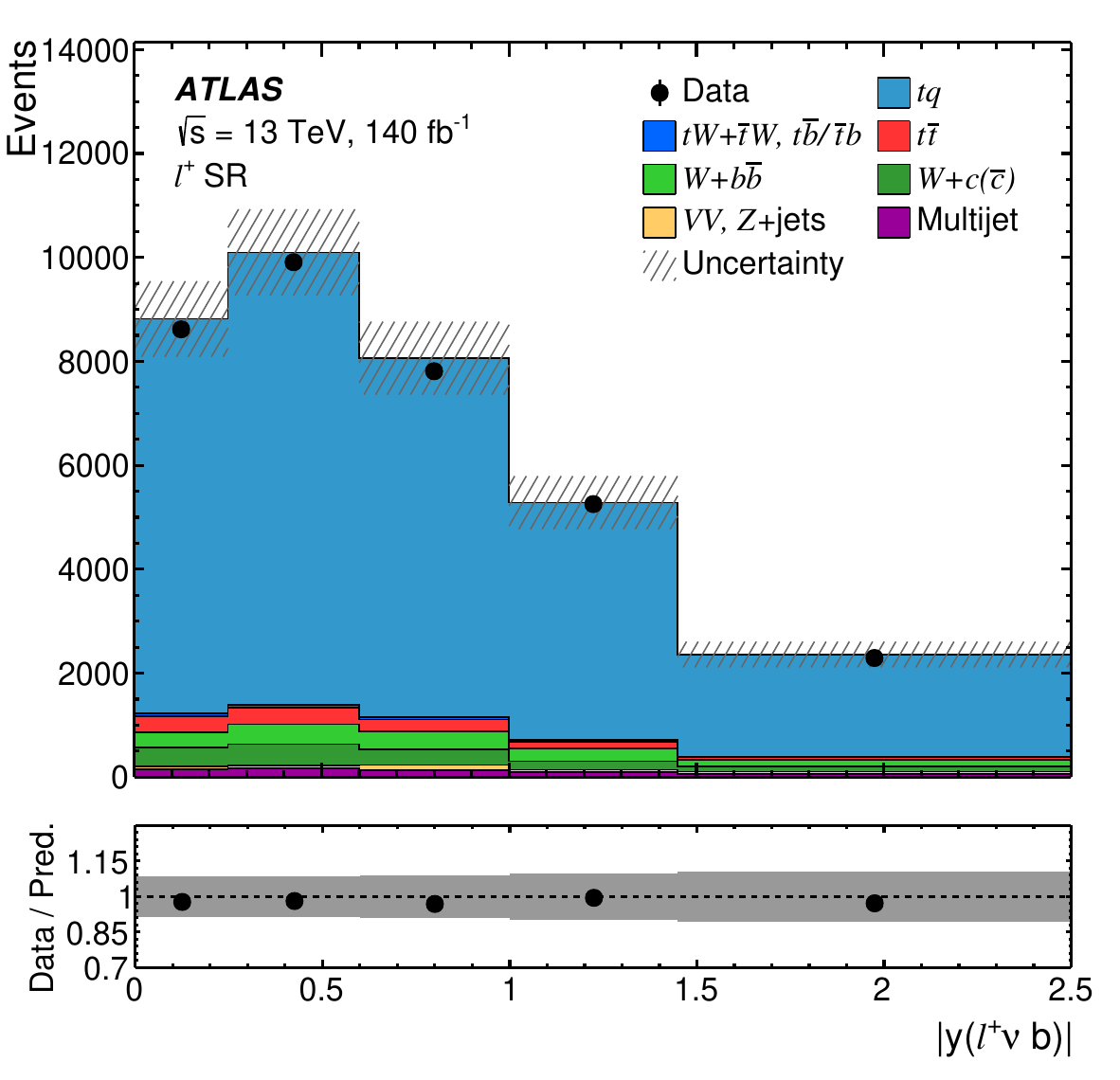}
\label{fig:DataMCcomprapSRp}
}
\subfigure[]{
\includegraphics[width=0.45\textwidth]{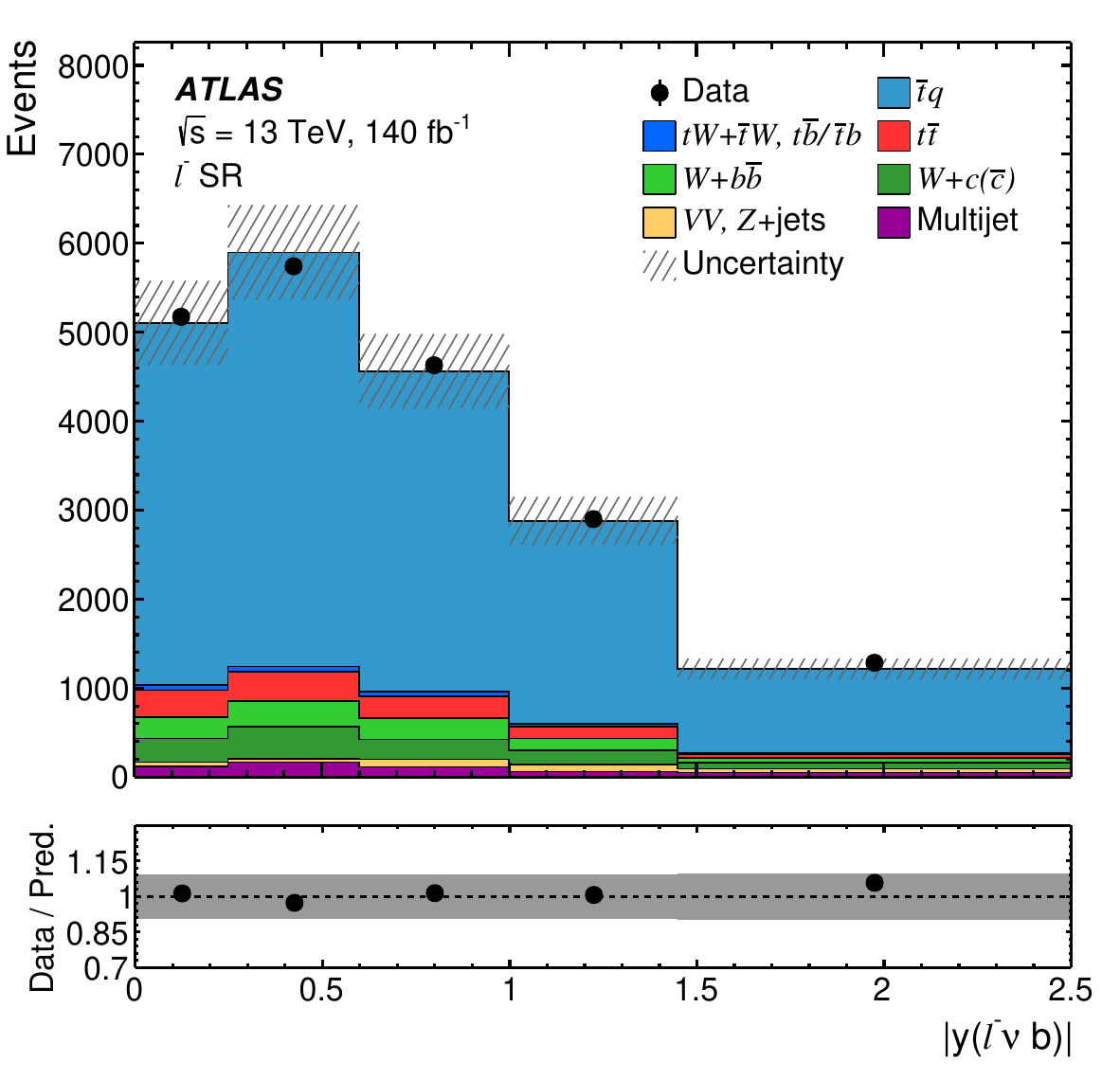}
\label{fig:DataMCcomprapSRn}
}
\caption{ Comparison of the measured data to the expected distributions estimated from MC simulation in the
\protect\subref{fig:DataMCcomppTSRp},\protect\subref{fig:DataMCcomprapSRp} $\ell^+$ SR and \protect\subref{fig:DataMCcomppTSRn},\protect\subref{fig:DataMCcomprapSRn} $\ell^-$ SR (i.e. the requirement $\Dnn > 0.93$ is imposed).
The distributions are shown for variables \protect\subref{fig:DataMCcomppTSRp} \pTtreco, \protect\subref{fig:DataMCcomppTSRn} \pTtbarreco,
\protect\subref{fig:DataMCcomprapSRp} \Raptreco and \protect\subref{fig:DataMCcomprapSRn} \Raptbarreco, utilising the binning optimised for the unfolding as described in \Sect{\ref{sec:inputs}}.
The uncertainty bands incorporate the MC statistical uncertainty and all systematic uncertainties.
The per-process scale factors are applied to the background processes.
}
\label{fig:DataMCcomp}
\end{figure}

\begin{table}[thb]
\sisetup{
table-align-uncertainty=true,
separate-uncertainty=true,
}
\caption{\label{tab:Yields} Expected event yields per process and observed event yields in the signal regions.
The given uncertainties are the rate uncertainties defined as the relative uncertainty in the post-fit event yields.
}
\begin{center}
\begin{tabular}{%
c
S[table-format=6.0,detect-weight]@{${}\pm{}$}S[table-format=5.0,detect-weight]
S[table-format=6.0,detect-weight]@{${}\pm{}$}S[table-format=5.0,detect-weight]
}
\toprule
Process & \multicolumn{2}{c}{$\ell^+$ SR} & \multicolumn{2}{c}{$\ell^-$ SR} \\ \midrule
$tq$ & 29740 & 590 & 15 & 15\\
$\bar{t}q$ & 7 & 7 & 15550 & 300\\
$t\bar{t}$ &  1020 & 40 & 1050 & 40\\
\tWschan& 200 & 20 & 210 & 21\\
$VV,Z$+jets &  290 & 60 & 300 & 60\\
$W+b\bar{b}$ &  1360 & 150 & 940 & 90 \\
$W+c(\bar{c})$ & 1340 & 270 & 1080 & 170\\
Multijet & 610 & 120 & 500 & 100 \\ \midrule
Total & 34580 & 680 & 19630 & 380 \\ \midrule
Observed &\multicolumn{2}{l}{\ \ \SI{33883}{}} &  \multicolumn{2}{l}{\ \ \SI{19730}{}}    \\
\bottomrule
\end{tabular}
\end{center}
\end{table}


\FloatBarrier

\section{Unfolding technique}
\label{sec:unfolding}

The unfolding aims to correct the measured distributions for resolution and efficiency effects caused by the detector.
D'Agostini's iterative approach~\cite{DAgostini:1994fjx}, implemented in RooUnfold~\cite{Adye:2011gm}, is used to unfold the distributions.
The distributions are unfolded to the parton level with stable top quarks, and extrapolated to the full kinematic phase space so that they can be easily compared with a variety of theory predictions.

The number of data events $N_j^\text{data}$ measured in bin $j$ of the distribution of the observable $X$ after detector reconstruction can be parameterised in the following way
\begin{equation*}
N_j^\text{data} = \sum_{k} M_{jk}\epsilon_k \mathcal{L}_\text{int}\Delta X_k \diff{\hat{\sigma}_k}{X_k} + \hat{B}_j.
\end{equation*}
Here $\mathcal{L}_\text{int}$ denotes the integrated luminosity and $\Delta X_k$ is the width of bin $k$ of the parton-level distribution of the observable $X$.
The migration matrix element $M_{jk}$ denotes the probability for a reconstructed event produced in bin $k$ to be reconstructed in bin $j$.
A strength of D'Agostini's approach is that the migration matrix is not inverted.
The nominal background $\hat{B}_j$ is the expected sum of background events in bin $j$.
The selection efficiency $\epsilon_k$ is defined as
\begin{equation*}
\epsilon_k = \frac{N_k^\text{sel}}{N_k^\text{prod}},
\end{equation*}
where $N_k^\text{prod}$ denotes the number of MC events produced in bin $k$ and $N_k^\text{sel}$ the number of events selected in bin $k$.
Since no fiducial region is defined on the final-state, $N_k^\text{sel}$ equals the number of events produced in bin $k$ that satisfy the event selection
at reconstruction level (cf.~\Sect{\ref{sec:selection}}) in an arbitrary bin $j$.


\FloatBarrier

\section{Binning and convergence of unfolding}
\label{sec:inputs}

The migration matrices and efficiencies used for the unfolding are determined with the nominal signal samples.
The binning of the distributions is optimised according to several criteria.
The statistical uncertainty in the data yields is kept at a moderate level, i.e. below 10\%, and the bin width is kept above the detector resolution.
The migration of events to off-diagonal bins is kept sufficiently low, i.e. below 50\%, to ensure the stability of the unfolding procedure.
The bin edges are synchronised between both signal regions to allow for the measurement of the cross-section ratio.

The obtained migration matrices are shown in the Appendix in \Fig{\ref{fig:MMs}}.
In both of the signal regions, a few data events, 7 in the $\ell^+$ SR and 8 in the $\ell^-$ SR, are measured with $\pTtortbarreco > \SI{500}{\gev}$.
The upper edge of the \pTtortbarreco distributions is not raised to incorporate all measured events because the differential cross-sections are normalised to the bin width.
Instead, a shadow bin with an upper edge of $\SI{2000}{\gev}$ is added to account for migration effects.
The cross-section is not measured in the shadow bin and it is therefore not shown in the following.

The unfolding procedure is regularised through the number of iterations.
The parton-level prediction for a given distribution is used as a biased prior and in each iteration replaced by the result of the previous iteration.
This procedure reduces the bias, i.e. the difference between the prediction and the true distribution, but leads to statistical fluctuations if too many iterations are performed.
The number of iterations used for unfolding is optimised by minimising the statistical correlation between the bins of the unfolded distributions.
For all distributions in all SRs, four iterations are chosen.

The unfolding procedure is validated to be robust against statistical fluctuations of the detector level distributions and changes in the shape of the signal process distributions.
The latter is tested by altering the shape of the signal process at both the parton and detector levels according to the difference between the expected and the observed distributions.
The altered detector level distributions are unfolded using the nominal migration matrix and efficiency and compared with the altered parton level distributions.
The difference between the unfolded results and the parton level, i.e. the non-closure, is found to be below 2\% for all distributions.


\FloatBarrier

\section{Systematic and statistical uncertainties}
\label{sec:uncertainties}

Several sources of systematic uncertainties affect the expected event yield from signal and
background processes, as well as the distribution's shapes.
The systematic uncertainties are divided into two major categories.
Experimental uncertainties are associated with detector limitations, calibrations and the reconstruction of the four-momenta of
final-state objects: electrons, muons, untagged jets, \btagged jets, and $\MET$.
The second category of uncertainties is related to the modelling of simulated processes.

\subsection{Sources of uncertainties}
The uncertainty in the integrated luminosity of the combined 2015--2018 data sample is 0.83\%~\cite{DAPR-2021-01}.
Scale factors are applied to simulated events to correct for reconstruction, identification,
isolation and trigger performance differences between data and detector simulation for electrons and muons~\cite{EGAM-2018-01,MUON-2018-03,EGAM-2021-02,MUON-2022-01}.
The net effect of the probability of charge misidentification of reconstructed electrons is a migration of events at the
level of \(6\times10^{-4}\) from the $\ell^+$ SR to the $\ell^-$ SR (cf.~\cite{TOPQ-2019-02}).
The uncertainty in the jet energy scale (JES)~\cite{JETM-2018-05} is decomposed into a set of 30 uncorrelated components.
Sources of uncertainty contributing to the JES uncertainty sources include the $\eta$ intercalibration of forward jets within $0.8< |\eta| < 4.5$ with those in the
central barrel region ($|\eta| < 0.8$), pile-up modelling, jet flavour composition and response, differences between jets induced by \bquarks and those from gluons or light-quarks,
single-particle response, detector modelling, non-closure, and effects of jets not fully contained within the calorimeter.
The uncertainty of the jet energy resolution (JER)~\cite{JETM-2018-05} comprises thirteen orthogonal components which account
for jet $\pt$- and $\eta$-dependent differences between simulation and data.
The uncertainty in the efficiency to satisfy the JVT requirement for pile-up suppression is also considered~\cite{PERF-2014-03}.
An uncertainty for \met is applied to account for a possible miscalibration of its soft-track component~\cite{PERF-2016-07}.
The \btag requirement made in the measurement requires consideration of uncertainties in the \btag efficiency of
true \bjets and in the mistagging rates of light-quark jets and \(c\)-jets.
The uncertainty in the \btag efficiency is decomposed into 45 orthogonal components~\cite{FTAG-2018-01}.
The uncertainty in the rate of mistagging \(c\)-jets as \bjets is decomposed into 20 orthogonal components~\cite{FTAG-2020-08}.
The uncertainty in the misidentification rate of light-quark jets is decomposed into 20 independent eigenvectors~\cite{FTAG-2019-02}.
To account for differences in the pile-up distribution between simulation and data, the pile-up profile
in the simulation is corrected to match the one in data.
The uncertainty associated with the correction factor is applied in the measurement as a variation
of the event weight.
In the following, these uncertainties are summarised as \textit{Experimental} uncertainty.

The rate uncertainties of the background processes are defined as described in \Sect{\ref{sec:background}} and summarised as \textit{Background rate} uncertainty.
Uncertainties in modelling parton showers and hadronisation are assigned to all top-quark production processes by comparing
the nominal samples with alternative samples for which \POWHEGBOX[v2] was interfaced to
\HERWIG[7.2.1]~\cite{Bahr:2008pv,Bellm:2015jjp} (for \ttbar production), \HERWIG[7.1.6] (for $\tW$ and $t\bar{b}/\bar{t}b$ production) or
\HERWIG[7.3.1] (for \tqtbarq production) instead of \PYTHIA[8.230].
Uncertainties related to the choice of the renormalisation scale \(\muR\) and the factorisation scale \(\muF\) for the matrix-element
calculations are evaluated by varying the scales independently by factors of
\(2\) and \(0.5\), separately for each of the top-quark production processes and
for \Wjets production.
The uncertainty in matching the NLO matrix elements to the parton shower when generating
\(t\tbar\), \tW and \tqtbarq events is evaluated according to the prescription in \Refn{\cite{Hoche:2021mkv}}.
The uncertainty in the choice of the $h_\mathrm{damp}$ parameter for the \(t\tbar\) and $\tW$ event
generation is estimated by using additional samples produced with the $h_\mathrm{damp}$
parameter set to $3\times m_t$.
Uncertainties in the amount of initial-state and final-state radiation are assessed for the
top-quark production processes by varying the parameter \texttt{Var3c} of the A14 parton-shower tune
within the uncertainties of the tune and, for final-state radiation, by varying \(\muR\) by factors of $0.5$ and $2.0$. The two variations are handled independently.
The uncertainty due to the scheme for removing the interference of the \tW process with
\(\ttbar\) production is evaluated by comparing the nominal sample, using the diagram-removal scheme,
with a sample produced with an alternative scheme (diagram subtraction)~\cite{Frixione:2008yi}.
Uncertainties in the PDFs are evaluated for the top-quark production processes using the
PDF4LHC15 prescription with 30 eigenvectors~\cite{Butterworth:2015oua}.
The uncertainty in the shape of the multijet background is evaluated by modifying the selection
criteria for jet-electron and non-prompt-muon candidates, thereby generating templates for the systematically varied distributions.
In the following plots, these uncertainties are summarised as \textit{Signal modelling} and \textit{Background modelling}.

\subsection{Evaluation of uncertainties}
To evaluate the impact of the experimental uncertainties, the rate uncertainties of the background processes and uncertainties related to the modelling of background processes,
the expected event yield is estimated from the systematically varied distributions at reconstruction level.
If an uncertainty is not defined for a process, the nominal prediction is used for the estimate.
The nominal background estimate is then subtracted and the resulting distributions are unfolded with the nominal migration matrix and efficiency.
The difference between the unfolded result and the unfolded nominal signal distribution per bin is taken as the uncertainty.
For uncertainties related to the modelling of the signal process, the systematically varied signal distribution is unfolded with the nominal migration matrix and efficiency
and any difference with the systematically varied prediction at parton level is taken as the uncertainty.

The statistical uncertainty of the unfolded data is determined via pseudo-experiments.
Each bin of the data distribution is fluctuated according to the expected statistical uncertainty to construct a pseudo-experiment.
The pseudo-experiment is then unfolded with the nominal migration matrix and efficiency and the standard deviation of all unfolded pseudo-experiments is taken as the data's statistical uncertainty per bin.

To evaluate the statistical uncertainty due to the limited size of the MC signal sample, each bin of the migration matrix and efficiency is fluctuated according to a Gaussian distribution
whose spread corresponds to the expected uncertainty due to the number of MC events in that bin.
The nominal signal is unfolded with the fluctuated corrections and the standard deviation of all unfolded pseudo-experiments is taken as the MC statistical uncertainty.

One-sided uncertainties, such as the evaluation of the parton shower, are symmetrised by taking the full uncertainty with opposite sign as down variation.
The total uncertainty in the unfolded cross-section is calculated per bin by adding the individual uncertainties in quadrature.
The contributions from the individual uncertainties with positive signs are added to the total up variation
and the contributions with negative signs are added to the total down variation.
If both the up and the down variation of an uncertainty have a positive (negative) sign, the larger contribution is added to the total up (down) variation and the contribution to the total down (up) variation is set to zero.

The uncertainties in the ratio of the differential cross-sections are evaluated after the unfolding, taking into account correlations between the uncertainties in the measurements.

The statistical uncertainties and the rate uncertainties in the $W+b\bar{b}$ and $W+c(\bar{c})$ background processes are decorrelated for $tq$ and $\bar{t}q$ production and therefore added in quadrature.

The statistical uncertainty in the single components of the JES, JER and modelling related uncertainties are evaluated with the bootstrap method~\cite{ATL-PHYS-PUB-2021-011} as these uncertainties show fluctuations in all measured distributions.
It was found that the fluctuations can be attributed to the statistical uncertainty of the systematic uncertainties, but the quoted uncertainties are consistent with the central values obtained with the bootstrap method.
The statistical uncertainties of the JES and JER uncertainties are larger than those of the modelling related uncertainties and, expressed in terms of the fractional uncertainty in the measured cross-sections, reach magnitudes of about $\pm0.5\%$ for the \Rapt or \Raptbar distributions and about $\pm1.0\%$ for the \pTt or \pTtbar distributions.
The statistical uncertainties may be of interest for future combinations and are published in the HEPData entry for this analysis.

Summaries of the statistical and systematic uncertainties in all measured differential cross-sections are presented in \Figs{\ref{fig:CommonSys}}{\ref{fig:CommonSysNorm}}.
The contributions of the different categories of uncertainties with a further breakdown of the experimental uncertainties are provided in the Appendix in \Tabrange{\ref{tab:CommonSubSyspTSRp}}{\ref{tab:CommonSubSysRatiorap}}.
The uncertainties in the absolute and normalised differential cross-sections are dominated by the signal-modelling-related and experimental uncertainties.
The statistical uncertainties become significant in the high-\pT region of the absolute differential cross-section distributions, in the normalised differential cross-sections and in the cross-section ratios.
For the latter, this can be attributed to the cancellation of systematic uncertainties.

\begin{figure}[htbp]
\centering
\subfigure[]{
\includegraphics[width=0.45\textwidth]{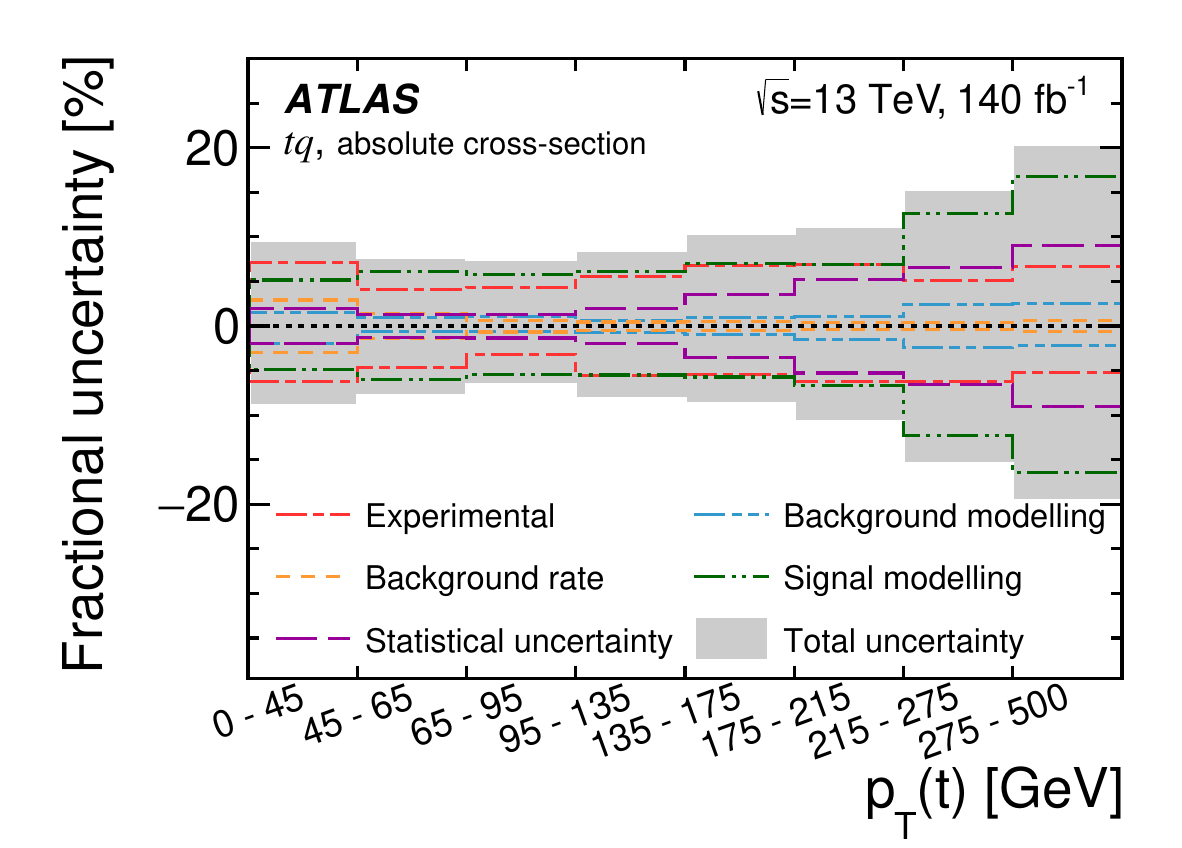}
\label{fig:CommonSyspTSRp}
}
\subfigure[]{
\includegraphics[width=0.45\textwidth]{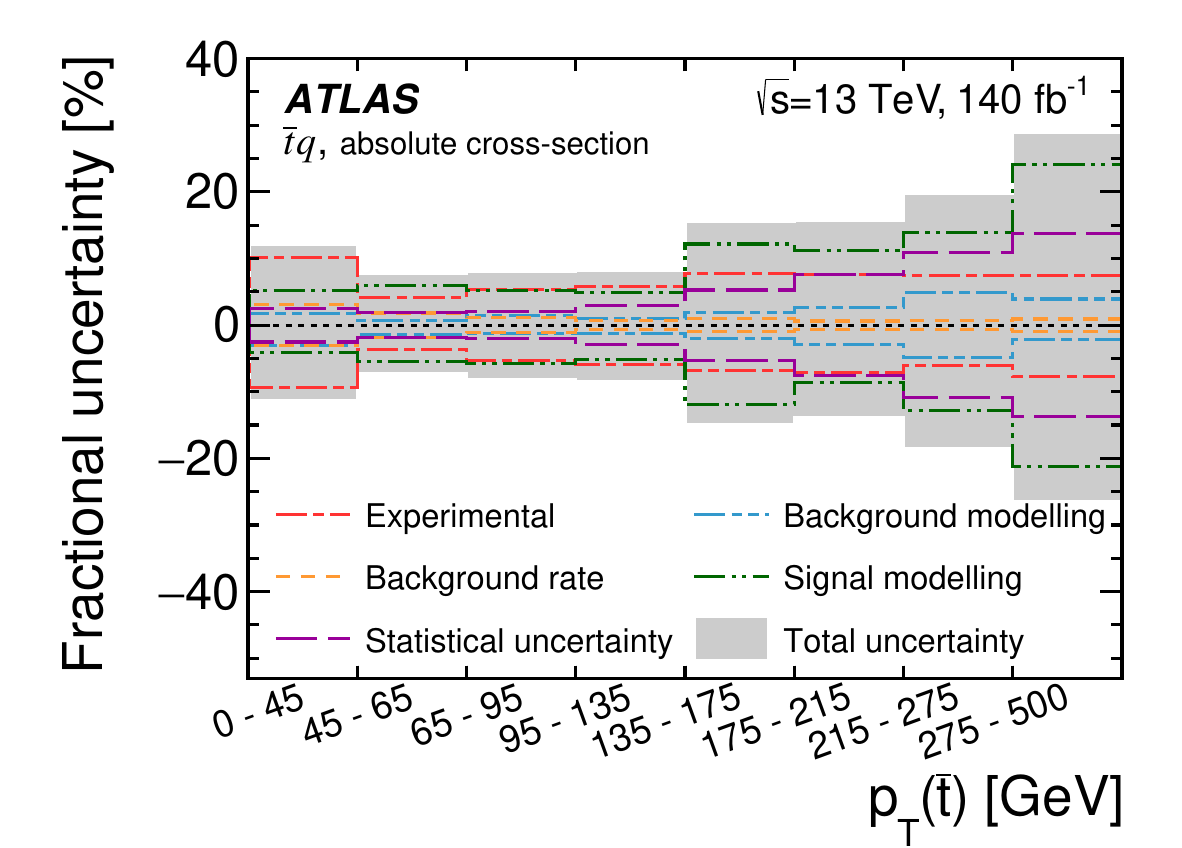}
\label{fig:CommonSyspTSRn}
} \\
\subfigure[]{
\includegraphics[width=0.45\textwidth]{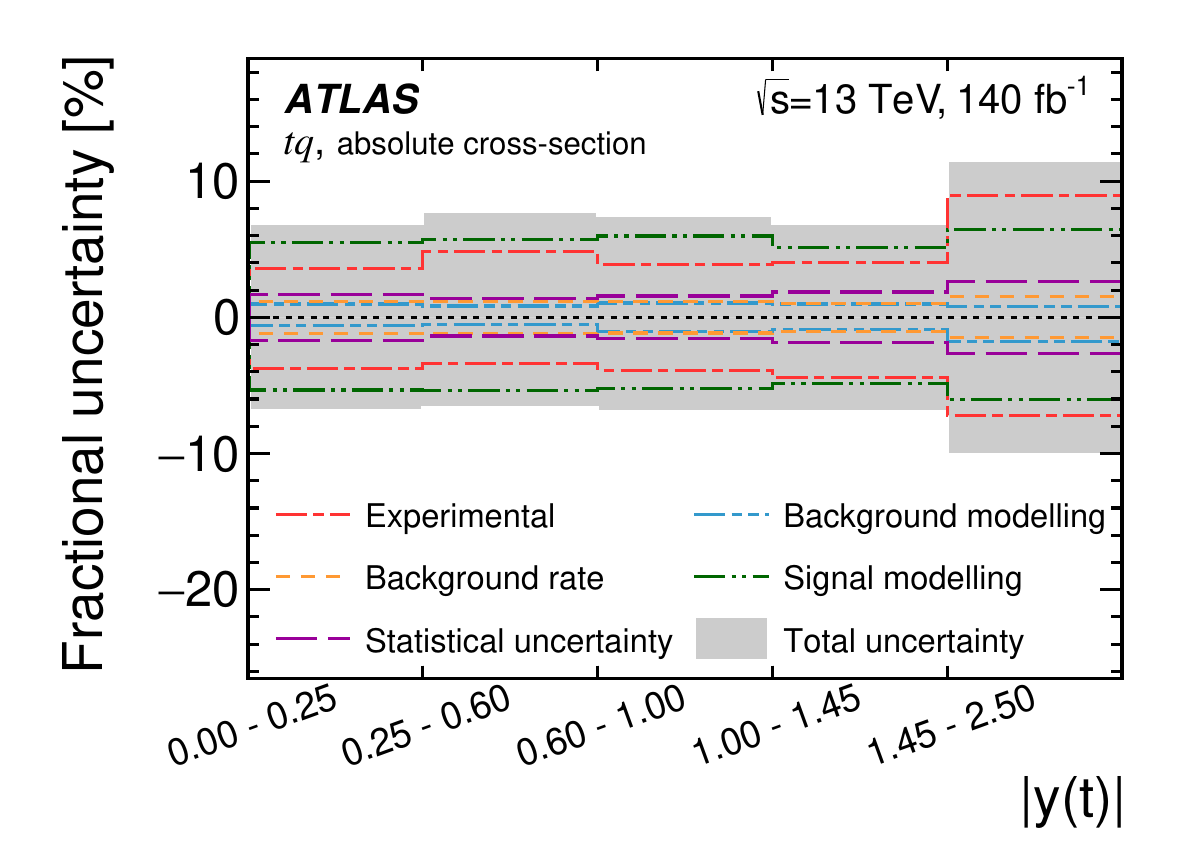}
\label{fig:CommonSysrapSRp}
}
\subfigure[]{
\includegraphics[width=0.45\textwidth]{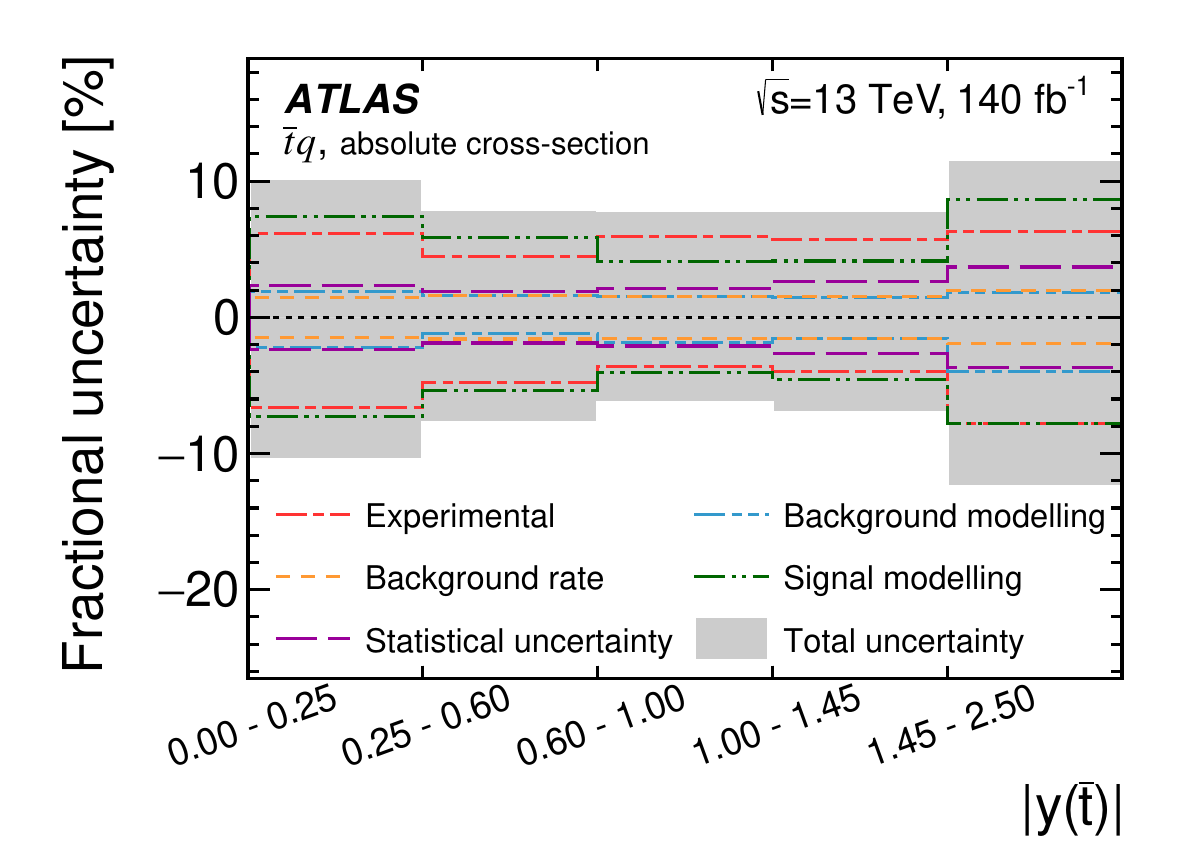}
\label{fig:CommonSysrapSRn}
} \\
\subfigure[]{
\includegraphics[width=0.45\textwidth]{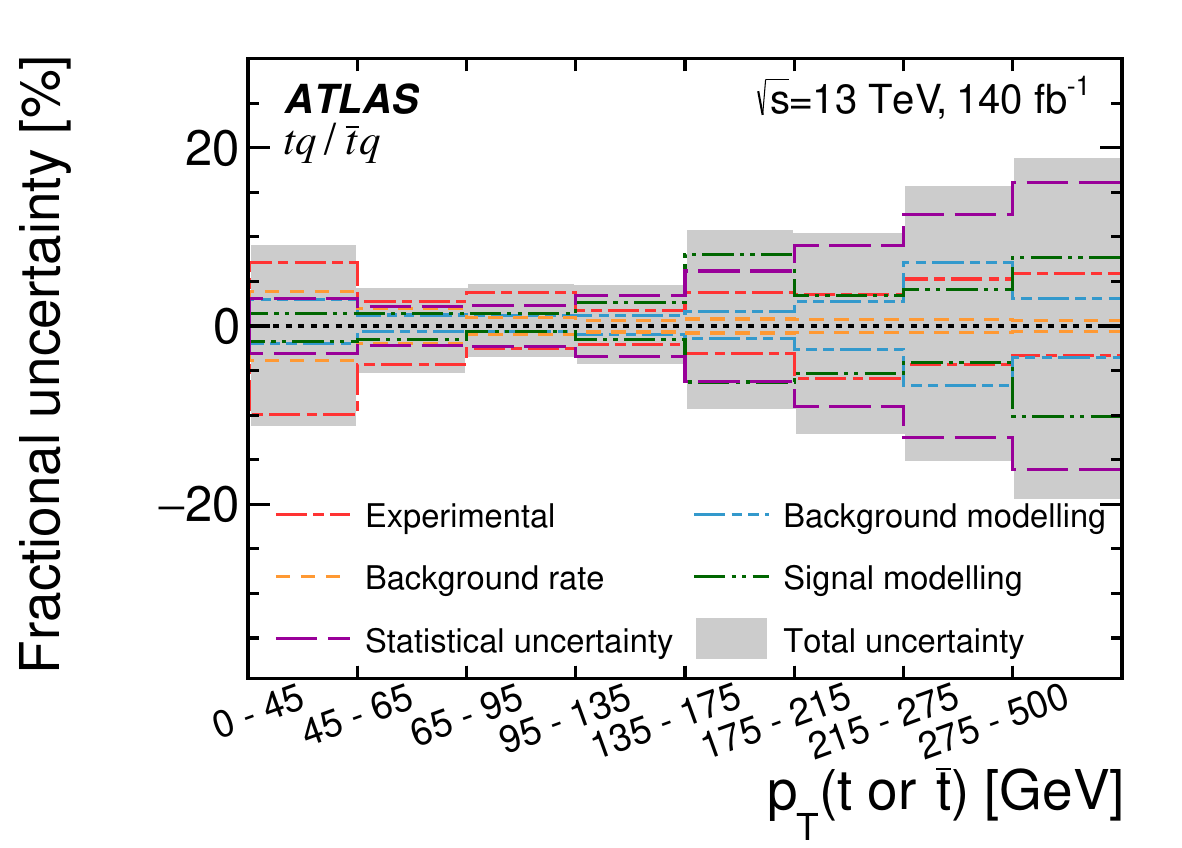}
\label{fig:CommonSysRatiopT}
}
\subfigure[]{
\includegraphics[width=0.45\textwidth]{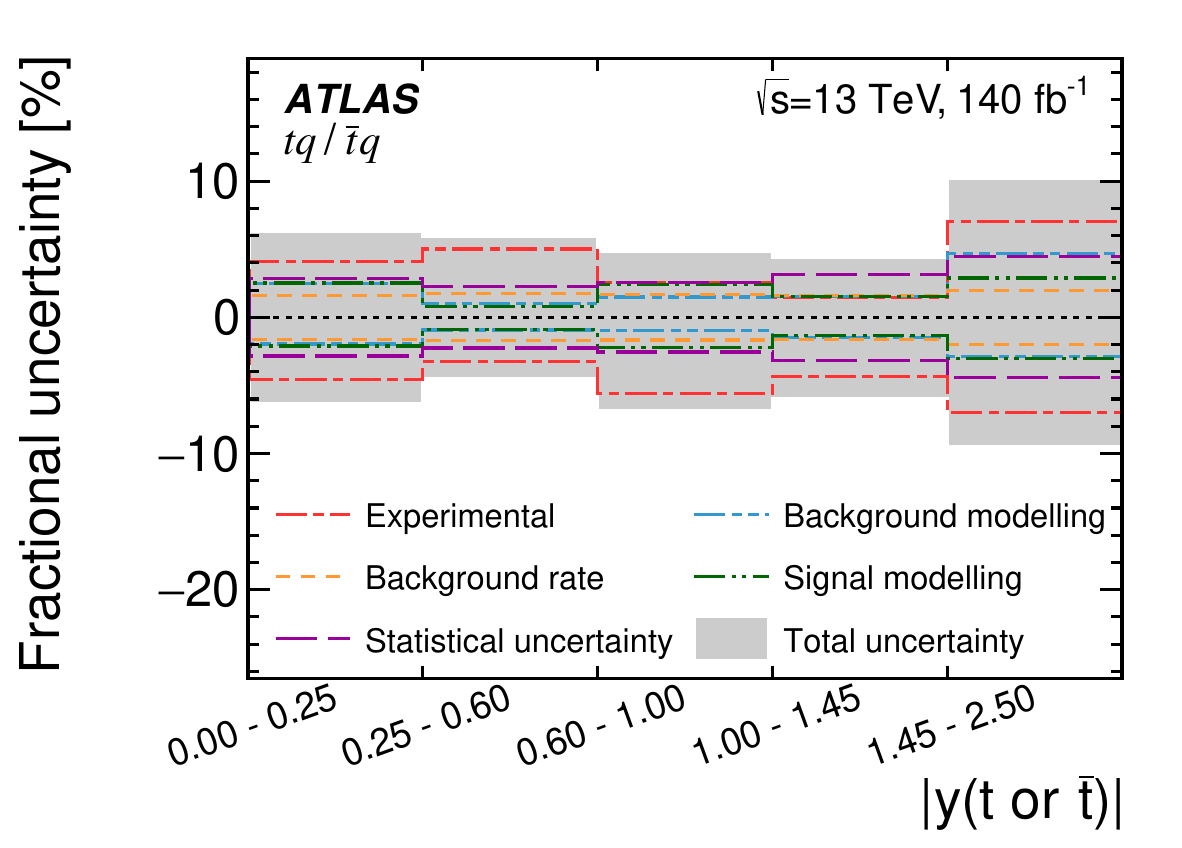}
\label{fig:CommonSysRatiorap}
}
\caption{ Summary of the uncertainties in the absolute differential \tqortbarq production cross-sections measured as a function of \protect\subref{fig:CommonSyspTSRp} \pTt,
\protect\subref{fig:CommonSyspTSRn} \pTtbar, \protect\subref{fig:CommonSysrapSRp} \Rapt and \protect\subref{fig:CommonSysrapSRn} \Raptbar and on the
ratio of the differential \tqtbarq production cross-sections measured as a function of \protect\subref{fig:CommonSysRatiopT} $p_\text{T}(\tortbar)$
and \protect\subref{fig:CommonSysRatiorap} $|y(\tortbar)|$.
Similar sources of uncertainties are grouped using their quadratic sum.
The shaded band indicates the total uncertainty, defined as the quadratic sum of all sources of uncertainties.
}
\label{fig:CommonSys}
\end{figure}

The uncertainties related to the rate and modelling of background processes are minor uncertainties for all measured cross-sections.
Their contributions differ between the $tq$ and $\bar{t}q$ production cross-sections as the fraction of background events differ between the signal regions.

\begin{figure}[htbp]
\centering
\subfigure[]{
\includegraphics[width=0.45\textwidth]{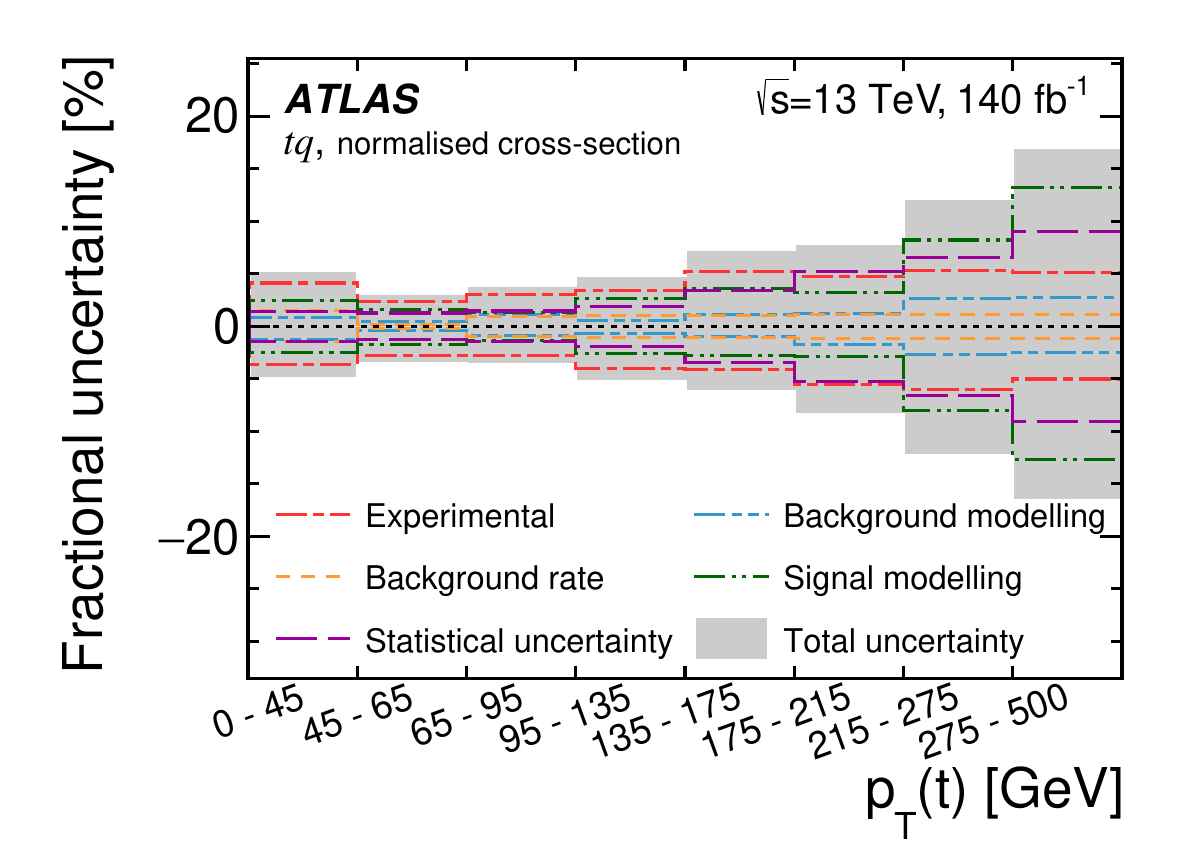}
\label{fig:CommonSysNormpTSRp}
}
\subfigure[]{
\includegraphics[width=0.45\textwidth]{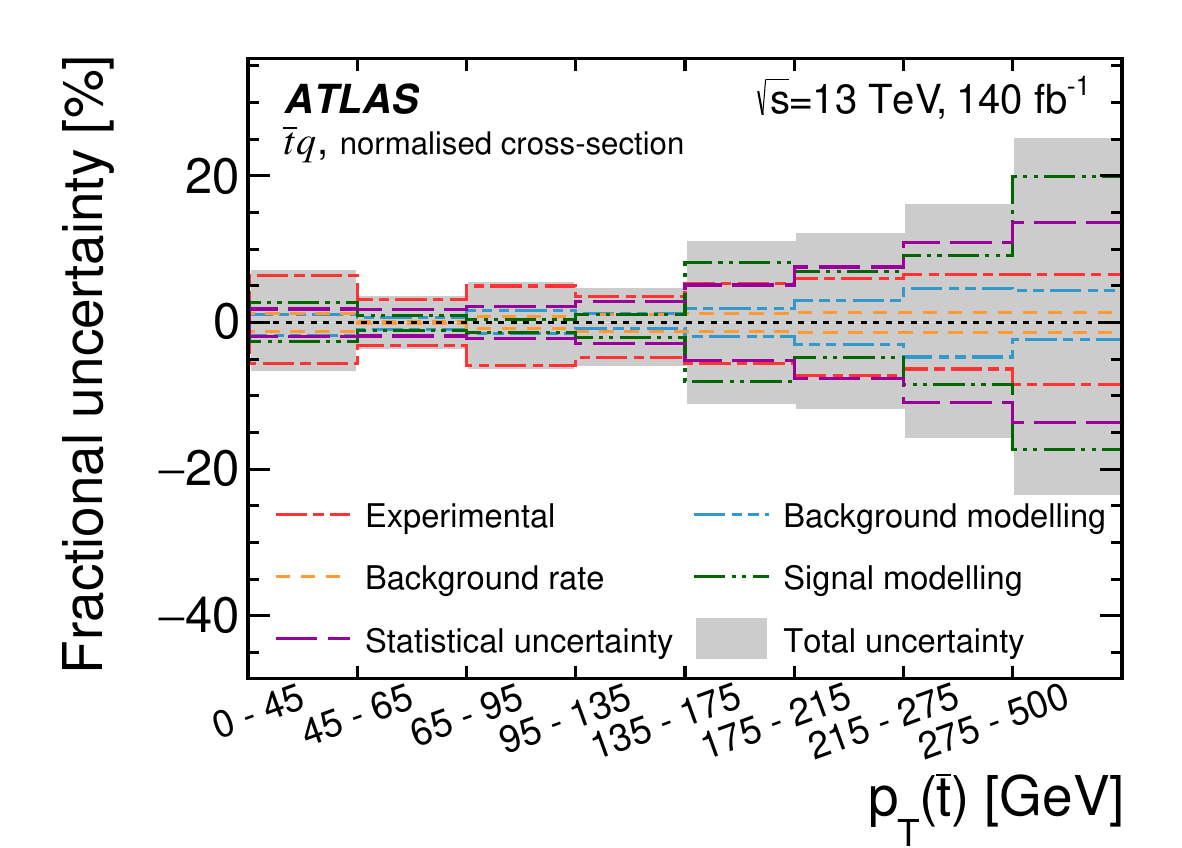}
\label{fig:CommonSysNormpTSRn}
} \\
\subfigure[]{
\includegraphics[width=0.45\textwidth]{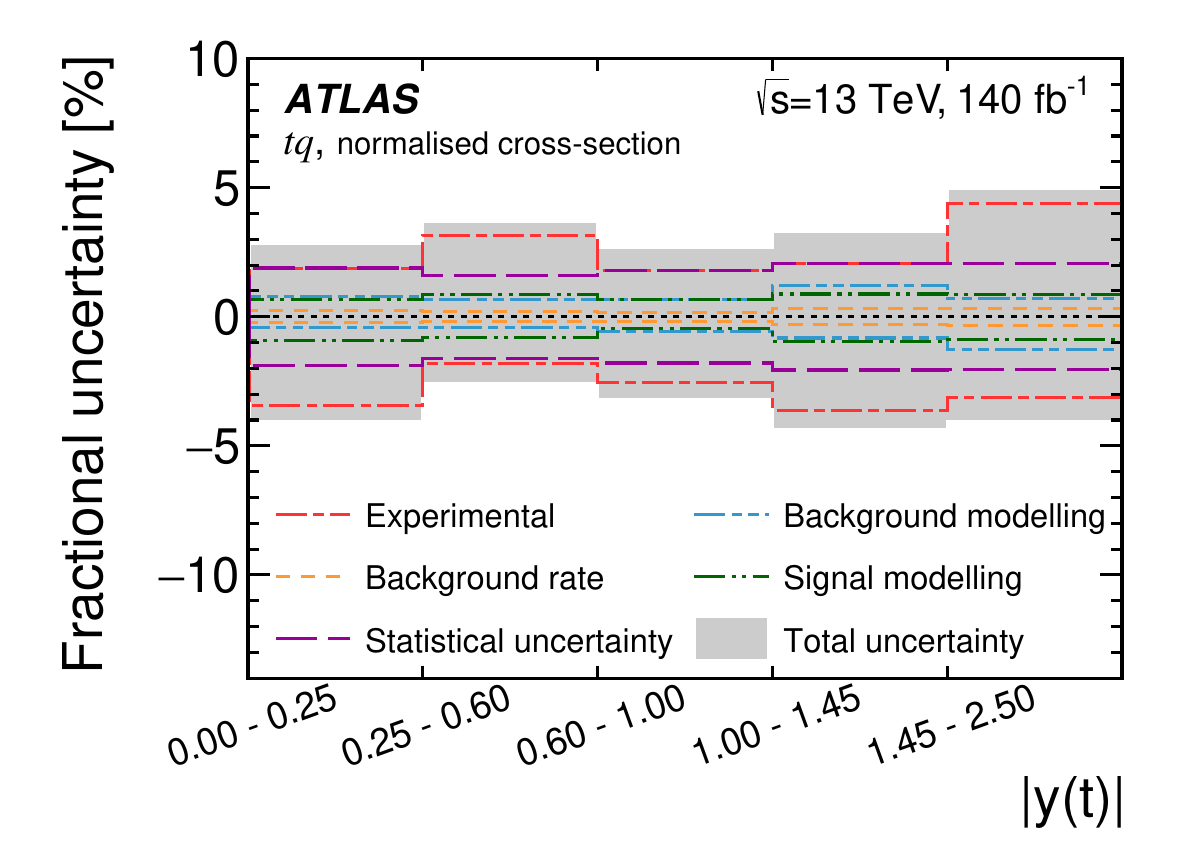}
\label{fig:CommonSysNormrapSRp}
}
\subfigure[]{
\includegraphics[width=0.45\textwidth]{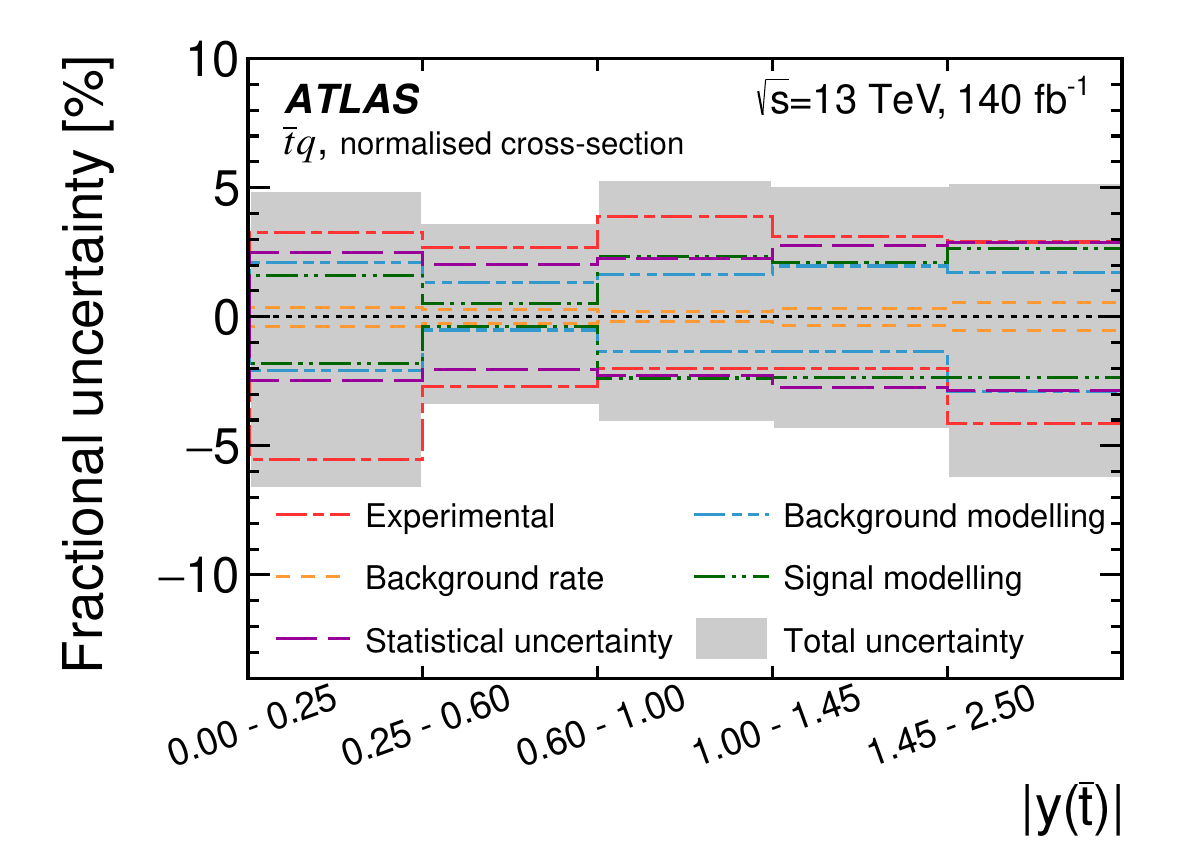}
\label{fig:CommonSysNormrapSRn}
}
\caption{ Summary of the uncertainties in the normalised differential \tqortbarq production cross-sections measured as a function of \protect\subref{fig:CommonSysNormpTSRp} \pTt,
\protect\subref{fig:CommonSysNormpTSRn} \pTtbar, \protect\subref{fig:CommonSysNormrapSRp} \Rapt and \protect\subref{fig:CommonSysNormrapSRn} \Raptbar.
Similar sources of uncertainties are grouped using their quadratic sum.
The shaded band indicates the total uncertainty, defined as the quadratic sum of all sources of uncertainties.
}
\label{fig:CommonSysNorm}
\end{figure}


\newpage
\FloatBarrier

\section{Results}
\label{sec:results}

Differential \tqtbarq production cross-sections are measured as a function of \pTt or \pTtbar and \Rapt or \Raptbar.
The absolute differential cross-sections are presented in \Figrange{\ref{fig:TotalResultsMCFMDDIS}}{\ref{fig:TotalResultsAllGenerators}},
the normalised differential cross-sections are shown in \Figrange{\ref{fig:NormalisedResultsMCFM_DDIS}}{\ref{fig:NormalisedResultsAllGenerators}}
and the ratios of the cross-sections are shown in \Fig{\ref{fig:ResultsRatio}}.
All measurement results are presented together with the theoretical predictions.
The numerical values of the measured cross-sections are given in the Appendix in \Tabrange{\ref{tab:ResultsptSRp}}{\ref{tab:ResultsrapRatio}}
and the $\chi^2$ probabilities for all theory predictions are listed in \Tabs{\ref{tab:Chi2pT}}{\ref{tab:Chi2rap}}.
The $\chi^2$ value of a given theory prediction is calculated as
\begin{equation*}
\chi^2 = V_{N_b}^T \cdot \text{Cov}_{N_b}^{-1} \cdot V_{N_b},
\end{equation*}
where $N_b$ denotes the number of bins, $V_{N_b}$ denotes the vector of differences between the unfolded data and the prediction and
$\text{Cov}_{N_b}$ denotes the covariance matrix which incorporates all uncertainties in the measurement result described in \Sect{\ref{sec:uncertainties}}.

The covariance matrix is obtained by performing pseudo-experiments. Each bin
of the measured distribution at detector level is varied according to a Poisson distribution with the
number of observed events in this bin as the mean value. Gaussian-distributed shifts are coherently added
for the MC statistical uncertainty, each experimental uncertainty and each background related uncertainty by scaling each
Poisson-fluctuated bin with the expected relative variation from the associated uncertainty effect.
The resulting distribution is unfolded with the nominal migration matrix and efficiency.
Signal modelling uncertainties are then coherently added with additional Gaussian-distributed shifts.
For the normalised cross-sections, $V_{N_b}$ and $\text{Cov}_{N_b}$ are replaced by $V_{N_b-1}$ and $\text{Cov}_{N_b-1}$ by discarding the last element of $V_{N_b}$ and the corresponding row and column of $\text{Cov}_{N_b}$.

The uncertainties in the theoretical predictions are added in quadrature on the diagonal elements of $\text{Cov}_{N_b}$.
For the predictions obtained with different PDF sets, the MC statistical uncertainty and the PDF uncertainty are considered
and the uncertainty in the predictions from different event generators additionally contains the uncertainty in the renormalisation and factorisation scale choice.
The uncertainty in the FO calculations comprises the uncertainty in the renormalisation and factorisation scale choice and the uncertainty in the value of \alphas.

Overall, a good agreement between the measured cross-sections and the theoretical predictions is observed.
The FO calculations with MCFM at LO do not agree with the measured \pTt and \pTtbar spectra as expected for a calculation missing additional gluon radiation.
The predictions at NLO and NNLO show very good agreement with the unfolded data, with the NNLO predictions being generally closer to the data and showing a better compatibility.
For the predictions obtained from different MC event generators, larger deviations between the predictions are only present in the high-\pT region of the absolute and differential cross-sections measured as a function of \pTt or \pTtbar,
but the deviations are mostly covered by the measurement uncertainties.

The predictions obtained with different PDF sets for the absolute differential cross-sections as a function of \Rapt or \Raptbar differ from each other by up to 10\%,
but all agree with the measured result within the measurement uncertainties.
The measured normalised differential production cross-sections as a function of \Rapt or \Raptbar agree very well with the different predictions, although some deviations are visible for the differential $\bar{t}q$ cross-section.
The $\chi^2$ probabilities of the different predictions for both the normalised and absolute cross-sections are very similar to each other for a given distribution,
except ABMP16, which has a noticeably lower probability than the other predictions for the $\bar{t}q$ production cross-sections binned in \Raptbar.

The predictions obtained with the ABMP16 PDF set overestimates the ratio of the differential cross-sections over the full measured \Rapt and \Raptbar ranges.
The other predictions are in excellent agreement with the measurement results.

At $\sqrt{s}=$\SI{8}{\TeV}~\cite{TOPQ-2015-05}, differential \tqtbarq production cross-sections were already measured separately as a function of \pTt or \pTtbar and \Rapt or \Raptbar.
In contrast, the present measurement is performed at a different centre-of-mass energy and, in addition, achieves substantially higher precision for both the absolute and normalised differential cross-sections.
Important factors for the improvement are the larger number of data events and a reduction of the contribution from background processes and their associated uncertainties.
The greatest improvement is achieved for the normalised differential cross-sections as a function of \Rapt or \Raptbar, where this measurement reaches a precision of \SI{3}{\%} to \SI{6}{\%} compared with \SI{7}{\%} to \SI{15}{\%}.
A previous analysis by CMS using a partial Run 2 data sample~\cite{CMS-TOP-17-023} measured combined \tqtbarq production cross-sections.
This analysis provides the first separate measurement of differential \tqtbarq production at $\sqrt{s}=$\SI{13}{\TeV} and reaches a higher precision, as it utilises the full Run 2 data sample.

In both measurements discussed above, the differential cross-sections as a function of \pTt or \pTtbar were measured in a range of up to \SI{300}{\GeV}.
This analysis measures the differential cross-sections up to \SI{500}{\GeV}, providing a significant extension of the probed \pTtortbar region.

\begin{figure}[htbp]
\centering
\subfigure[]{
\includegraphics[width=0.45\textwidth]{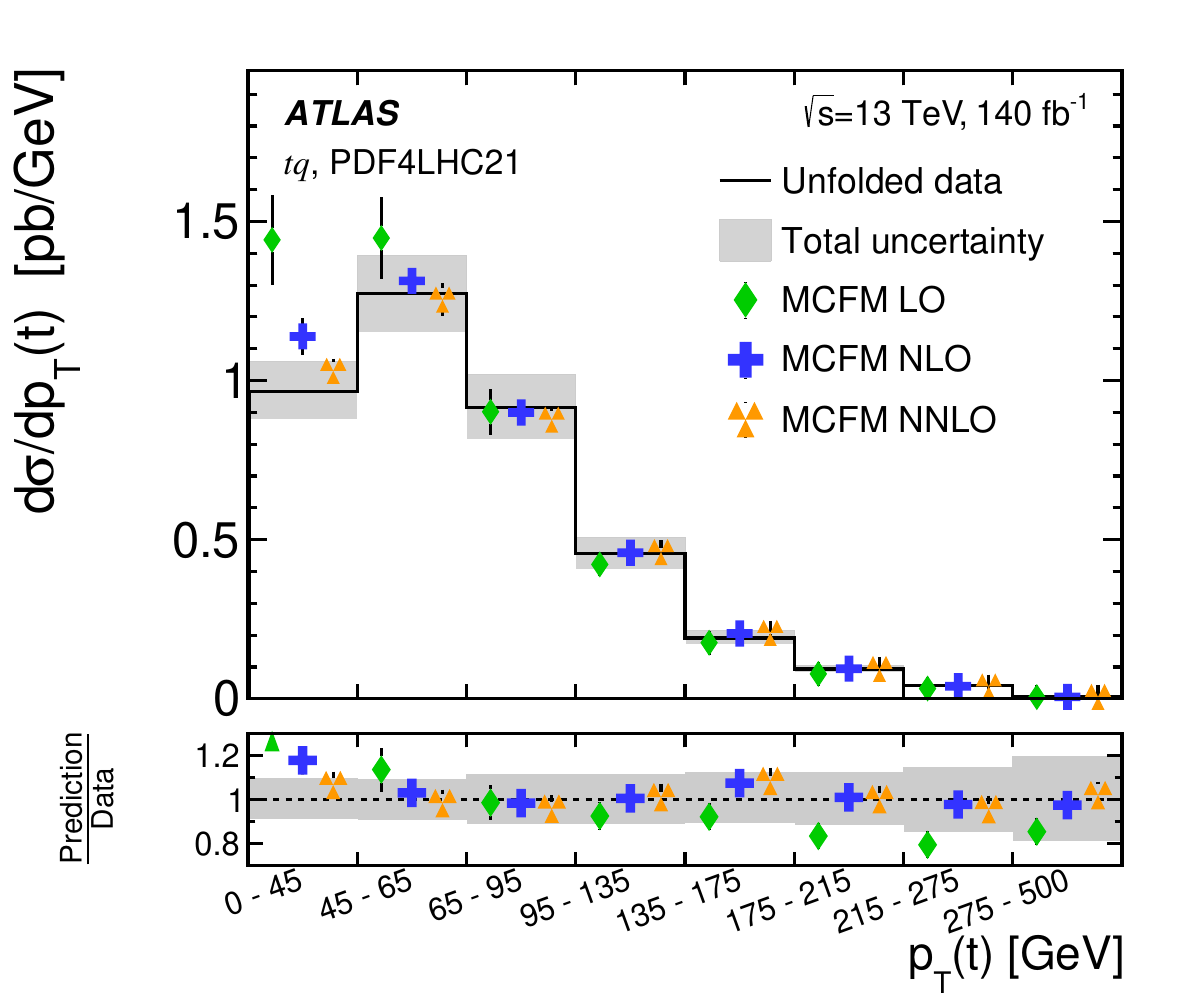}
\label{fig:TotalResultpTSRpMCFMDDIS}
}
\subfigure[]{
\includegraphics[width=0.45\textwidth]{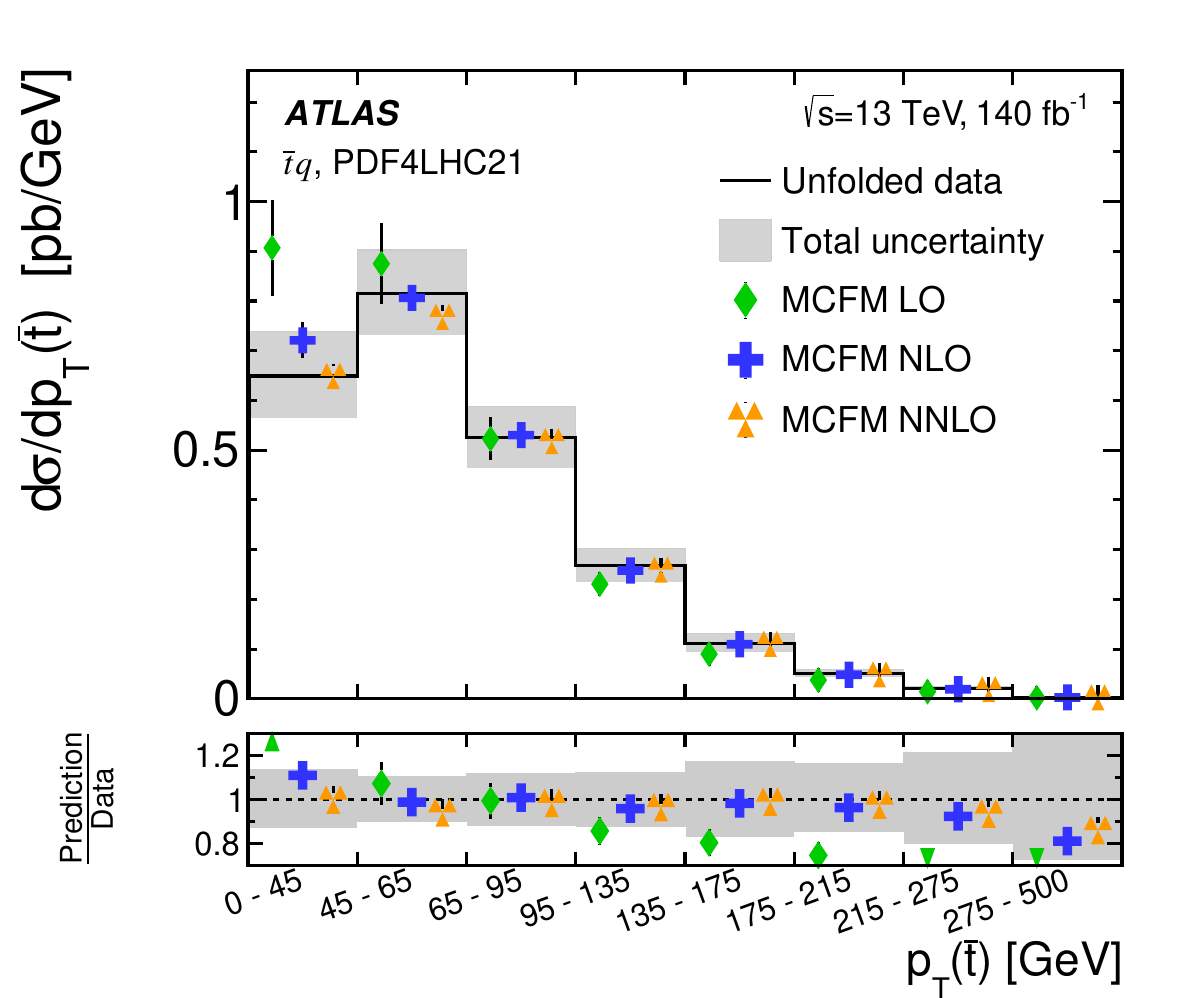}
\label{fig:TotalResultpTSRnMCFMDDIS}
} \\
\subfigure[]{
\includegraphics[width=0.45\textwidth]{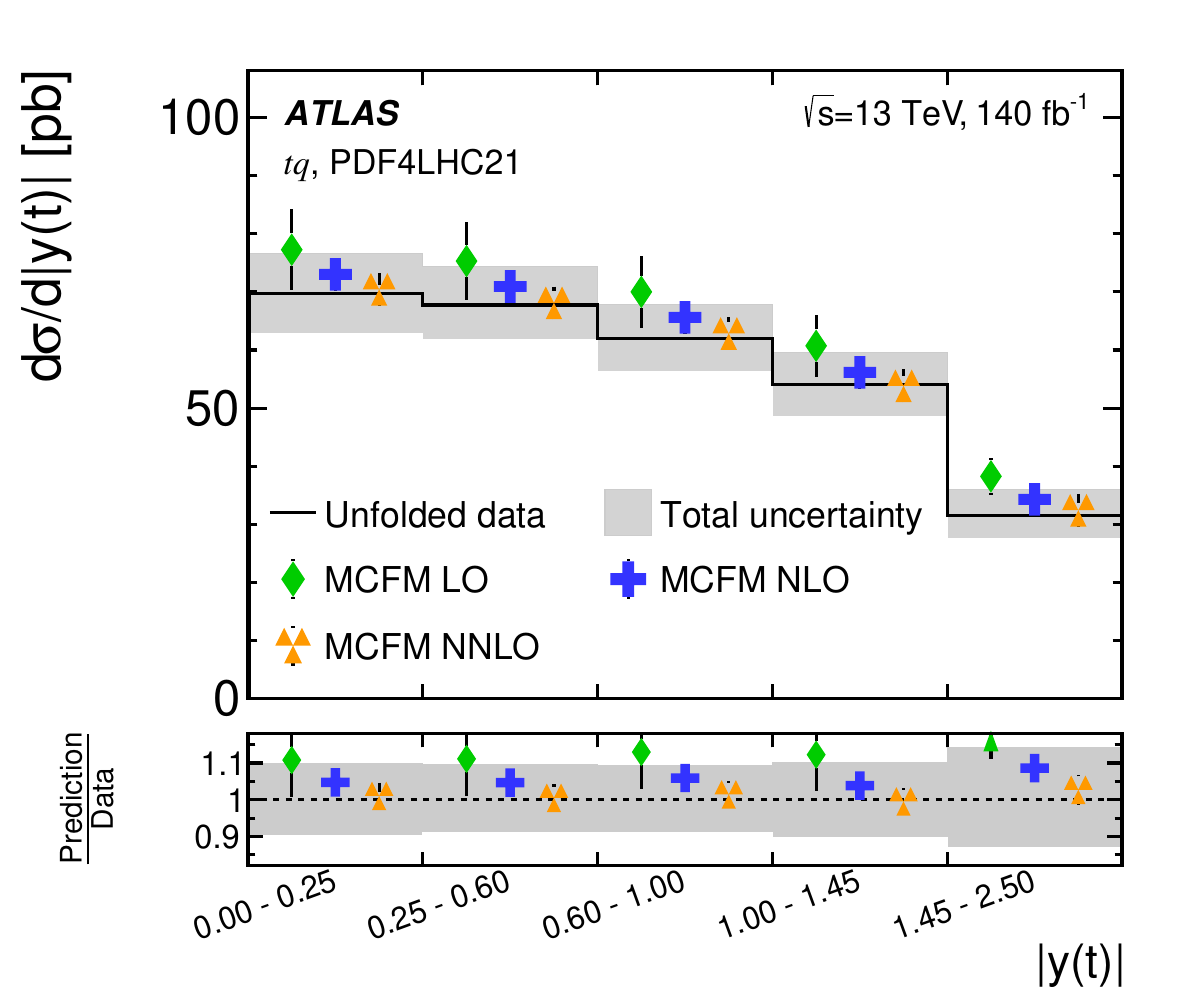}
\label{fig:TotalResultrapSRpMCFMDDIS}
}
\subfigure[]{
\includegraphics[width=0.45\textwidth]{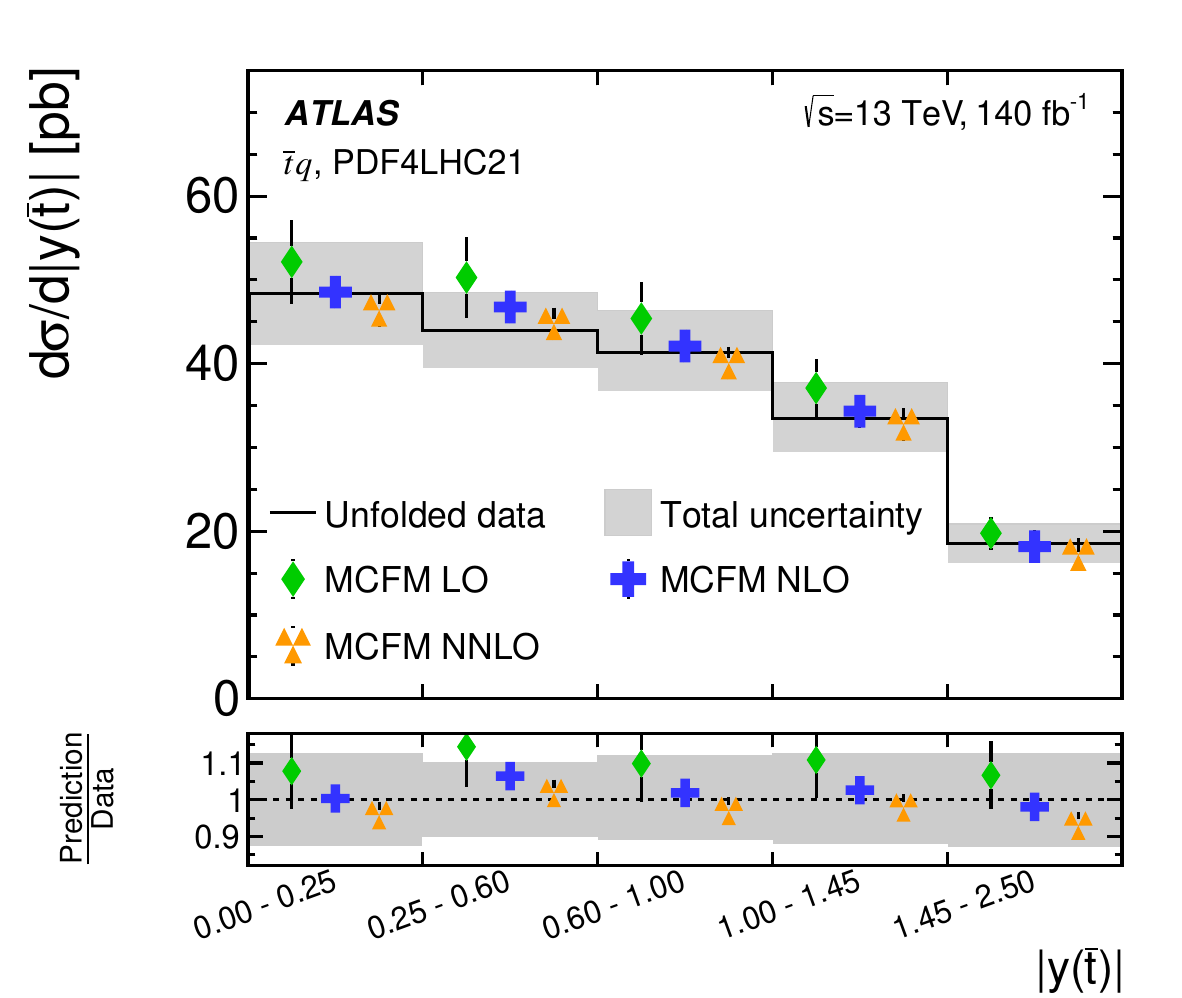}
\label{fig:TotalResultrapSRnMCFMDDIS}
}\\

\caption{Absolute unfolded differential \protect\subref{fig:TotalResultpTSRpMCFMDDIS} \protect\subref{fig:TotalResultrapSRpMCFMDDIS} $tq$ cross-sections and
\protect\subref{fig:TotalResultpTSRnMCFMDDIS} \protect\subref{fig:TotalResultrapSRnMCFMDDIS} $\bar{t}q$ cross-sections as a function of
\protect\subref{fig:TotalResultpTSRpMCFMDDIS} \protect\subref{fig:TotalResultpTSRnMCFMDDIS} \pTt or \pTtbar and
\protect\subref{fig:TotalResultrapSRpMCFMDDIS} \protect\subref{fig:TotalResultrapSRnMCFMDDIS} \Rapt or \Raptbar.
The shaded band indicates the total measurement uncertainty.
The cross-sections are compared with the theoretical predictions from FO calculations done with MCFM at LO, NLO and NNLO.
The uncertainty in the theoretical predictions comprises the uncertainty in the renormalisation and factorisation scale choice and the uncertainty in the value of \alphas.
}
\label{fig:TotalResultsMCFMDDIS}
\end{figure}

\begin{figure}[htbp]
\centering
\subfigure[]{
\includegraphics[width=0.45\textwidth]{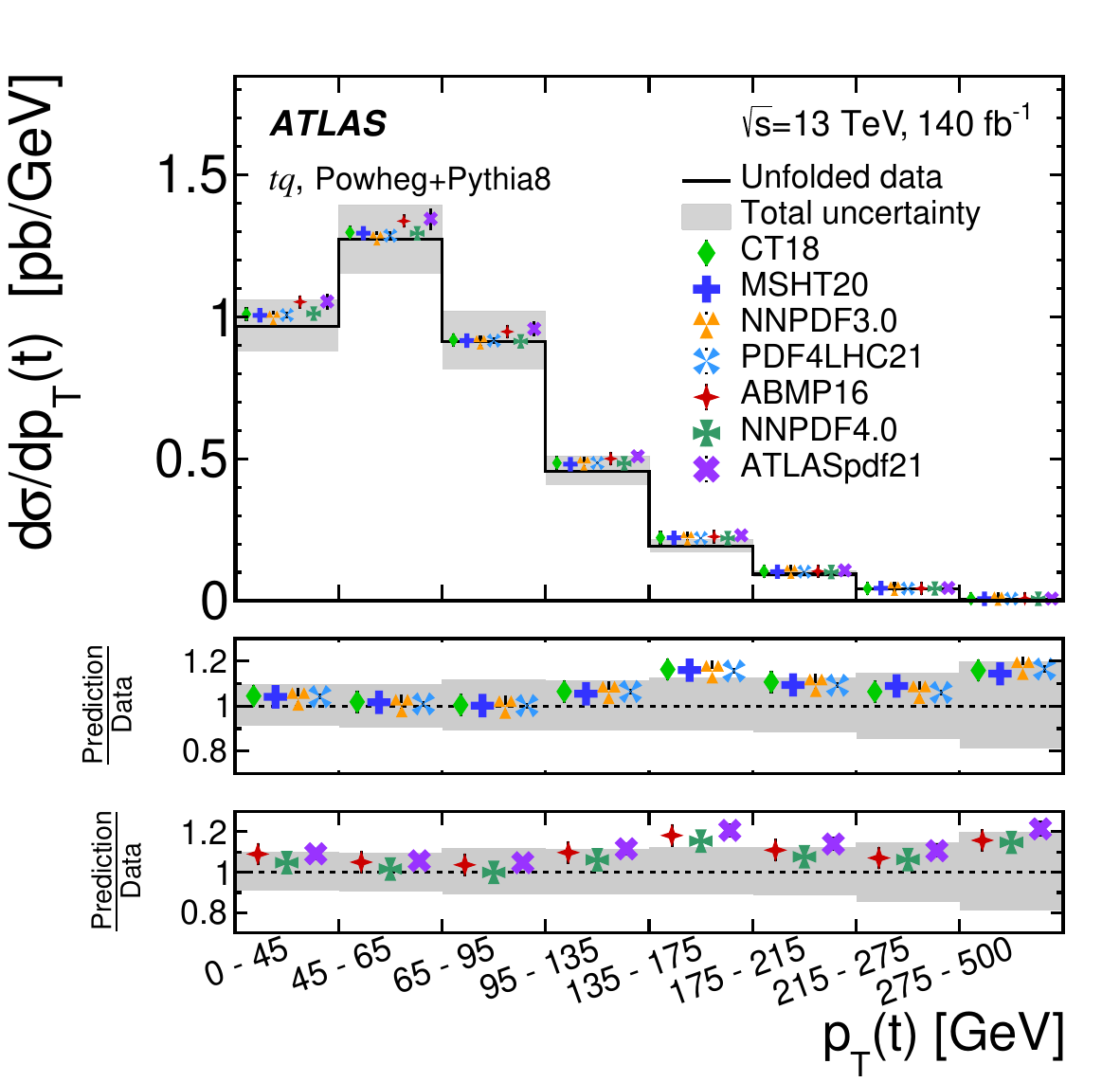}
\label{fig:TotalResultpTSRpAllPDFs}
}
\subfigure[]{
\includegraphics[width=0.45\textwidth]{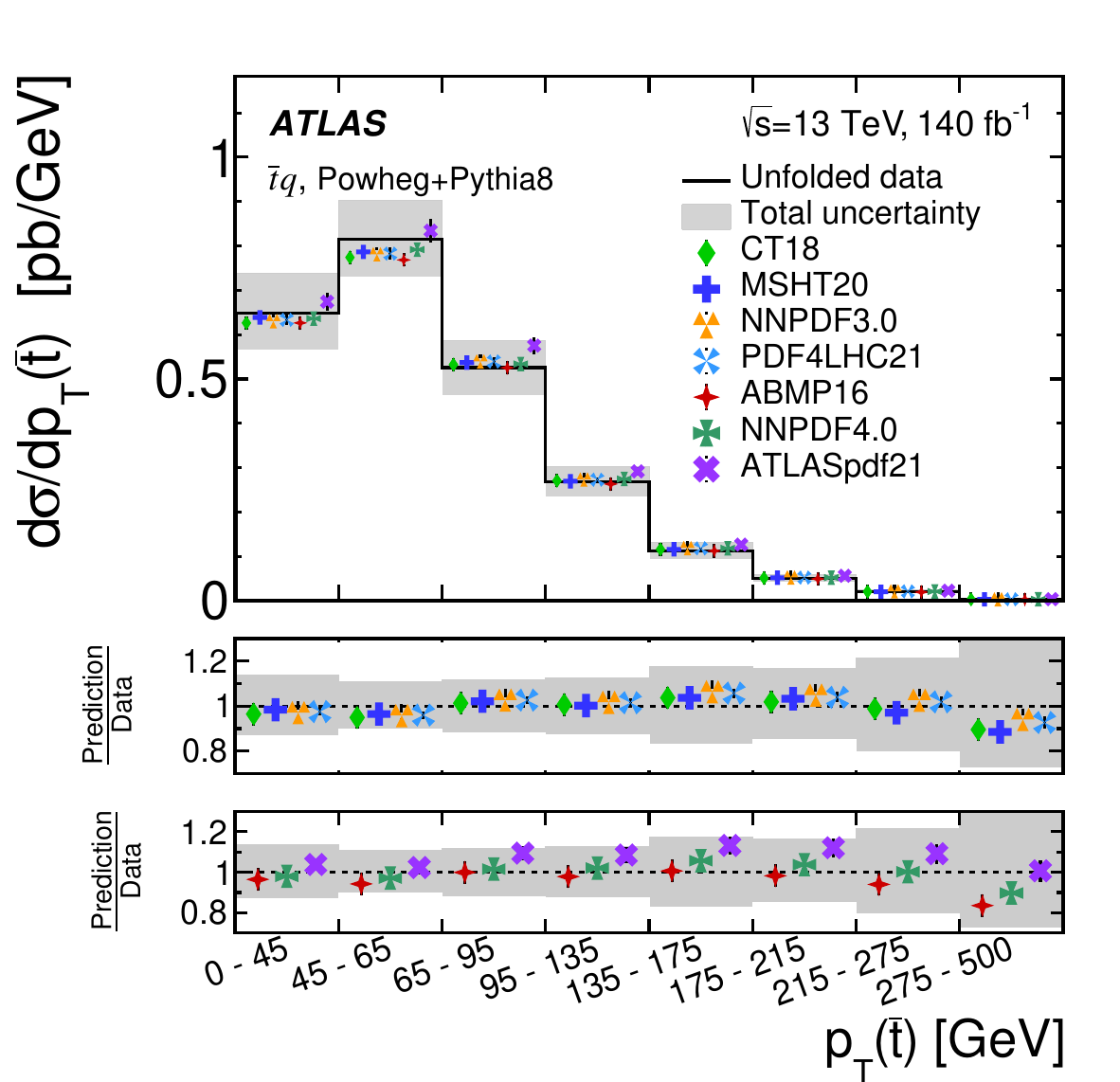}
\label{fig:TotalResultpTSRnAllPDFs}
} \\
\subfigure[]{
\includegraphics[width=0.45\textwidth]{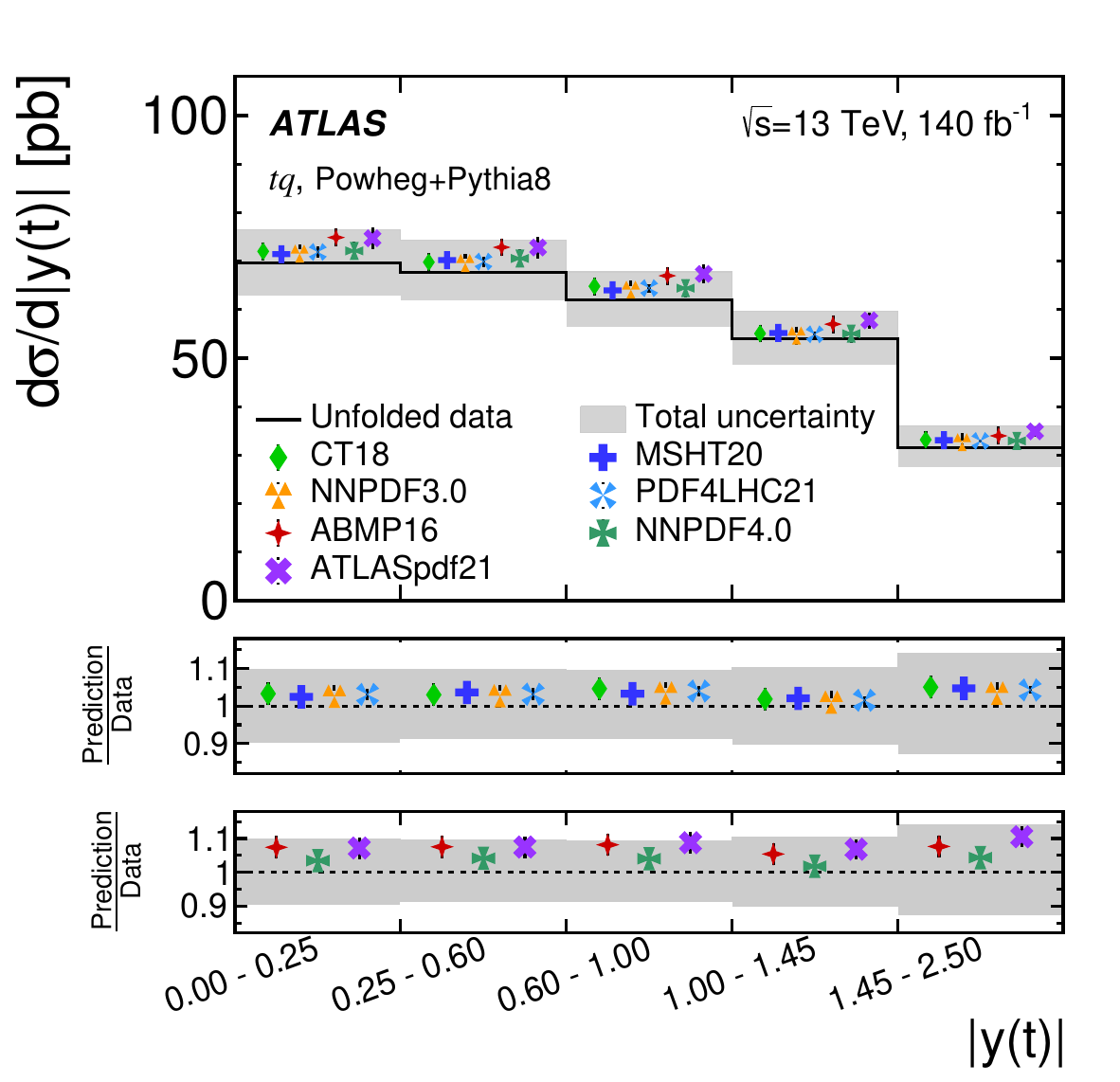}
\label{fig:TotalResultrapSRpAllPDFs}
}
\subfigure[]{
\includegraphics[width=0.45\textwidth]{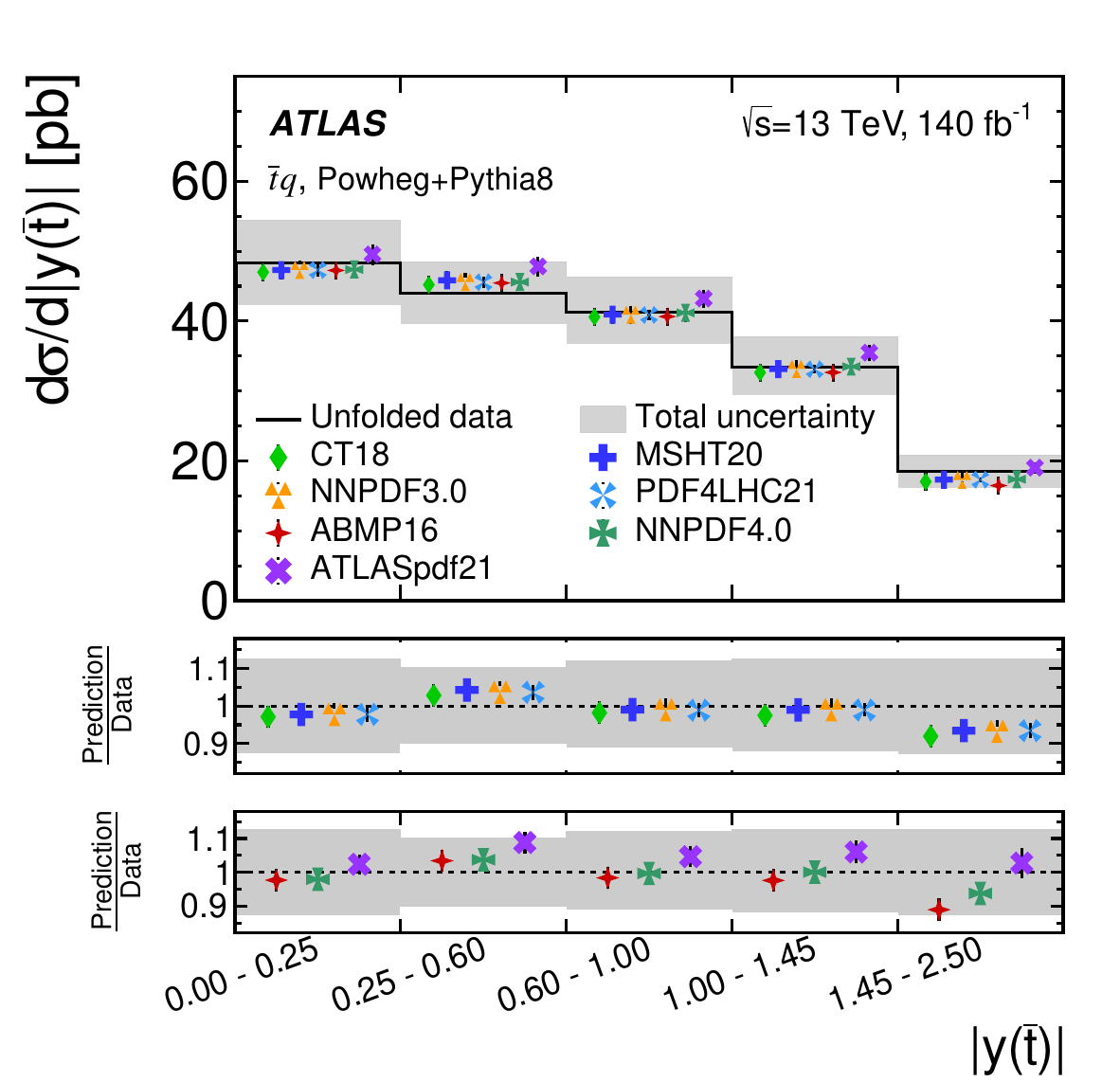}
\label{fig:TotalResultrapSRnAllPDFs}
}\\
\caption{Absolute unfolded differential \protect\subref{fig:TotalResultpTSRpAllPDFs} \protect\subref{fig:TotalResultrapSRpAllPDFs} $tq$ cross-sections and
\protect\subref{fig:TotalResultpTSRnAllPDFs} \protect\subref{fig:TotalResultrapSRnAllPDFs} $\bar{t}q$ cross-sections as a function of
\protect\subref{fig:TotalResultpTSRpAllPDFs} \protect\subref{fig:TotalResultpTSRnAllPDFs} \pTt or \pTtbar and
\protect\subref{fig:TotalResultrapSRpAllPDFs} \protect\subref{fig:TotalResultrapSRnAllPDFs} \Rapt or \Raptbar.
The shaded band indicates the total measurement uncertainty.
The cross-sections are compared with the predictions from \POWHEG+\PYTHIA[8] with different PDF sets.
The uncertainty in the theoretical predictions comprises the MC statistical uncertainty and the PDF uncertainty.
}
\end{figure}

\begin{figure}[htbp]
\centering
\subfigure[]{
\includegraphics[width=0.45\textwidth]{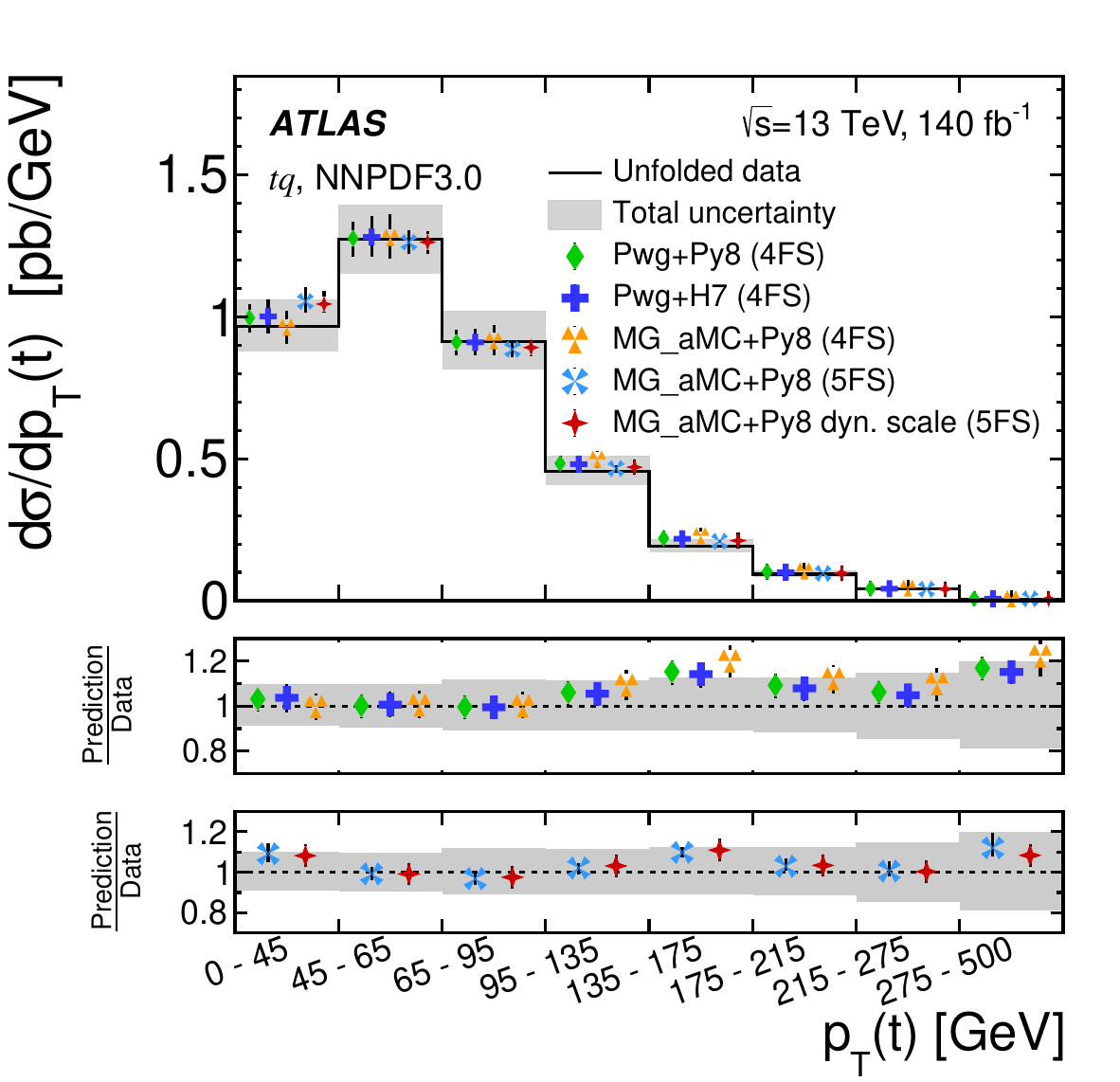}
\label{fig:TotalResultpTSRpAllGenerators}
}
\subfigure[]{
\includegraphics[width=0.45\textwidth]{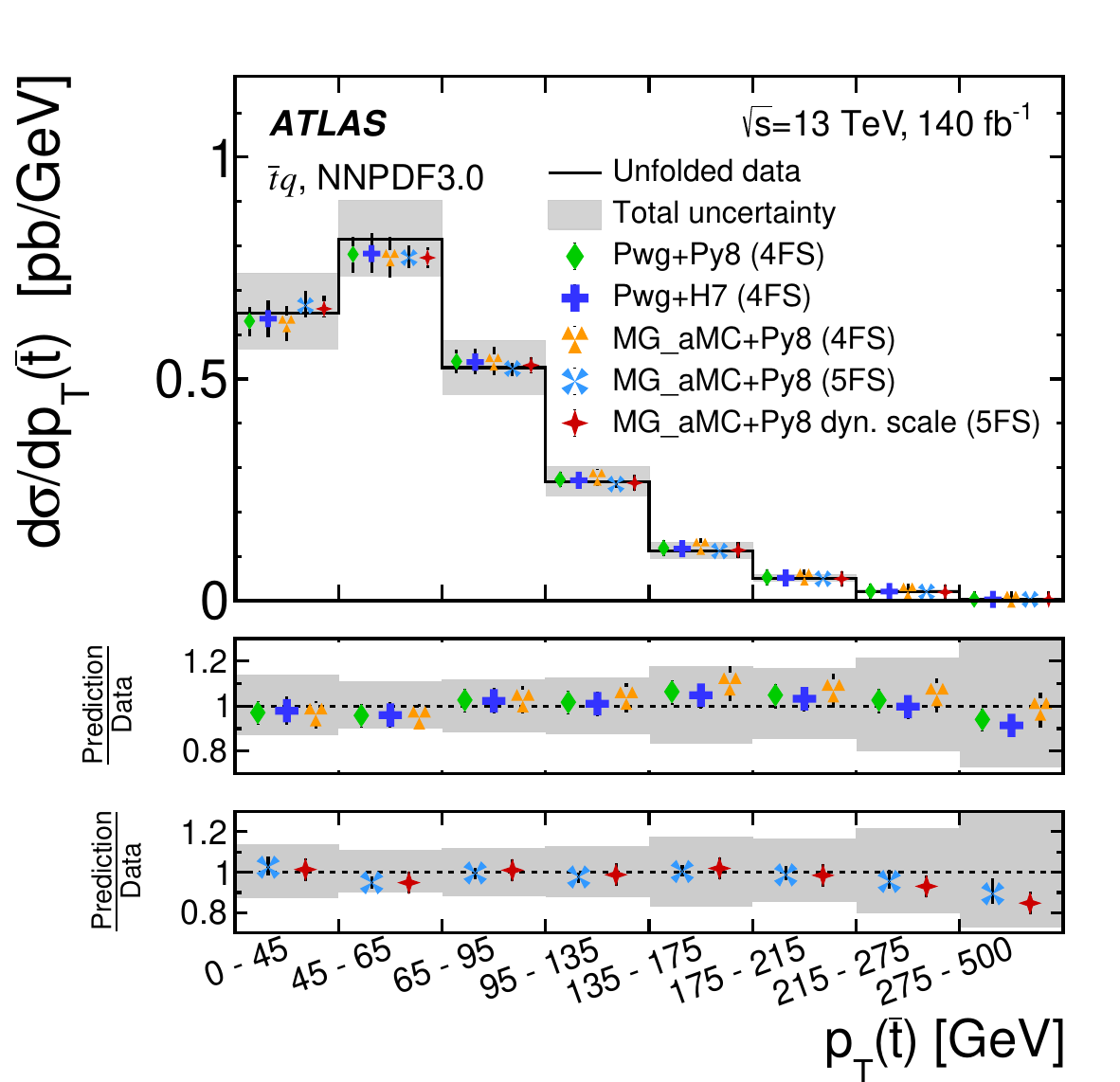}
\label{fig:TotalResultpTSRnAllGenerators}
} \\
\subfigure[]{
\includegraphics[width=0.45\textwidth]{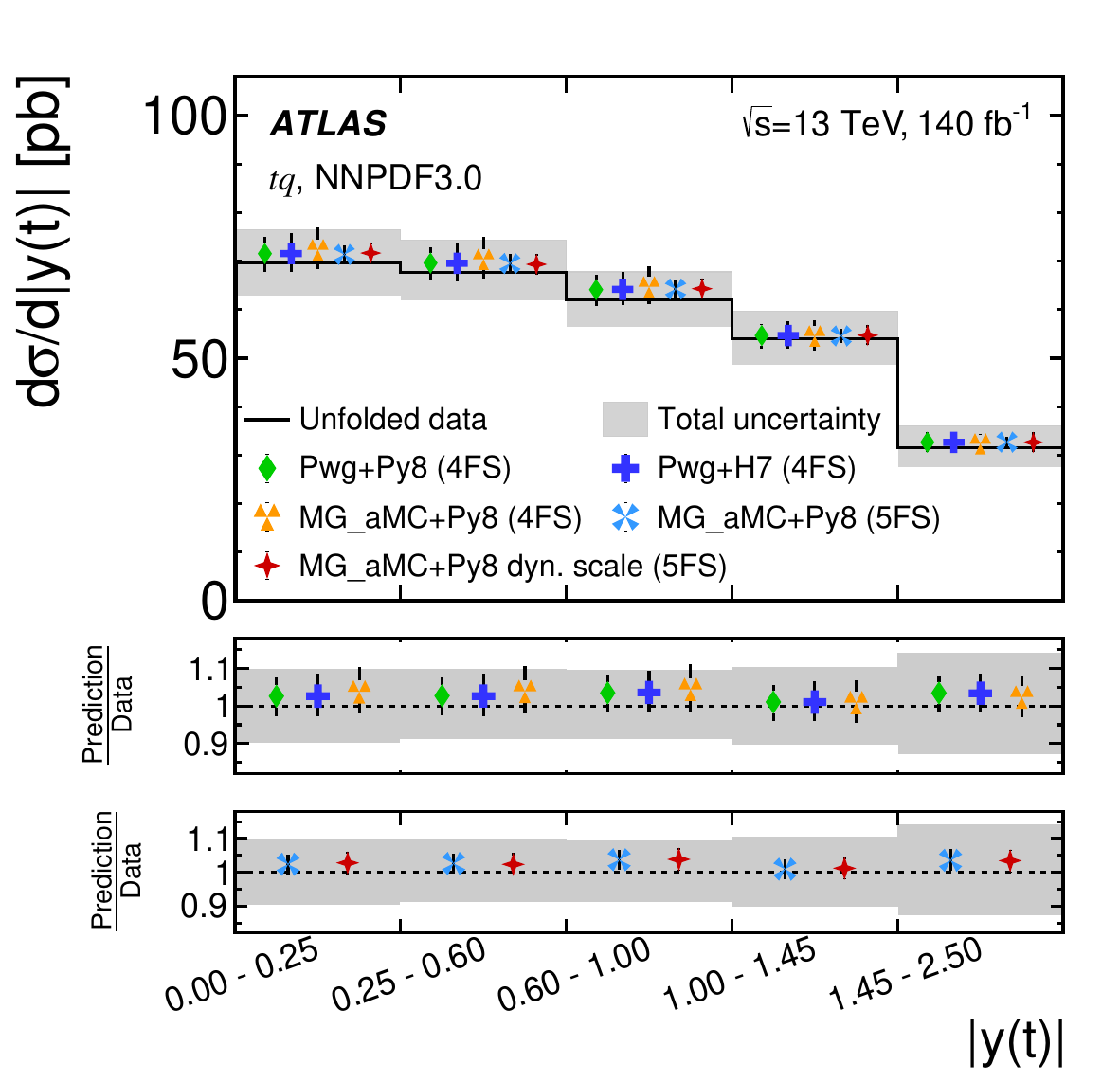}
\label{fig:TotalResultrapSRpAllGenerators}
}
\subfigure[]{
\includegraphics[width=0.45\textwidth]{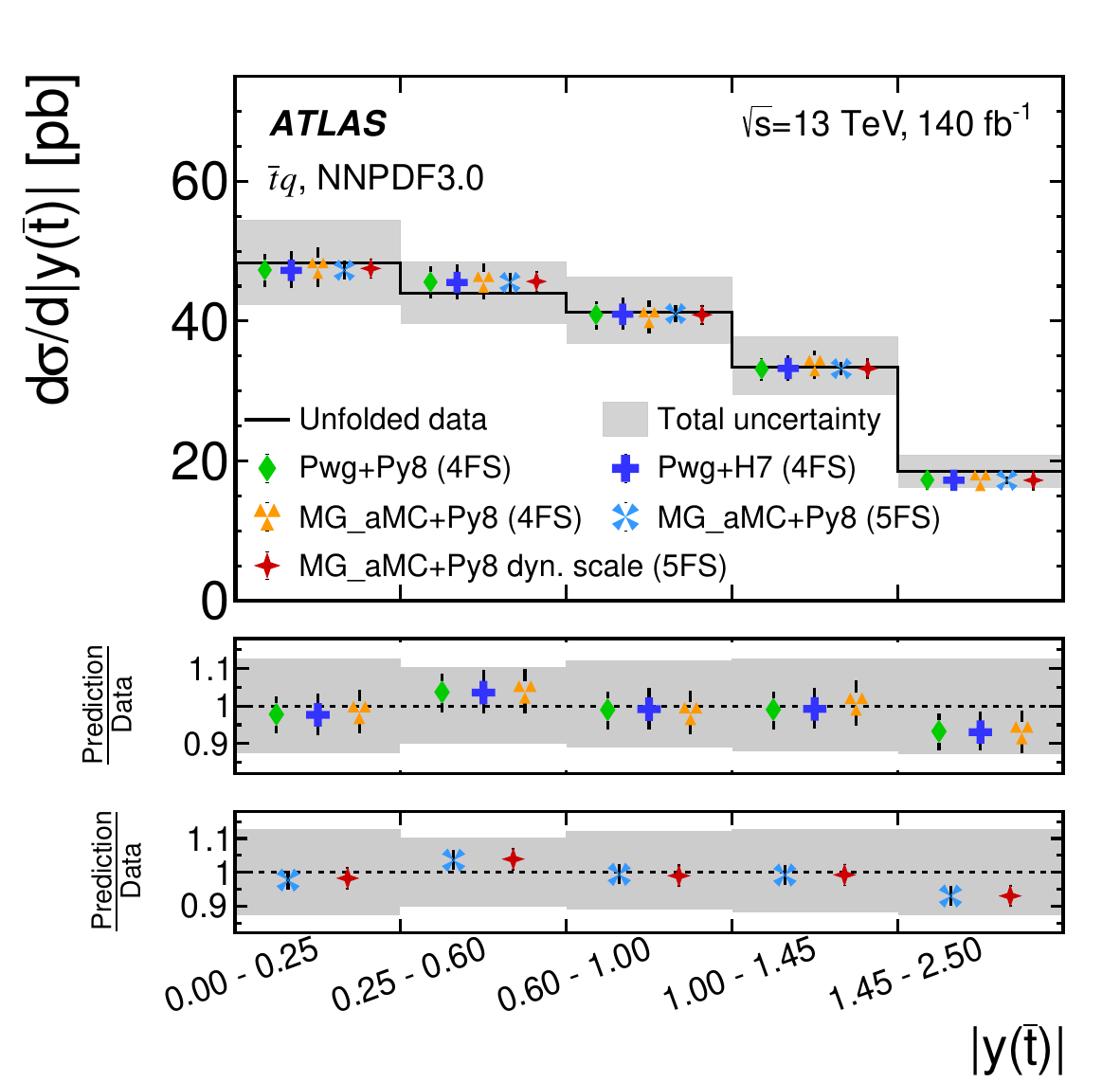}
\label{fig:TotalResultrapSRnAllGenerators}
}\\
\caption{Absolute unfolded differential \protect\subref{fig:TotalResultpTSRpAllGenerators} \protect\subref{fig:TotalResultrapSRpAllGenerators} $tq$ cross-sections and
\protect\subref{fig:TotalResultpTSRnAllGenerators} \protect\subref{fig:TotalResultrapSRnAllGenerators} $\bar{t}q$ cross-sections as a function of
\protect\subref{fig:TotalResultpTSRpAllGenerators} \protect\subref{fig:TotalResultpTSRnAllGenerators} \pTt or \pTtbar and
\protect\subref{fig:TotalResultrapSRpAllGenerators} \protect\subref{fig:TotalResultrapSRnAllGenerators} \Rapt or \Raptbar.
The shaded band indicates the total measurement uncertainty.
The cross-sections are compared with the predictions from different MC event generators and parton shower programs.
All predictions are generated with the \NNPDF[3.0] PDF set.
The uncertainty in the theoretical predictions comprises the MC statistical uncertainty, the uncertainty in the renormalisation and factorisation scale choice and the PDF uncertainty.
}
\label{fig:TotalResultsAllGenerators}
\end{figure}

\begin{figure}[htbp]
\centering
\subfigure[]{
\includegraphics[width=0.45\textwidth]{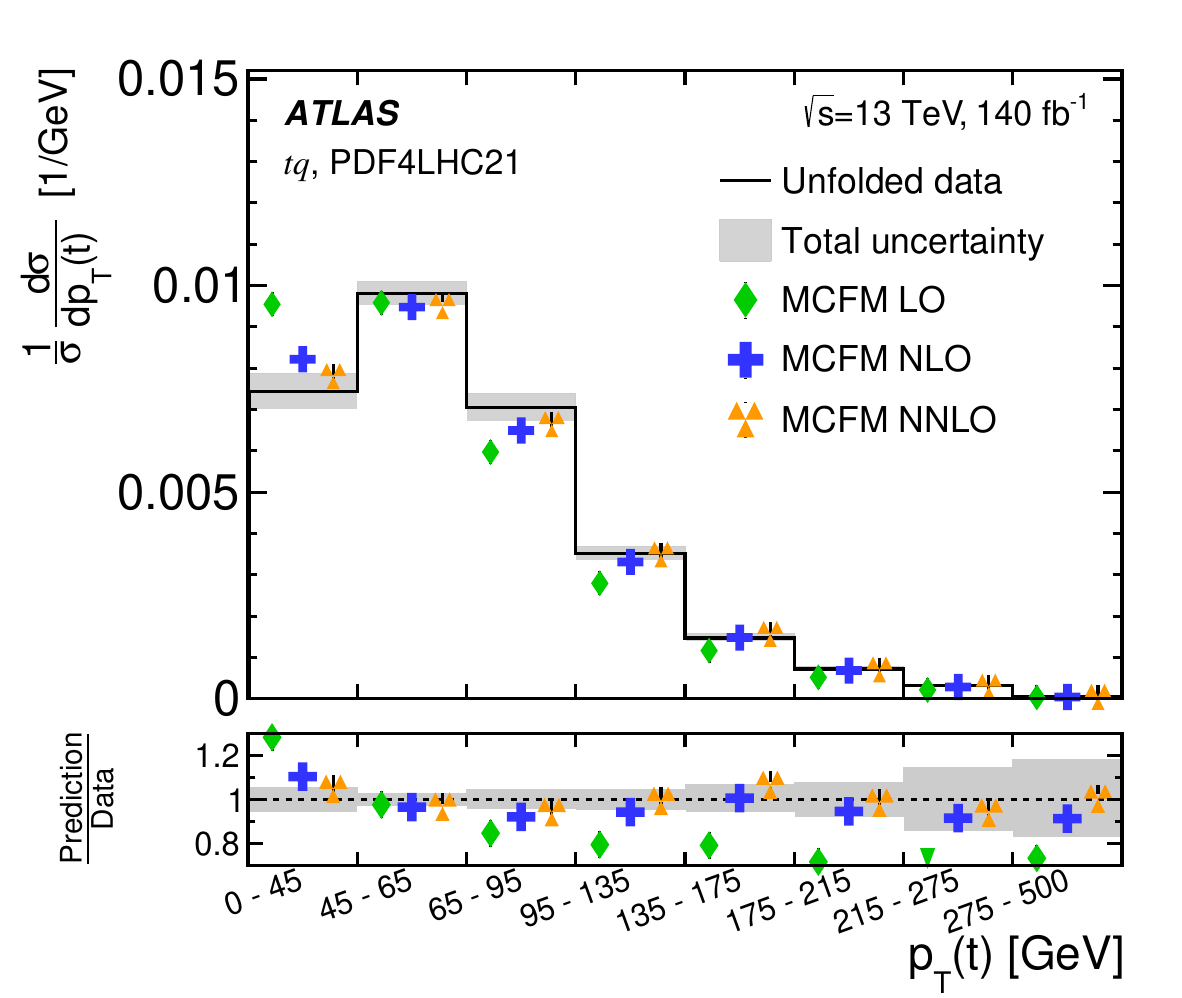}
\label{fig:NormalisedResultpTSRpMCFM_DDIS}
}
\subfigure[]{
\includegraphics[width=0.45\textwidth]{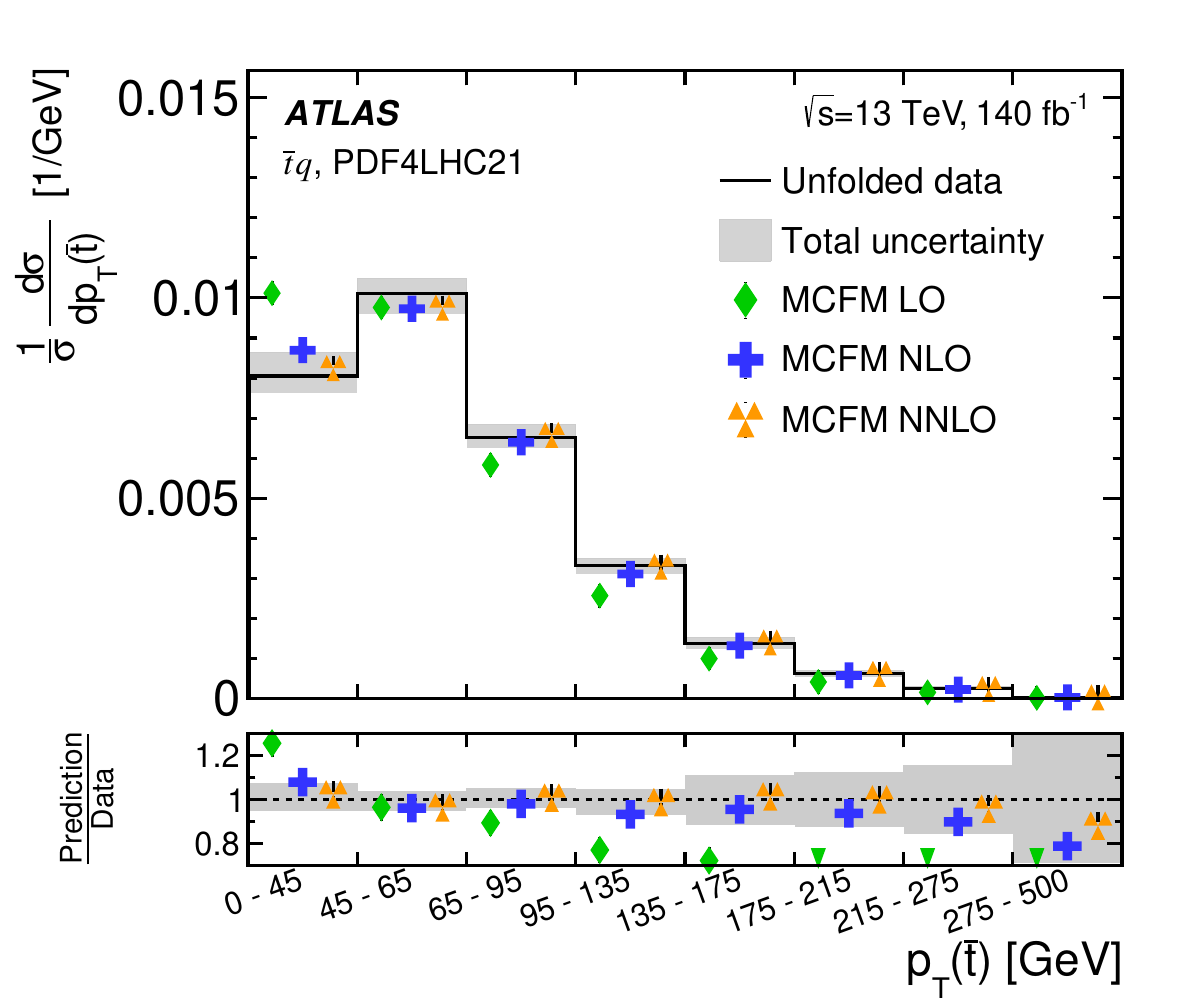}
\label{fig:NormalisedResultpTSRnMCFM_DDIS}
}\\
\subfigure[]{
\includegraphics[width=0.45\textwidth]{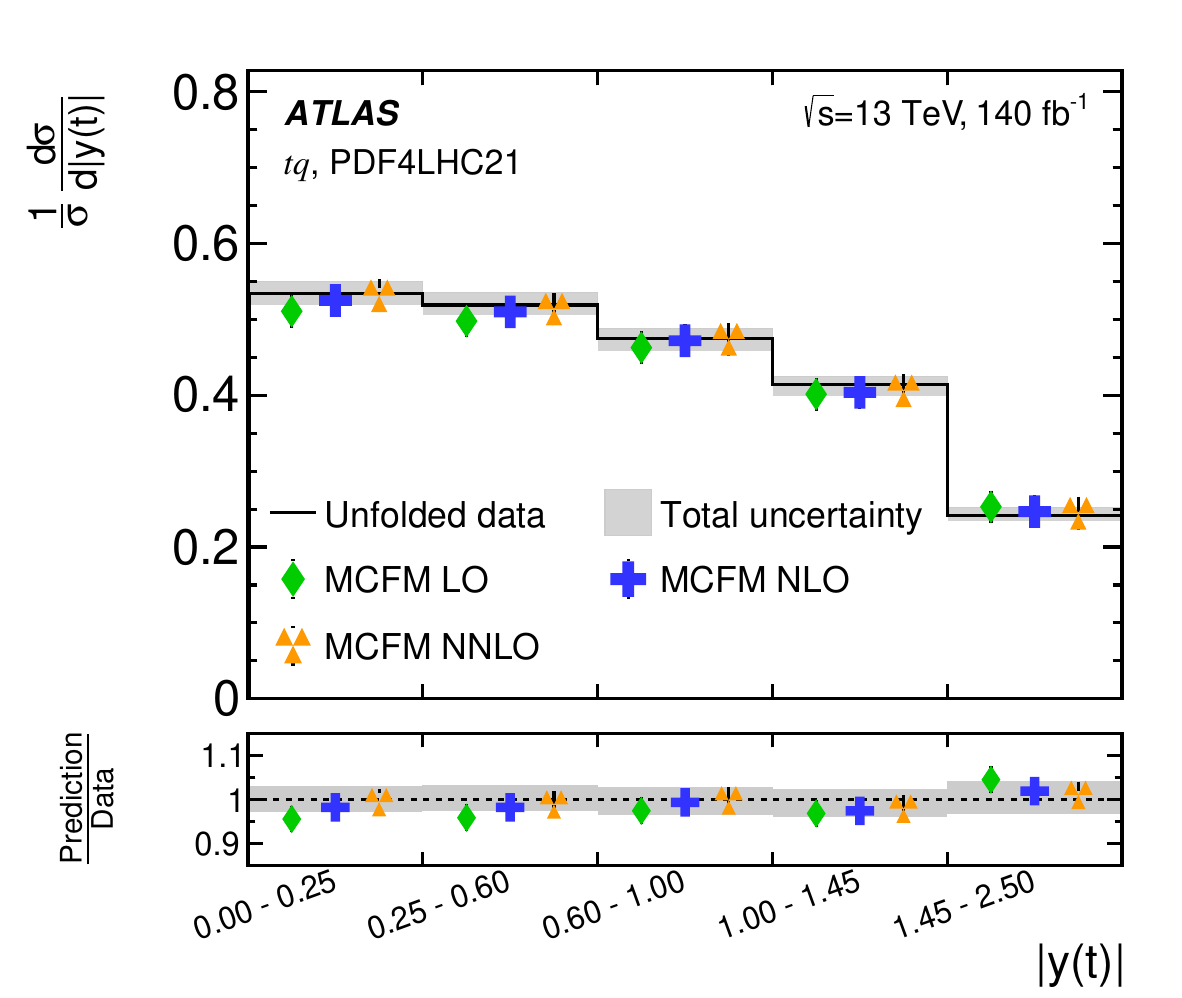}
\label{fig:NormalisedResultrapSRpMCFM_DDIS}
}
\subfigure[]{
\includegraphics[width=0.45\textwidth]{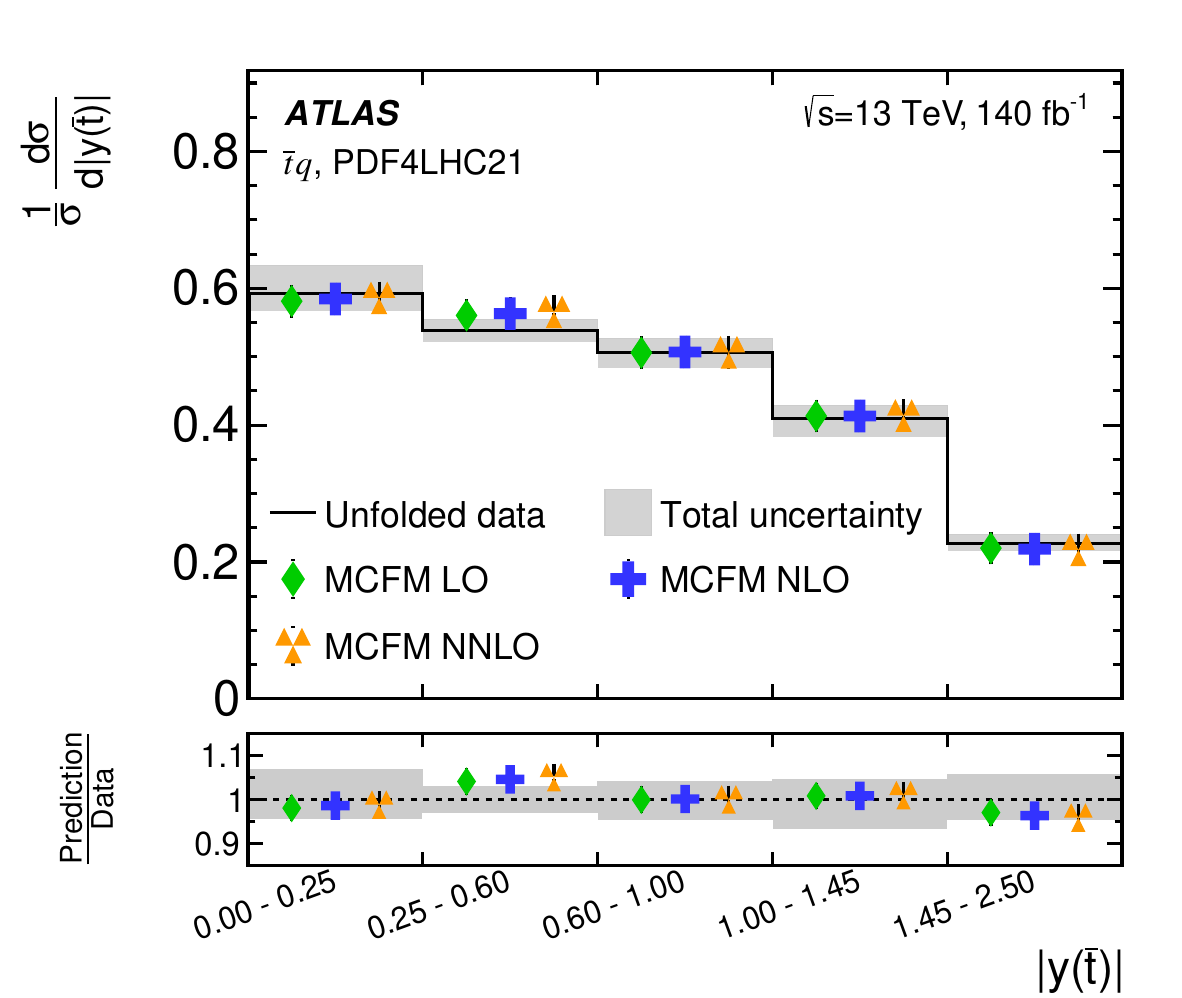}
\label{fig:NormalisedResultrapSRnMCFM_DDIS}
}
\caption{Normalised unfolded differential \protect\subref{fig:NormalisedResultpTSRpMCFM_DDIS} \protect\subref{fig:NormalisedResultrapSRpMCFM_DDIS} $tq$ cross-sections and
\protect\subref{fig:NormalisedResultpTSRnMCFM_DDIS} \protect\subref{fig:NormalisedResultrapSRnMCFM_DDIS} $\bar{t}q$ cross-sections as a function of
\protect\subref{fig:NormalisedResultpTSRpMCFM_DDIS} \protect\subref{fig:NormalisedResultpTSRnMCFM_DDIS} \pTt or \pTtbar and
\protect\subref{fig:NormalisedResultrapSRpMCFM_DDIS} \protect\subref{fig:NormalisedResultrapSRnMCFM_DDIS} \Rapt or \Raptbar.
The shaded band indicates the total measurement uncertainty.
The cross-sections are compared with the theoretical predictions from FO calculations done with MCFM at LO, NLO and NNLO.
The uncertainty in the theoretical predictions from FO calculations comprises the uncertainty in the renormalisation and factorisation scale choice and the uncertainty in the value of \alphas.
}
\label{fig:NormalisedResultsMCFM_DDIS}
\end{figure}

\begin{figure}[htbp]
\centering
\subfigure[]{
\includegraphics[width=0.45\textwidth]{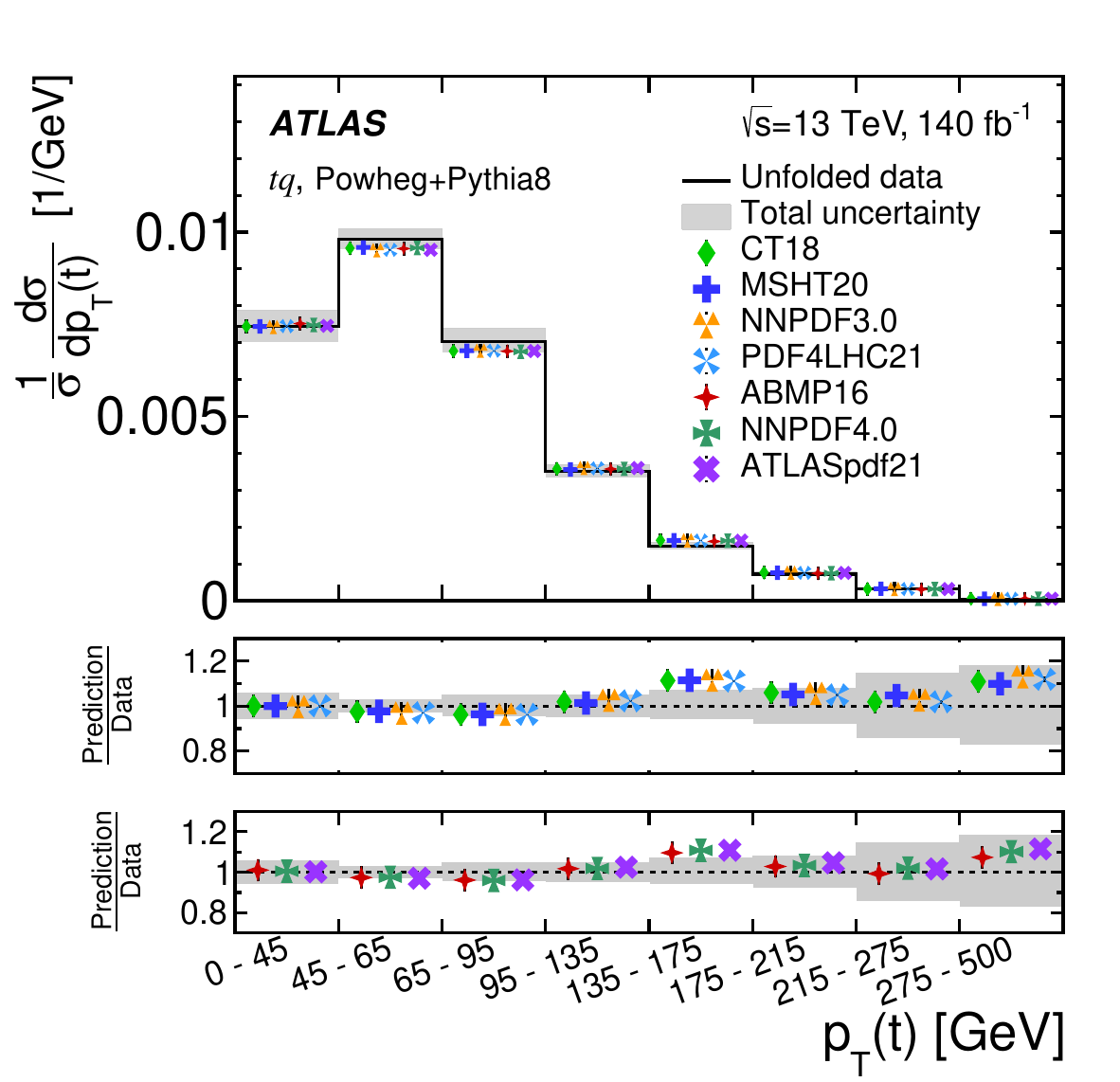}
\label{fig:NormalisedResultpTSRpAllPDFs}
}
\subfigure[]{
\includegraphics[width=0.45\textwidth]{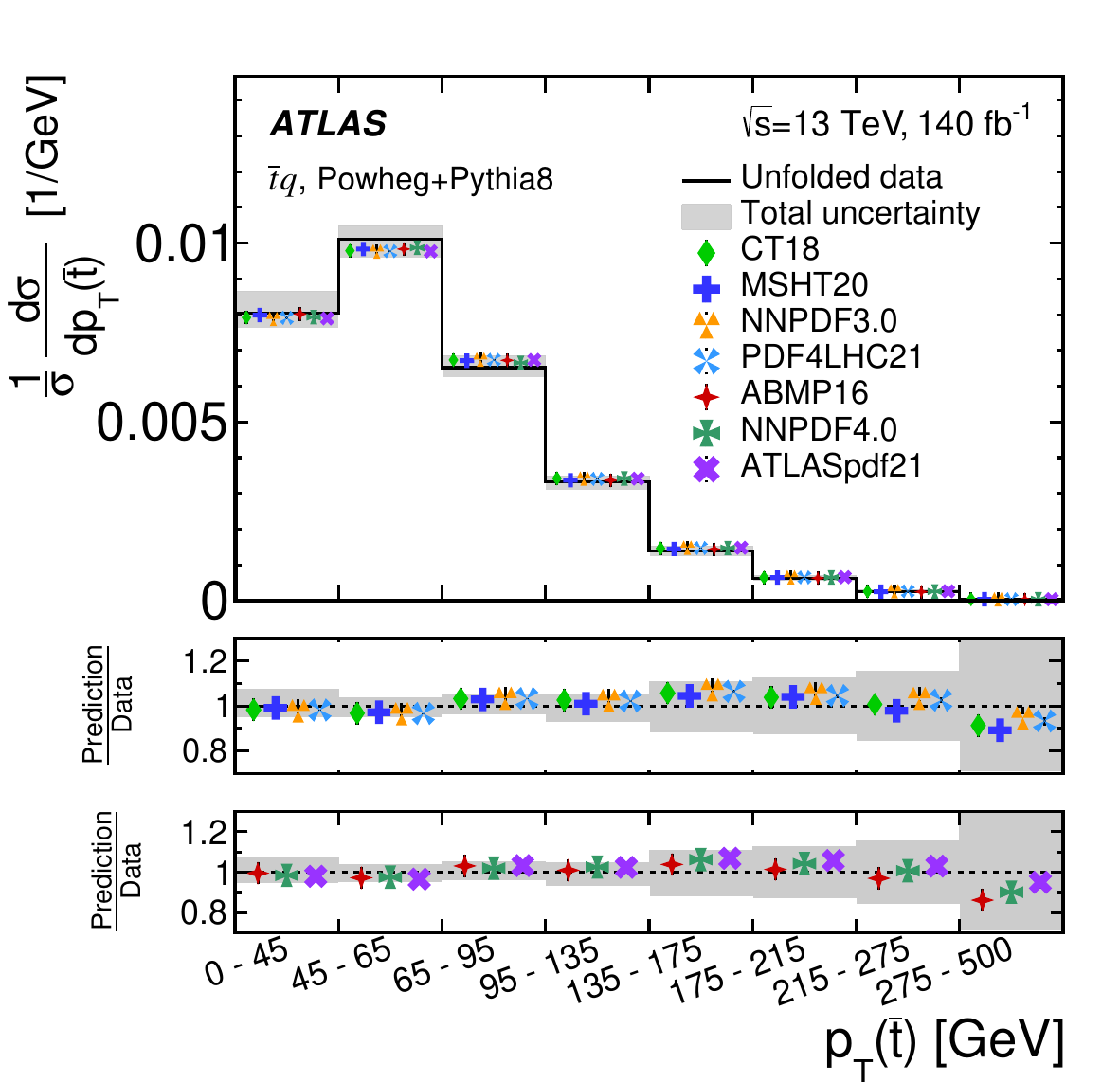}
\label{fig:NormalisedResultpTSRnAllPDFs}
}\\
\subfigure[]{
\includegraphics[width=0.45\textwidth]{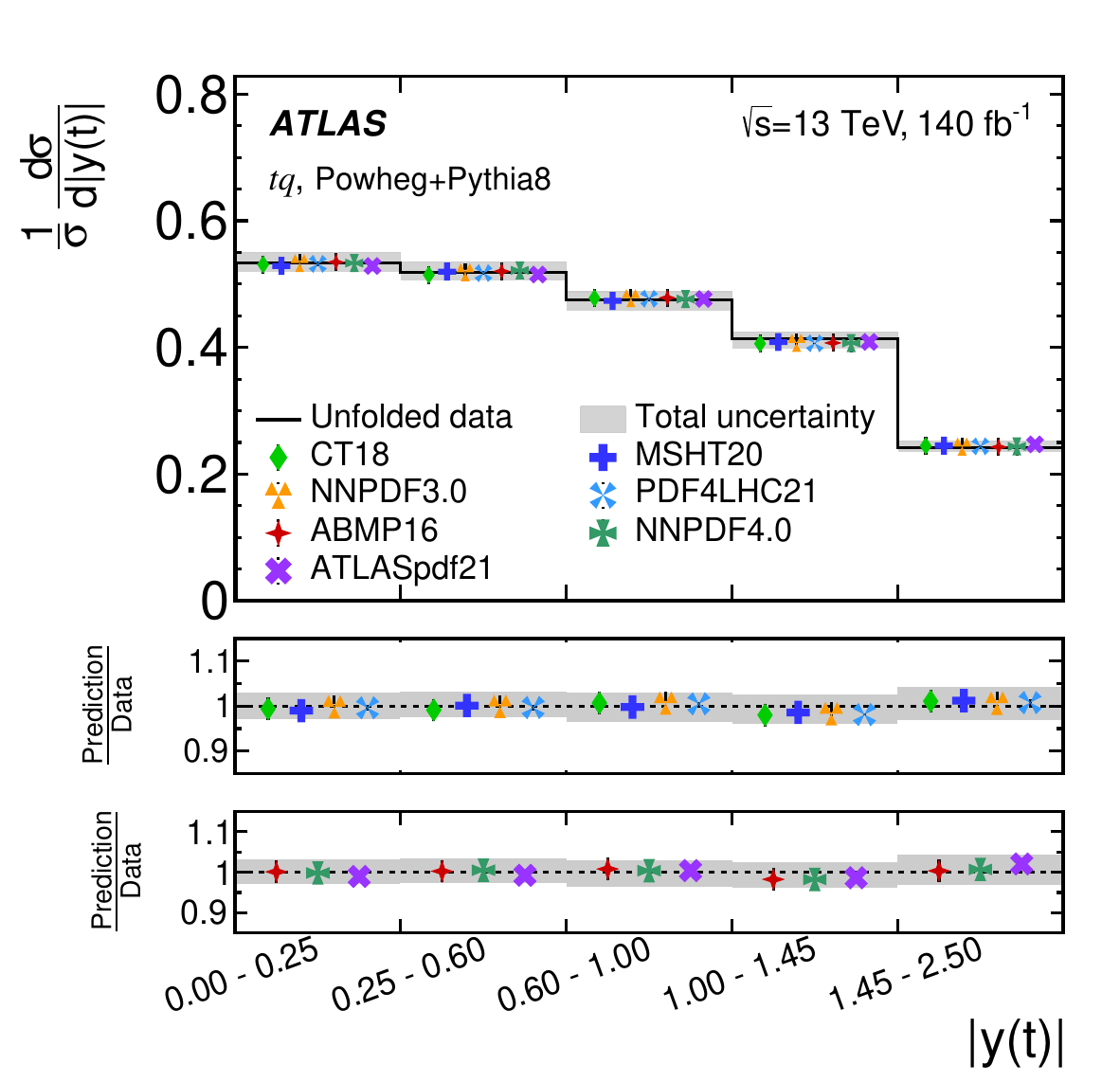}
\label{fig:NormalisedResultrapSRpAllPDFs}
}
\subfigure[]{
\includegraphics[width=0.45\textwidth]{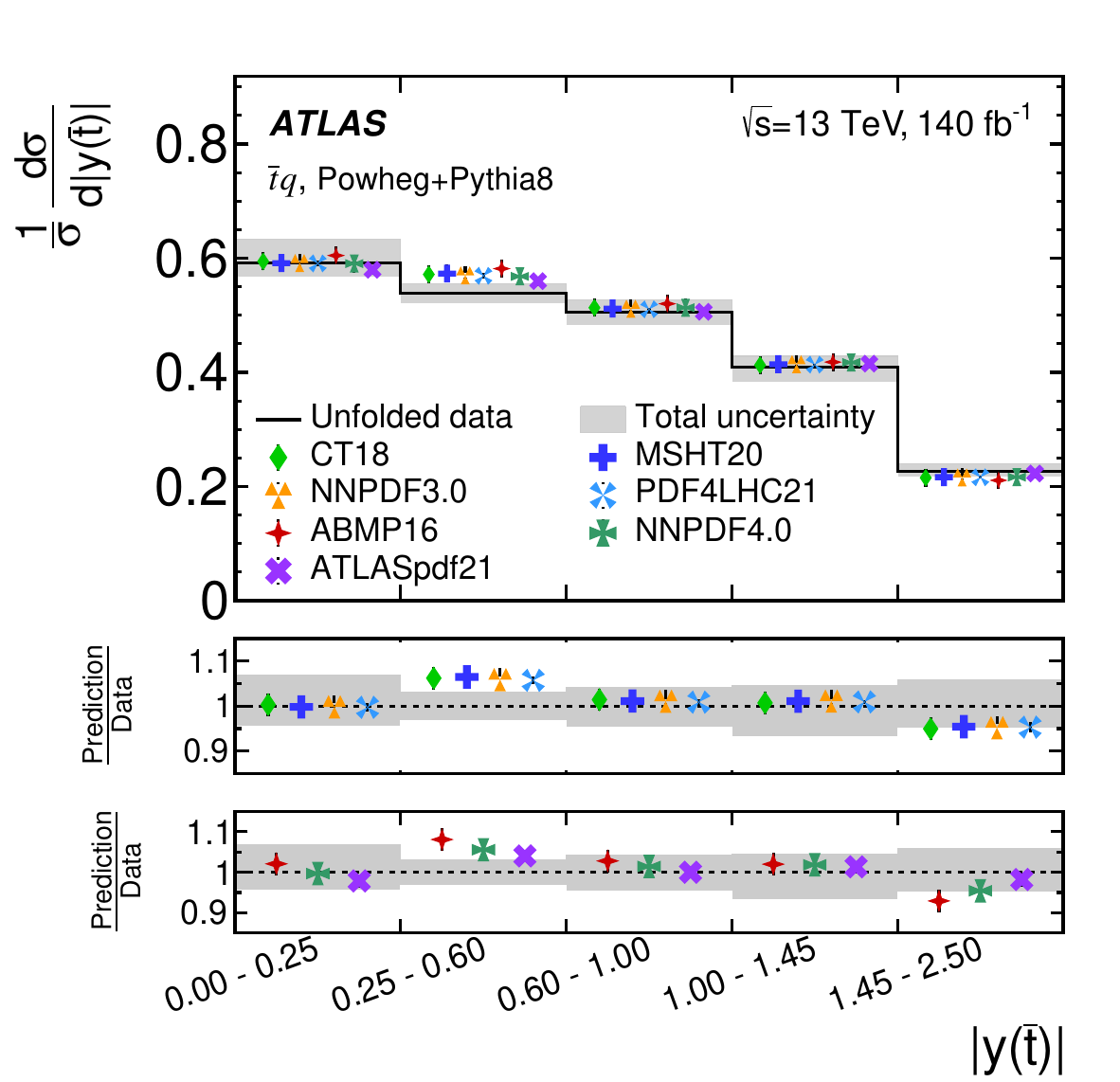}
\label{fig:NormalisedResultrapSRnAllPDFs}
}
\caption{Normalised unfolded differential \protect\subref{fig:NormalisedResultpTSRpAllPDFs} \protect\subref{fig:NormalisedResultrapSRpAllPDFs} $tq$ cross-sections and
\protect\subref{fig:NormalisedResultpTSRnAllPDFs} \protect\subref{fig:NormalisedResultrapSRnAllPDFs} $\bar{t}q$ cross-sections as a function of
\protect\subref{fig:NormalisedResultpTSRpAllPDFs} \protect\subref{fig:NormalisedResultpTSRnAllPDFs} \pTt or \pTtbar and
\protect\subref{fig:NormalisedResultrapSRpAllPDFs} \protect\subref{fig:NormalisedResultrapSRnAllPDFs} \Rapt or \Raptbar.
The shaded band indicates the total measurement uncertainty.
The cross-sections are compared with the predictions from \POWHEG+\PYTHIA[8] with different PDF sets.
The uncertainty in the theoretical predictions comprises the MC statistical uncertainty and the PDF uncertainty.
}
\end{figure}

\begin{figure}[htbp]
\centering
\subfigure[]{
\includegraphics[width=0.45\textwidth]{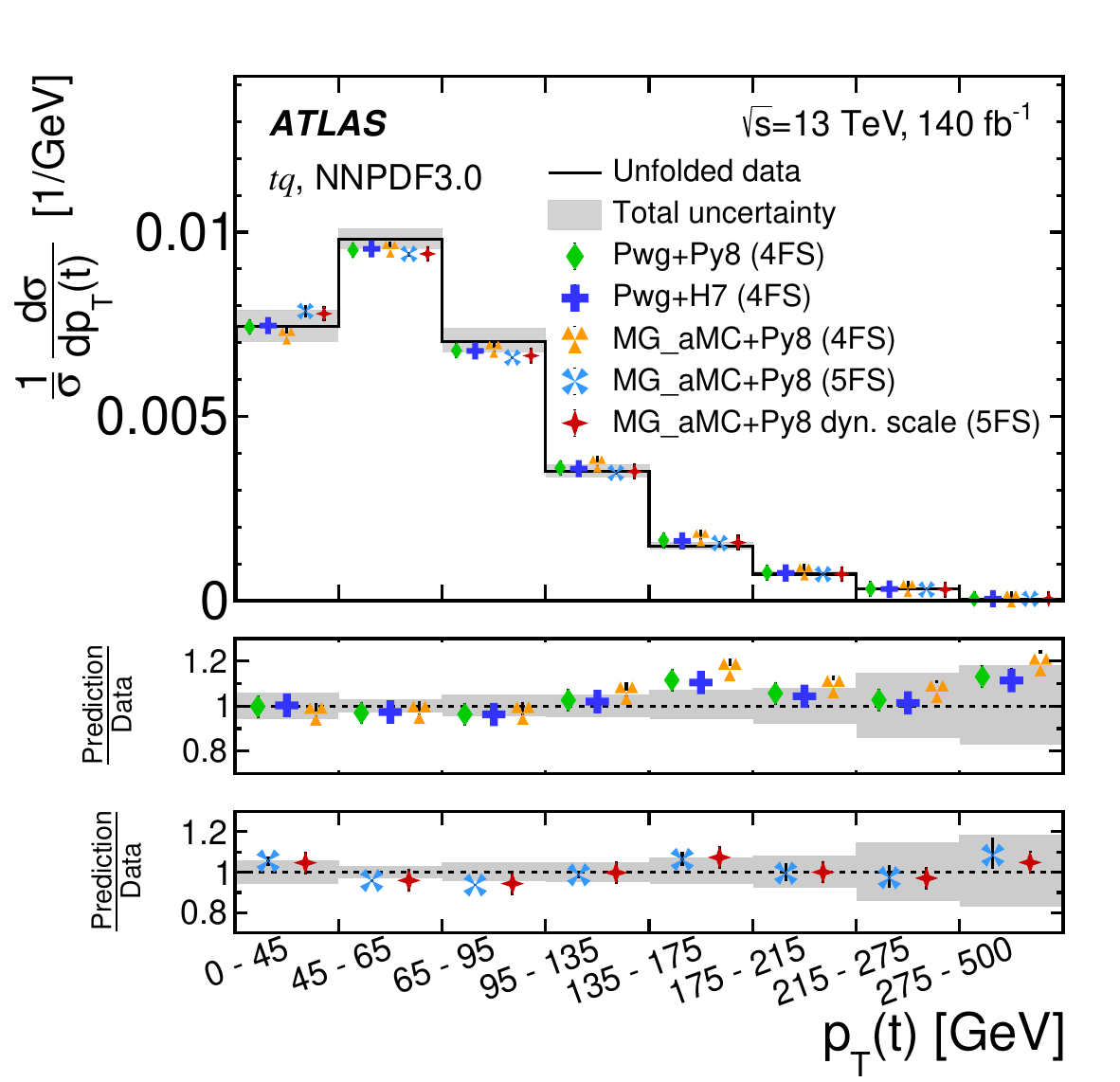}
\label{fig:NormalisedResultpTSRpAllGenerators}
}
\subfigure[]{
\includegraphics[width=0.45\textwidth]{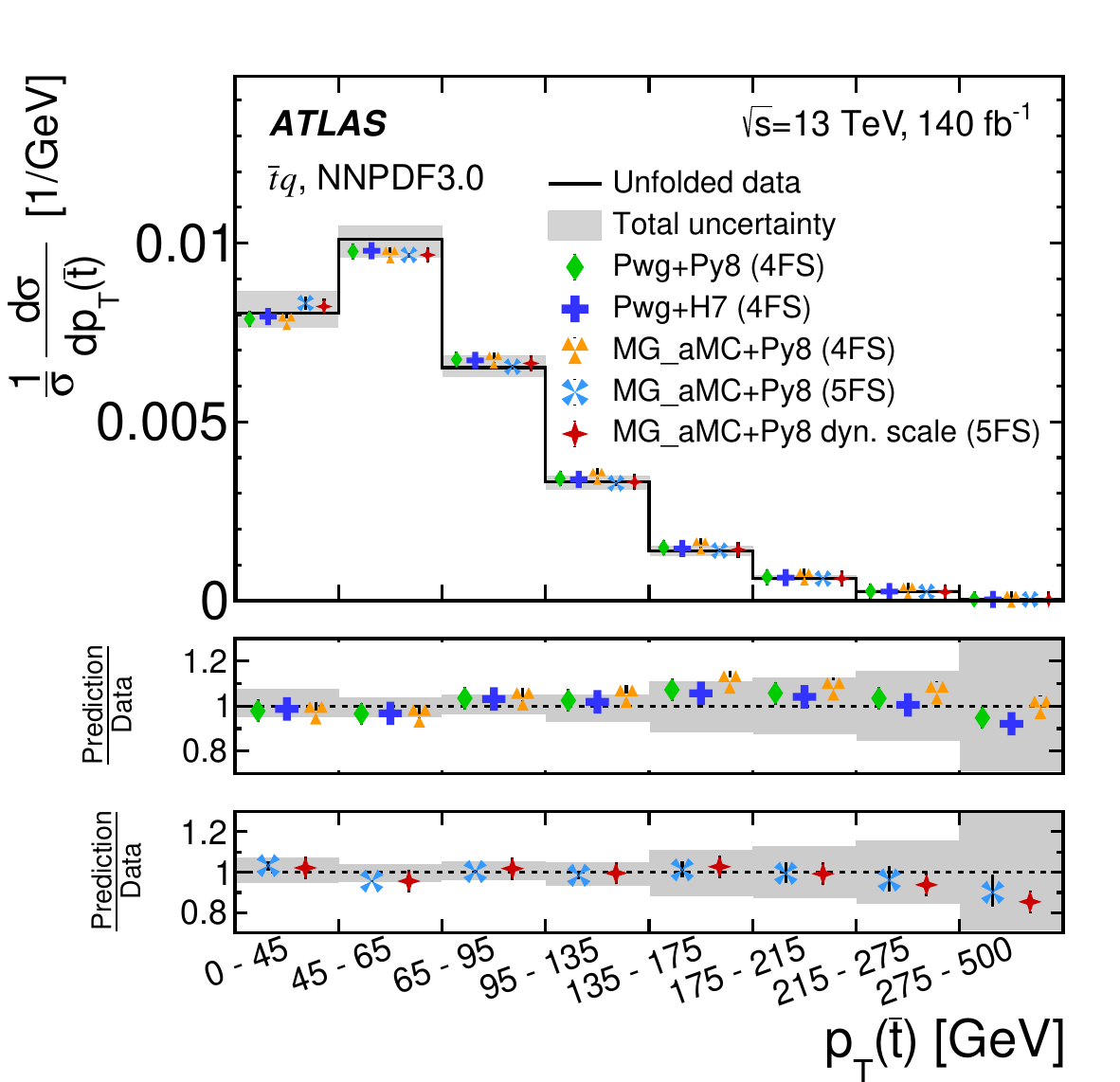}
\label{fig:NormalisedResultpTSRnAllGenerators}
}\\
\subfigure[]{
\includegraphics[width=0.45\textwidth]{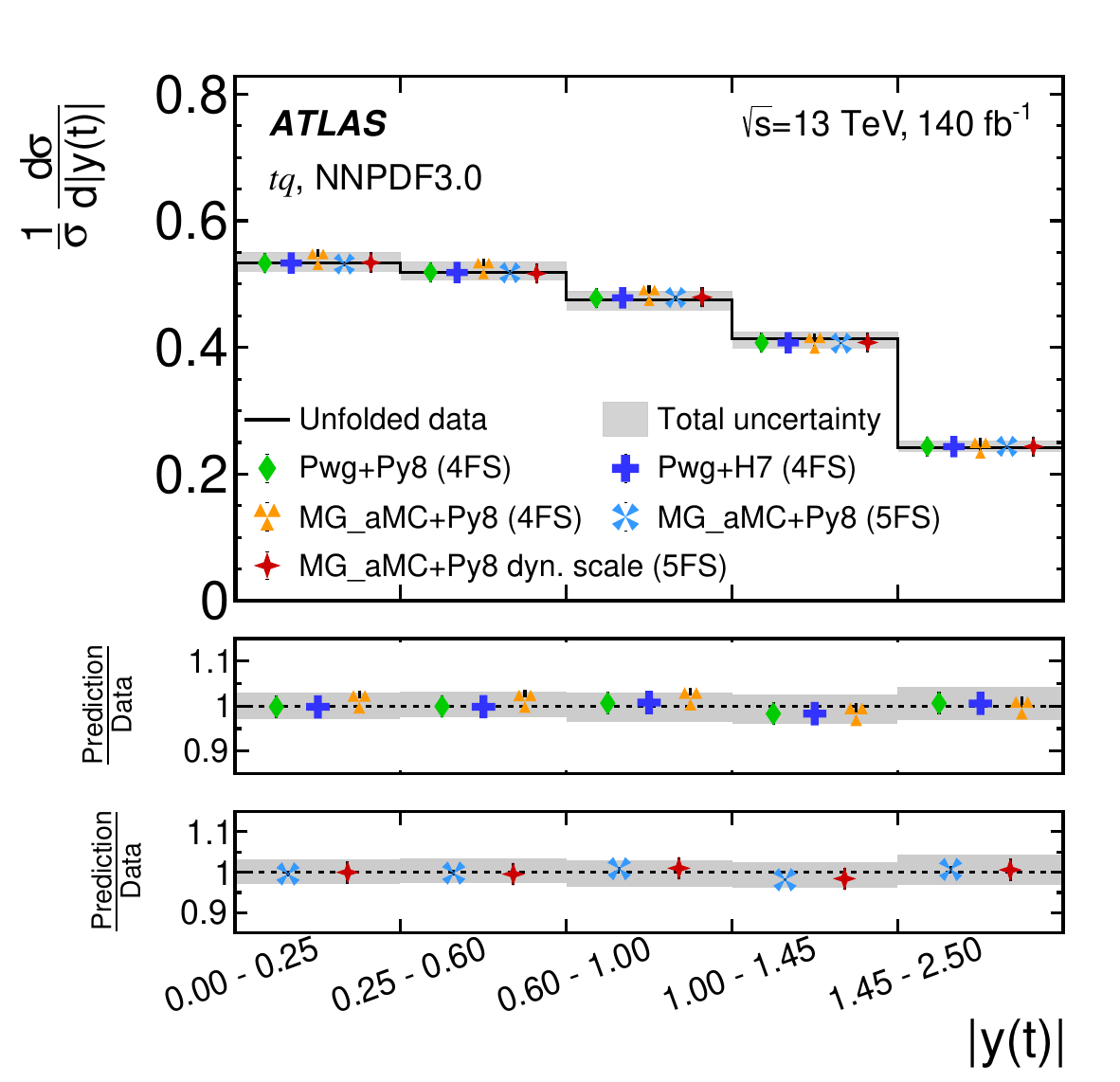}
\label{fig:NormalisedResultrapSRpAllGenerators}
}
\subfigure[]{
\includegraphics[width=0.45\textwidth]{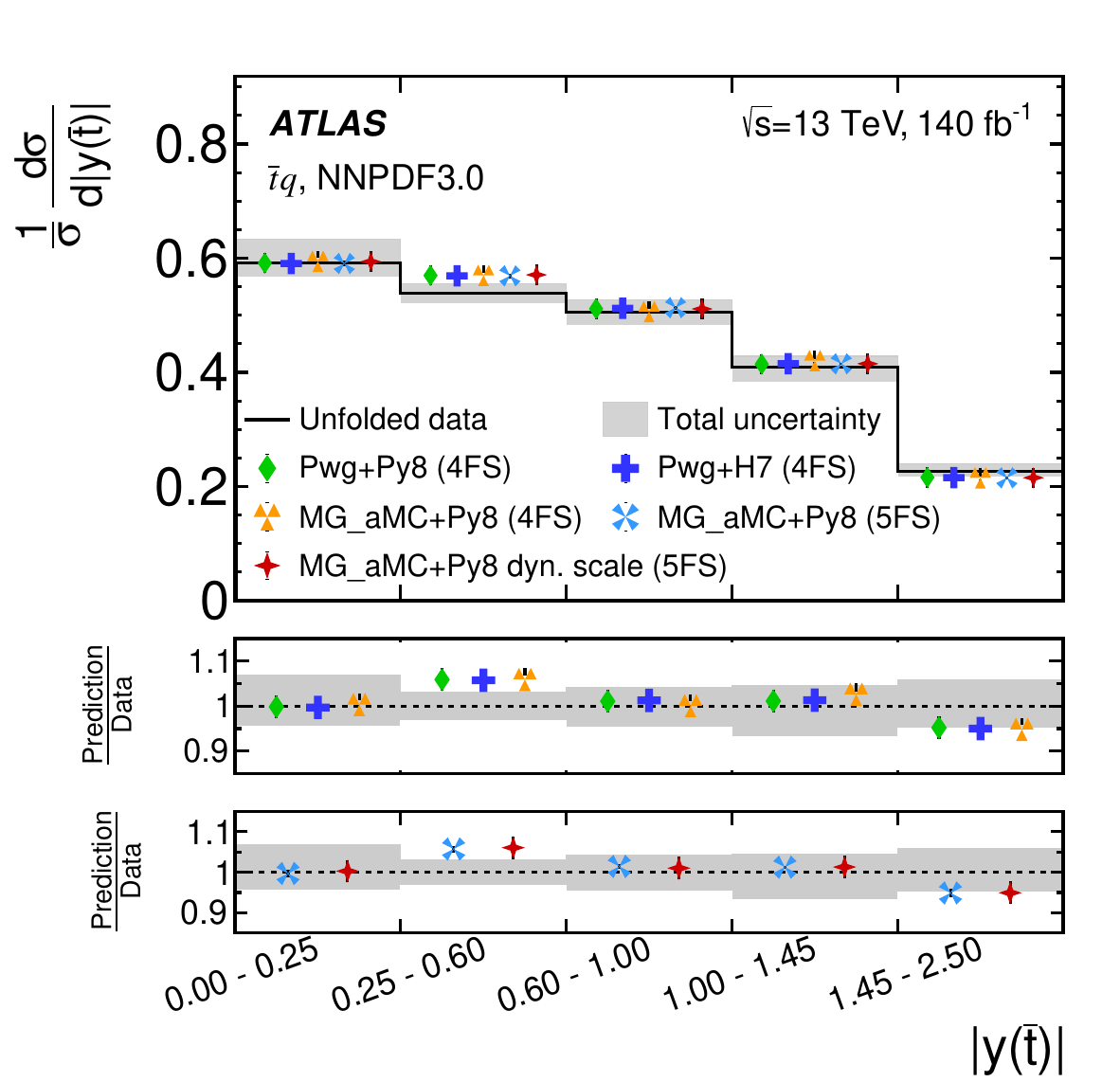}
\label{fig:NormalisedResultrapSRnAllGenerators}
}
\caption{Normalised unfolded differential \protect\subref{fig:NormalisedResultpTSRpAllGenerators} \protect\subref{fig:NormalisedResultrapSRpAllGenerators} $tq$ cross-sections and
\protect\subref{fig:NormalisedResultpTSRnAllGenerators} \protect\subref{fig:NormalisedResultrapSRnAllGenerators} $\bar{t}q$ cross-sections as a function of
\protect\subref{fig:NormalisedResultpTSRpAllGenerators} \protect\subref{fig:NormalisedResultpTSRnAllGenerators} \pTt or \pTtbar and
\protect\subref{fig:NormalisedResultrapSRpAllGenerators} \protect\subref{fig:NormalisedResultrapSRnAllGenerators} \Rapt or \Raptbar.
The shaded band indicates the total measurement uncertainty.
The cross-sections are compared with the predictions from different MC event generators and parton shower programs.
All predictions are generated with the \NNPDF[3.0] PDF set.
The uncertainty in the theoretical predictions comprises the MC statistical uncertainty, the uncertainty in the renormalisation and factorisation scale choice and the PDF uncertainty.
}
\label{fig:NormalisedResultsAllGenerators}
\end{figure}

\begin{figure}[htbp]
\centering
\vspace*{-8pt}
\subfigure[]{
\includegraphics[trim={0 0.5cm 0 1cm},width=0.45\textwidth]{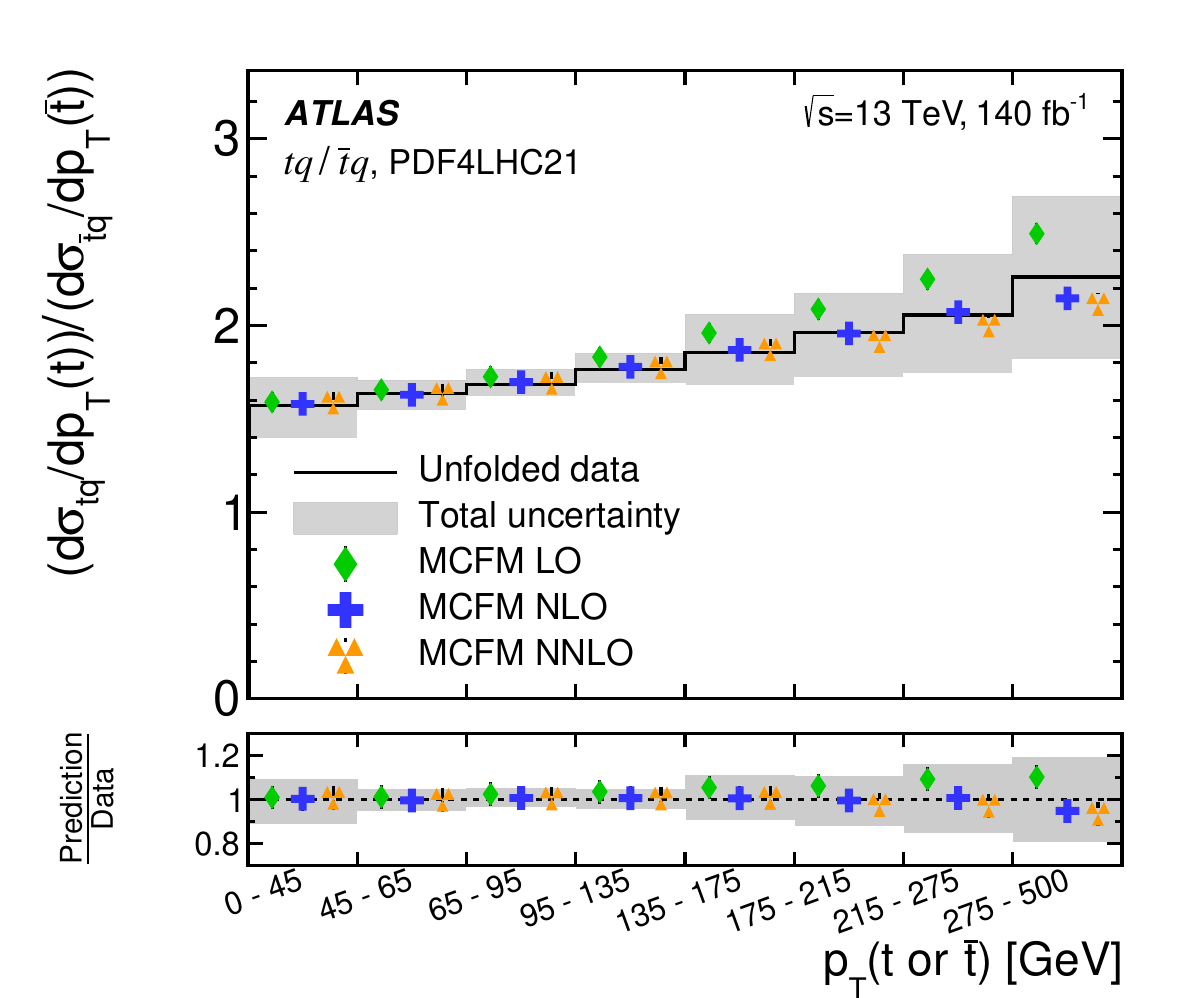}
\label{fig:ResultpTRatioMCFMDDIS}
}
\subfigure[]{
\includegraphics[trim={0 0.5cm 0 1cm},width=0.45\textwidth]{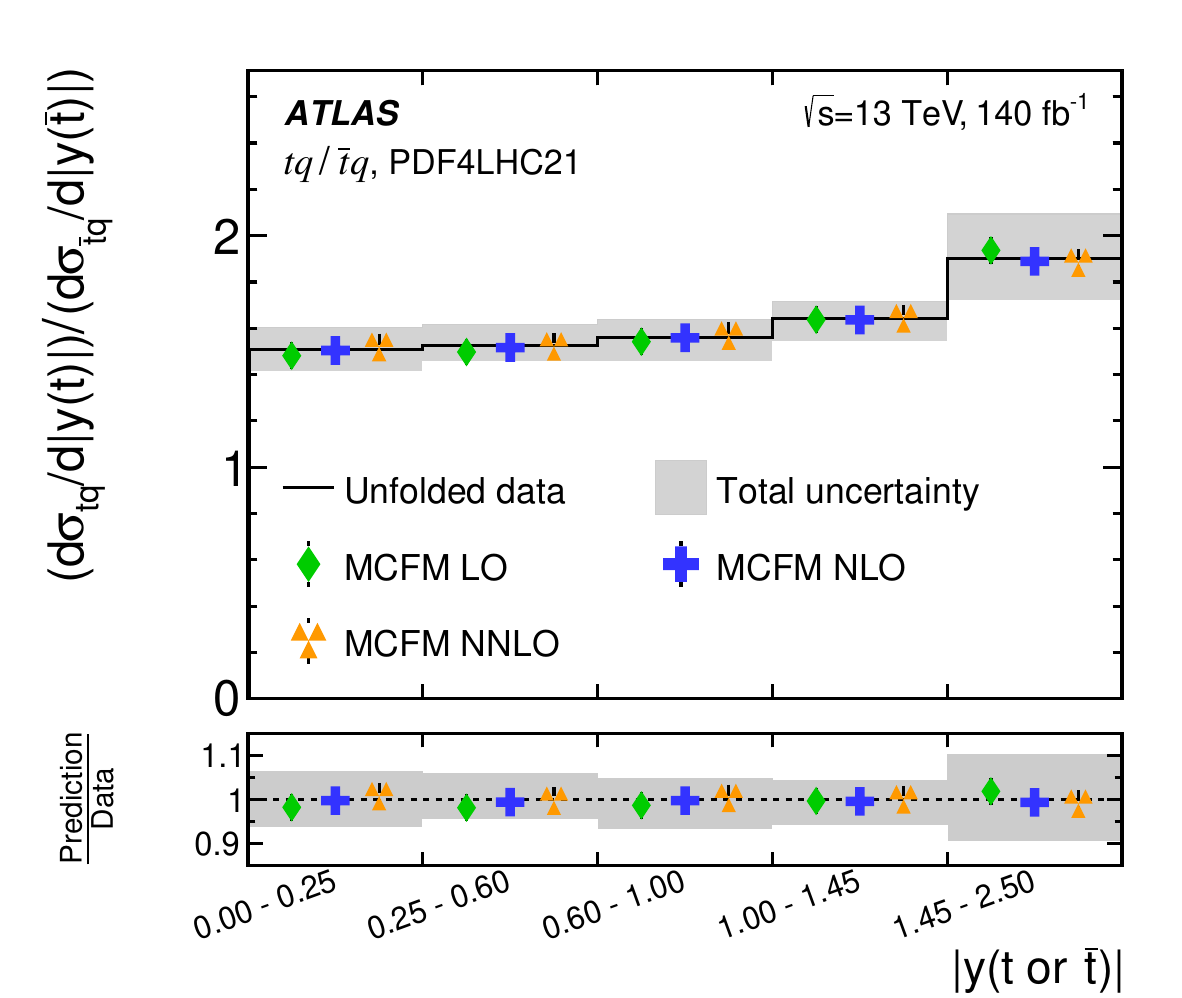}
\label{fig:ResultrapRatioMCFMDDIS}
}\\
\subfigure[]{
\includegraphics[trim={0 0.5cm 0 1cm},width=0.45\textwidth]{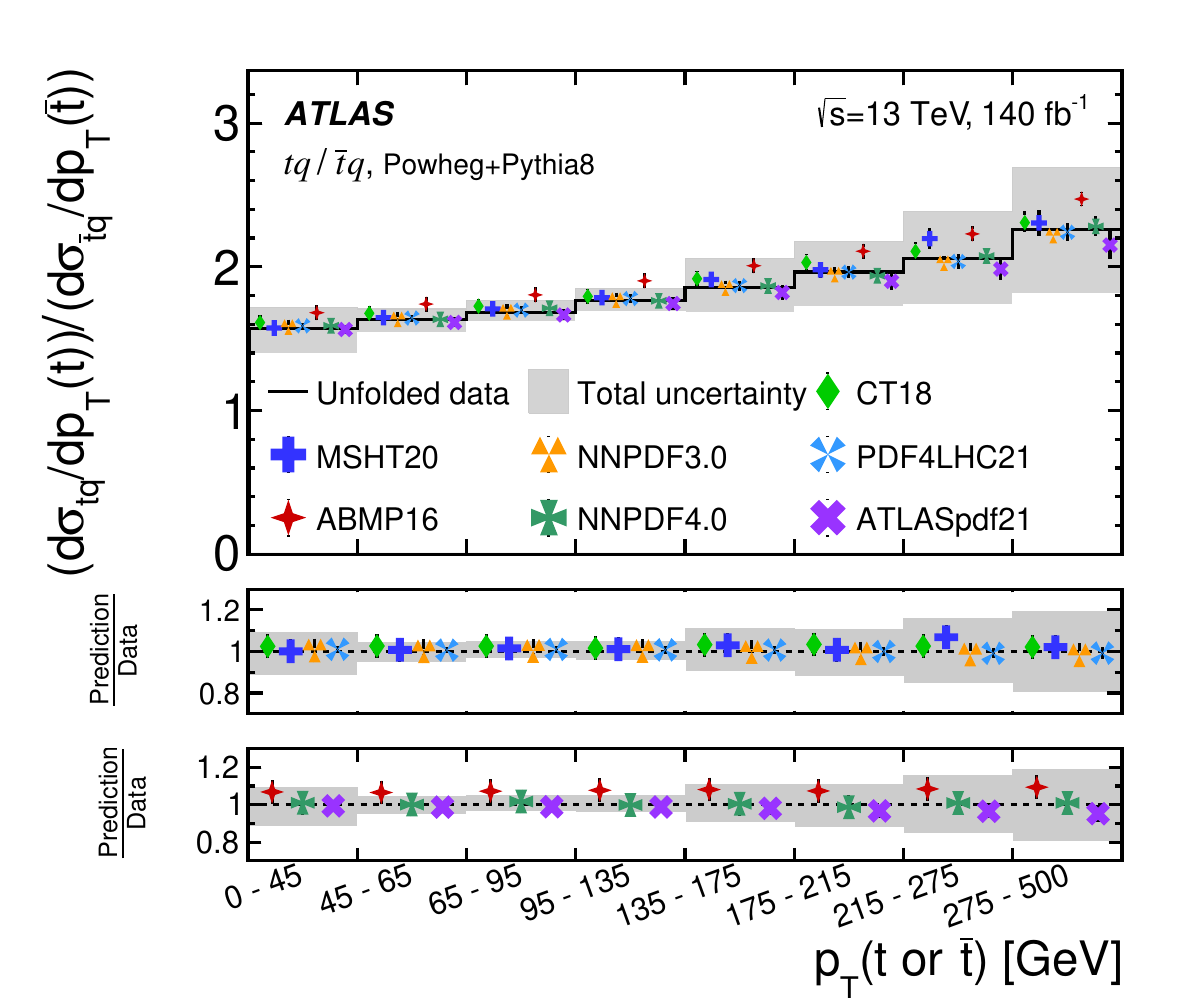}
\label{fig:ResultpTRatioAllPDFs}
}
\subfigure[]{
\includegraphics[trim={0 0.5cm 0 1cm},width=0.45\textwidth]{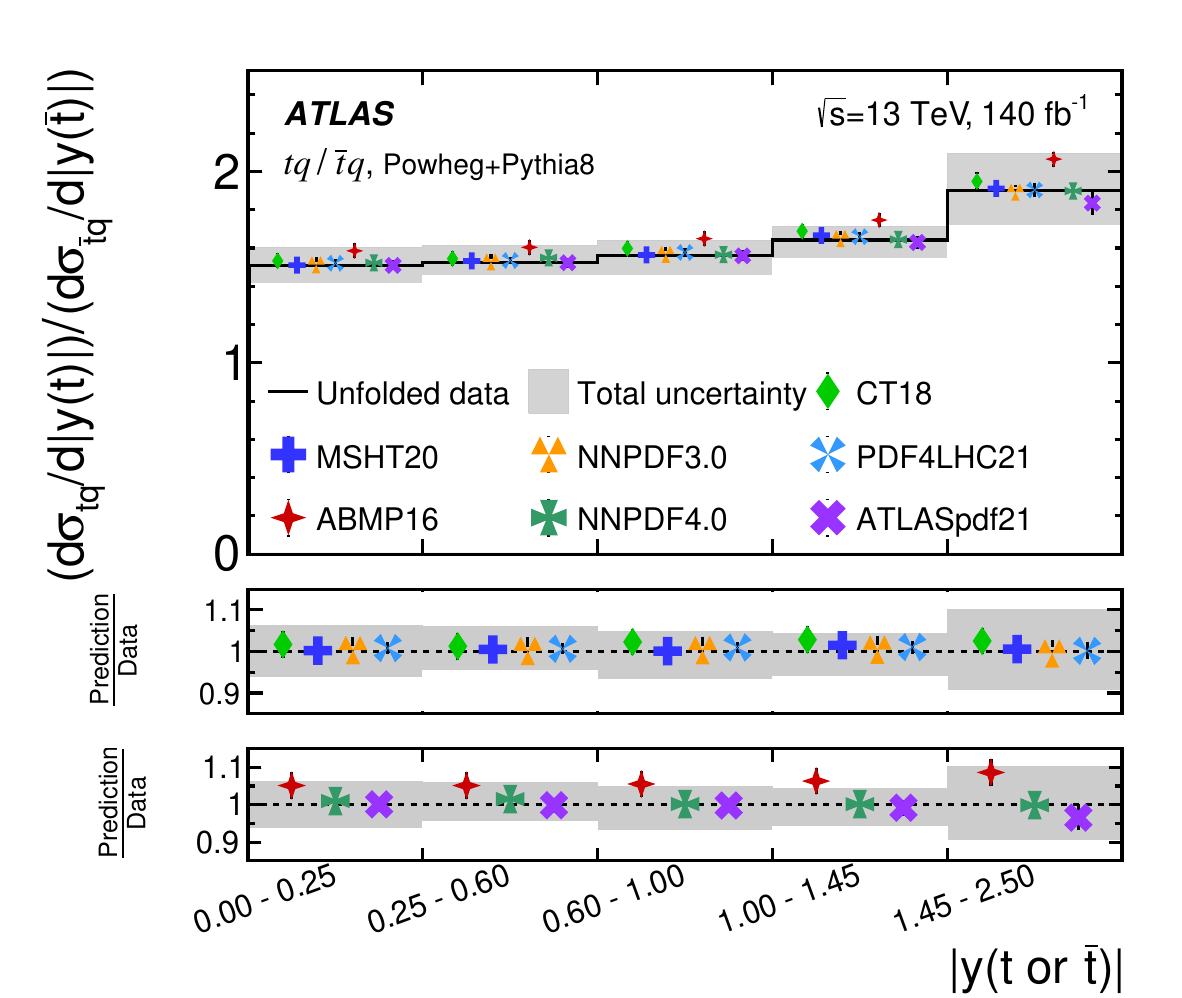}
\label{fig:ResultrapRatioAllPDFs}
}\\
\subfigure[]{
\includegraphics[trim={0 0.5cm 0 1cm},width=0.45\textwidth]{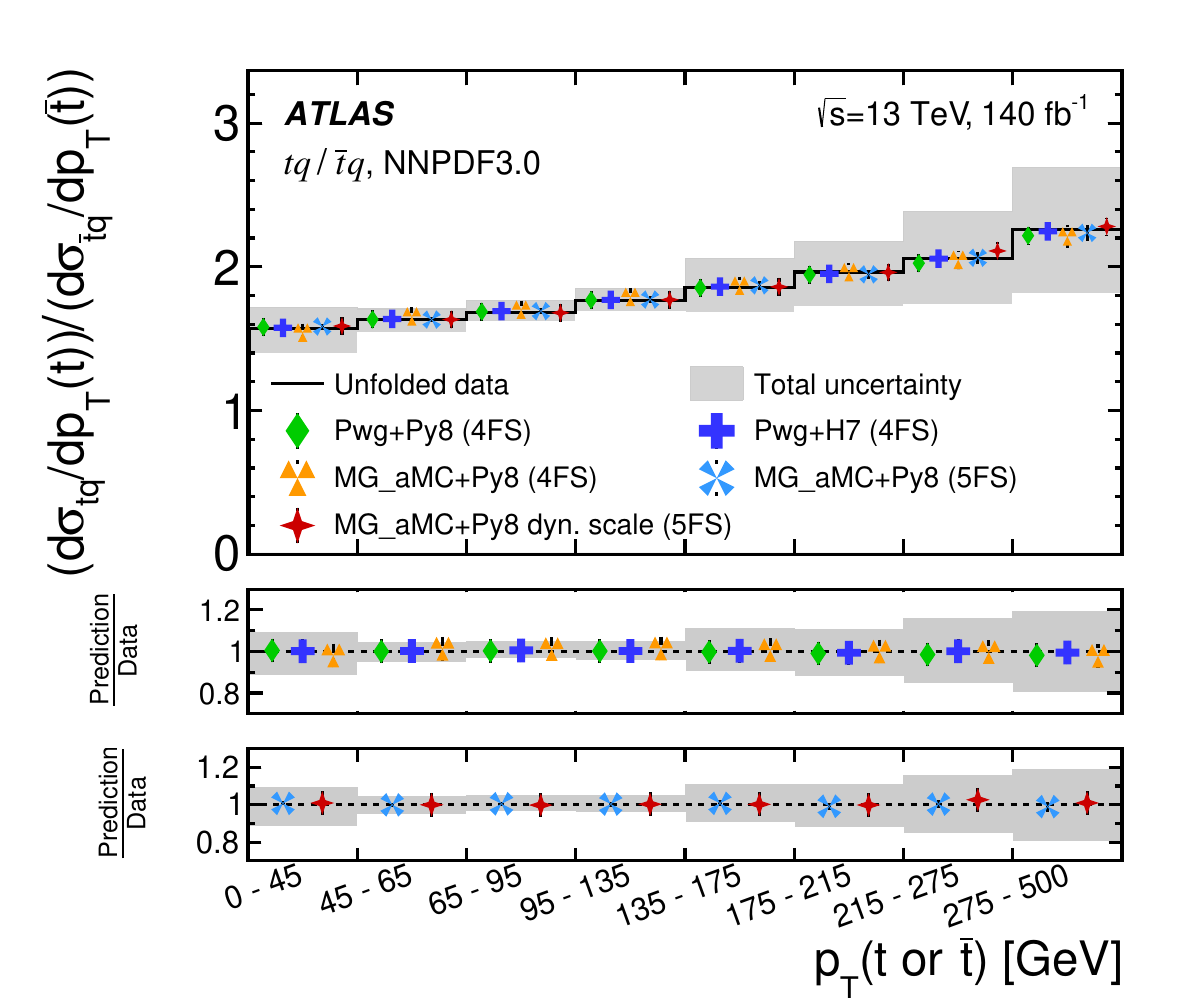}
\label{fig:ResultpTRatioAllGenerators}
}
\subfigure[]{
\includegraphics[trim={0 0.5cm 0 1cm},width=0.45\textwidth]{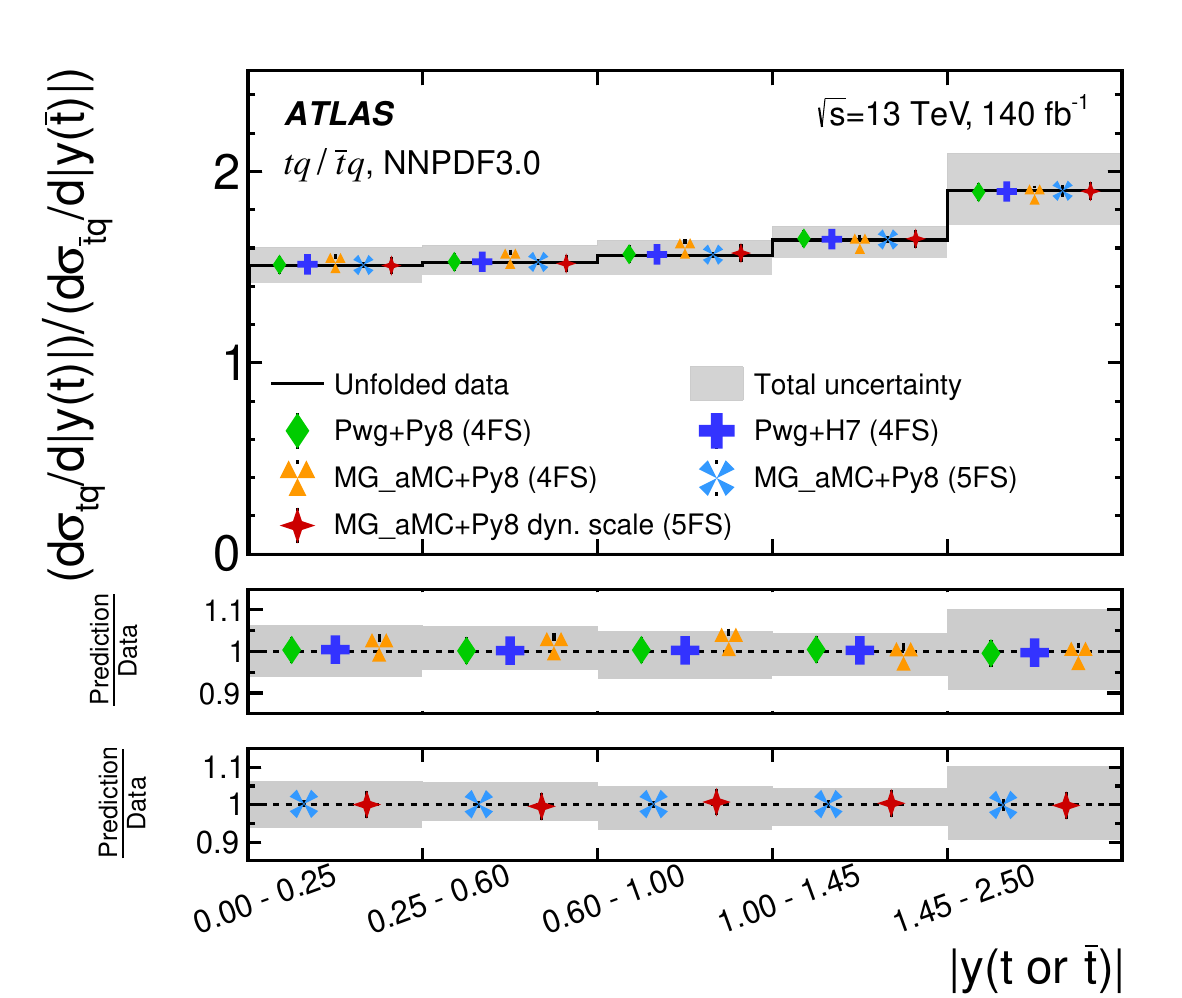}
\label{fig:ResultrapRatioAllGenerators}
}\\
\caption{Ratio of the unfolded differential cross-sections as a function of
\protect\subref{fig:ResultpTRatioMCFMDDIS} \protect\subref{fig:ResultpTRatioAllPDFs} \protect\subref{fig:ResultpTRatioAllGenerators} \pTtortbar and
\protect\subref{fig:ResultrapRatioMCFMDDIS} \protect\subref{fig:ResultrapRatioAllPDFs} \protect\subref{fig:ResultrapRatioAllGenerators} \Raptortbar.
The shaded band indicates the total measurement uncertainty.
The cross-sections are compared with theoretical predictions obtained from
\protect\subref{fig:ResultpTRatioMCFMDDIS} \protect\subref{fig:ResultrapRatioMCFMDDIS} FO calculations done with MCFM at LO, NLO and NNLO,
\protect\subref{fig:ResultpTRatioAllPDFs} \protect\subref{fig:ResultrapRatioAllPDFs} \POWHEG+\PYTHIA[8] with different PDF sets and
\protect\subref{fig:ResultpTRatioAllGenerators} \protect\subref{fig:ResultrapRatioAllGenerators} different MC event generators and parton shower programs with the \NNPDF[3.0] PDF set.
The uncertainty in the FO calculations comprises the uncertainty in the renormalisation and factorisation scale choice and the uncertainty in the value of \alphas.
The uncertainty in the predictions obtained with different PDF sets comprises the MC statistical uncertainty and the PDF uncertainty.
The uncertainty in the predictions from different event generators comprises the MC statistical uncertainty, the uncertainty in the renormalisation and factorisation scale choice and the PDF uncertainty.
}
\label{fig:ResultsRatio}
\end{figure}

\begin{table}[thb]
\caption{\label{tab:Chi2pT} $\chi^2$ probabilities of the predictions for the absolute and normalised differential \tqortbarq production cross-sections and the ratio of the cross-sections as a function of \pTt, \pTtbar and \pTtortbar respectively.
The uncertainties in the measured cross-sections and in the predictions are considered.
The number of degrees of freedom (NDF) is listed for each distribution.
}
\begin{center}
\resizebox{0.9\textwidth}{!}{
\begin{tabular}{l|c|c|c|c|c}
\toprule
Prediction & $\diff{\sigma(tq)}{p_\text{T}(t)}$ & $\frac{1}{\sigma(tq)} \diff{\sigma(tq)}{p_\text{T}(t)}$ & $\diff{\sigma(\bar{t}q)}{p_\text{T}(\bar{t})}$ & $\frac{1}{\sigma(\bar{t}q)} \diff{\sigma(\bar{t}q)}{p_\text{T}(\bar{t})}$ & $\left(\diff{\sigma(tq)}{p_\text{T}(t)}\right) / \left(\diff{\sigma(\bar{t}q)}{p_\text{T}(\bar{t})}\right) $  \\ \midrule
& NDF=8 & NDF=7 & NDF=8 & NDF=7 & NDF=8\\ \midrule
\multicolumn{6}{c}{MCFM, with PDF4LHC21} \\ \hline
LO &0.03 &0.0 & 0.02 &0.0 & 0.99\\
NLO  &0.55 &0.50 & 0.83 &0.69  & 1.0\\
NNLO &0.70 &0.65 & 0.97 &0.89  & 1.0 \\  \hline
\multicolumn{6}{c}{Generators, with NNPDF3.0} \\ \hline
\POWHEG\PYTHIA[8] (4FS)  &0.78 &0.59 & 0.98 &0.91  &1.0\\
\MGaMC (4FS) &0.60 &0.19 & 0.96 &0.86 &1.0\\
\MGaMC (5FS)  &0.60 &0.60 & 0.96 &0.93 &1.0\\
\MGaMC dyn. scale (5FS) &0.68 &0.65 & 0.90 &0.84  &1.0\\
\POWHEG+\HERWIG[7] (4FS) &0.85 &0.66 & 0.99 &0.93 &1.0\\  \hline
\multicolumn{6}{c}{PDFs, with \POWHEG\PYTHIA[8] (4FS)} \\ \hline
CT18 &0.58 &0.57 & 0.97 &0.93 &1.0\\
{\textsc MSHT20}  &0.62 &0.60 &0.97 &0.93 & 1.0\\
PDF4LHC21  &0.63 &0.63 &0.96 &0.92 &1.0 \\
ABMP16  &0.64 &0.70 & 0.96 &0.90 & 0.34\\
NNPDF4.0  &0.59 &0.59 & 0.99 &0.96 &1.0\\
ATLASpdf21  &0.55 &0.62 & 0.91 &0.92 &1.0\\
\bottomrule
\end{tabular}}
\end{center}
\end{table}

\begin{table}[thb]
\caption{\label{tab:Chi2rap} $\chi^2$ probabilities of the predictions for the absolute and normalised differential \tqortbarq production cross-sections and the ratio of the cross-sections as a function of \Rapt, \Raptbar and \Raptortbar respectively.
The uncertainties in the measured cross-sections and in the predictions are considered.
The number of degrees of freedom (NDF) is listed for each distribution.
}
\begin{center}
\resizebox{0.9\textwidth}{!}{
\begin{tabular}{l|c|c|c|c|c}
\toprule
Prediction & $\diff{\sigma(tq)}{|y(t)|}$ & $ \frac{1}{\sigma(tq)} \diff{\sigma(tq)}{|y(t)|}$ & $\diff{\sigma(\bar{t}q)}{|y(\bar{t})|}$ & $ \frac{1}{\sigma(\bar{t}q)} \diff{\sigma(\bar{t}q)}{|y(\bar{t})|}$  & $\left(\diff{\sigma(tq)}{|y(t)|}\right) / \left(\diff{\sigma(\bar{t}q)}{|y(\bar{t})|}\right)$  \\ \midrule
& NDF=5 & NDF=4 & NDF=5 & NDF=4 & NDF=5\\ \midrule
\multicolumn{6}{c}{MCFM, with PDF4LHC21} \\ \hline
LO &0.65& 0.54 &0.82 &0.86 &0.99\\
NLO &0.97 &0.93 &0.68 &0.81 &1.0\\
NNLO & 1.0 &0.98 &0.81 &0.74 &1.0\\  \hline
\multicolumn{6}{c}{Generators, with NNPDF3.0} \\ \hline
\POWHEG\PYTHIA[8] (4FS) & 1.0 &0.98 &0.81 &0.61 &1.0\\
\MGaMC (4FS) & 0.99 &0.93 &0.81 &0.53 &0.99\\
\MGaMC (5FS) & 0.99 &0.97 &0.69 &0.61 &1.0\\
\MGaMC dyn. scale (5FS) & 0.99 & 0.97 &0.65 &0.57 &1.0\\
\POWHEG+\HERWIG[7] (4FS) & 1.0 & 0.98 &0.84 &0.62 &1.0 \\ \hline
\multicolumn{6}{c}{PDFs, with \POWHEG\PYTHIA[8] (4FS)} \\ \hline
CT18 & 0.98 &0.96 &0.68 &0.55 & 0.96\\
{\textsc MSHT20} &1.0 &1.0 &0.61 &0.52 & 1.0\\
PDF4LHC21 & 0.99 &0.98 &0.70 &0.65 &1.0\\
ABMP16 & 0.91 &0.98 &0.38 &0.18 & 0.21\\
NNPDF4.0 & 0.99 &0.99 &0.70 &0.61 &1.0\\
ATLASpdf21 & 0.92 &0.99 &0.64 &0.84 &1.0\\
\bottomrule
\end{tabular}}
\end{center}
\end{table}

%
%
%
%
%
%
%
%
%
%
%
%
%
%
%
%
%
%
%
%
%
%
%
%
%
%
%
%
%
%
%

%
%
%
%
%
%
%
%
%
%
%
%
%
%
%
%
%
%
%
%
%
%
%
%
%
%
%
%
%
%

%
%
%
%
%
%
%
%
%
%
%
%
%
%
%
%
%
%
%
%
%
%
%
%
%
%
%
%
%
%


\FloatBarrier

\subsection{Effective field theory interpretation}

The SMEFT provides a model-independent framework for
indirect searches for new physics. Within this framework, the SM is regarded as a low-energy
approximation of a more fundamental theory involving interactions at an energy scale $\Lambda$.
The impact of new physics is parameterised by higher-dimensional operators maintaining SM symmetries.
The effective dimension-6 Lagrangian is given by
\[
\mathcal{L}_{\mathrm{eff}}=\mathcal{L}_{\mathrm{SM}}+\sum_{i} \frac{C_{i}}{\Lambda^{2}} O_{i}
+ \mathrm{Hermitian\; conjugate},
\]
where $\mathcal{L}_{\mathrm{SM}}$ is the SM Lagrangian. The $O_i$ are effective dimension-6 operators
and the $C_i$ are the associated Wilson coefficients.

In this EFT interpretation, constraints on the contribution from the four-quark operator \Oqq~\cite{Buckley:2016} are set.
The relevant operators, expressed in the Warsaw basis~\cite{Grzadkowski_2010}, are
\begin{equation*}
\arraycolsep=1.4pt\def\arraystretch{2.2}
\begin{array}{r@{}l}
O_{qq}^{1(ijkl)} & = (\bar{q}_i \gamma^\mu q_j)(\bar{q}_k \gamma_\mu q_l) \ \ \ \mathrm{and} \\
O_{qq}^{3(ijkl)} & = (\bar{q}_i \gamma^\mu \tau^I q_j)(\bar{q}_k \gamma_\mu \tau^I q_l).
\end{array}
\end{equation*}
The \(q\) denote weak-isospin doublets and $ijkl\in {1,2,3}$ are the quark generation indices.
All contributing four-quark processes depend solely on a linear combination of Wilson coefficients
\begin{equation*}
C_{Qq}^{3,1} = \sum_{i=1,2} C_{qq}^{3(ii33)} + \frac{1}{6} C_{qq}^{1(i33i)} - \frac{1}{6} C_{qq}^{3(i33i)},
\end{equation*}
where $Q$ represents the doublet of the third quark generation,
therefore the four-quark interaction is fully characterised by \Oqq.
There are two other dimension-6 operators which can alter single-top production and interfere with the SM diagrams at LO, namely $O_{\Phi Q}^3$ and $O_{tW}$ \cite{TOPQ-2020-17}.
They are not considered in this interpretation, i.e. their respective Wilson coefficients are set to zero.
The operator \Oqq leads to non-SM single top-quark production, as illustrated in
\Fig{\ref{fig:Feynman_EFT_tchan}}, for example via the process \(b+u\rightarrow t+d\).
Other possible processes like \(d+u\rightarrow t+b\) and variants containing antiquarks and/or quarks of the second generation are considered in the EFT interpretation.
Top quarks produced in this way feature different angular distributions and transverse momentum spectra than those produced according to the SM.

\begin{figure}[htb]
\centering
\includegraphics[width=0.25\textwidth]{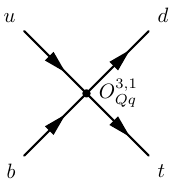}
\caption{Exemplary LO Feynman diagram of a four-quark contact interaction leading to the
production of a single top quark.}
\label{fig:Feynman_EFT_tchan}
\end{figure}

For large absolute values of \Cqq, the number of single top quarks and top antiquarks produced with a high transverse momentum is expected to increase.
Therefore, the high-\pT region of the spectra provides a high sensitivity to \Cqq.
Since the radiation of additional jets leads to increased jet-\pT, this analysis is not optimised for a high sensitivity to \Cqq, as the event selection requires candidate events to contain exactly two jets.
Nevertheless, the differential cross-sections as a function of \pTt or \pTtbar are measured for high transverse momenta of up to 500 \GeV, making an interpretation feasible.

The differential cross-sections as a function of \pTt or \pTtbar are parameterised by a polynomial of second degree in the EFT coefficient \Cqq
\begin{equation}\label{eq:Cqqdependency}
\arraycolsep=1.4pt\def\arraystretch{2.2}
\begin{array}{r@{}l}
\diff{\sigma_{(\tortbar)}}{p_\text{T}(\tortbar)}(\Cqq) & = \left(\diff{\sigma_{(\tortbar)}}{p_\text{T}(\tortbar)}\right)_\text{SM}\cdot\dxsecEFT,  \ \ \ \mathrm{with} \\
\dxsecEFT & = \left(1 + a_1 \cdot \Cqq + a_2 \cdot (\Cqq)^2\right).
\end{array}
\end{equation}

The fixed-order predictions obtained with MCFM at NNLO are used in $(\text{d}\sigma_{(\tortbar)}/\text{d}p_\text{T}(\tortbar))_\text{SM}$, which is the SM prediction.
The expected relative change of the cross-sections as a function of \Cqq is described by \dxsecEFT,
where the term linear in \Cqq covers the effect of the interference of SM and non-SM amplitudes and the term proportional to $(\Cqq)^2$ is entirely due to the four-quark operator.
It is obtained by unfolding the EFT MC samples using the nominal migration matrix and selection efficiency;
and the resulting cross-section predictions are normalised to the prediction obtained from unfolding the EFT MC sample generated with \Cqq set to zero.

The distributions from the EFT samples are unfolded to extract the parameterisations because a non-zero contribution from \Cqq alters the kinematic properties of the event, which impacts the selection efficiency.
It was validated that the impact is not primarily caused by the \Dnn requirement.
The selection efficiencies obtained for the EFT MC samples are shown in \Fig{\ref{fig:EFTselectionEff}}.
The decreased selection efficiencies for large positive values of \Cqq negate the effect of the increased cross-section predictions.
The magnitude of the effect is shown exemplarily for one bin of the differential $tq$ cross-section in \Fig{\ref{fig:EFTparameterisation}}, where the parameterisation of \dxsecEFT obtained with the approach explained above is compared with the parameterisation obtained from parton-level predictions.
The parameterisation obtained from parton-level predictions is not used for the interpretation.

\begin{figure}[!htbp]
\centering
\subfigure[]{
\includegraphics[width=0.45\textwidth]{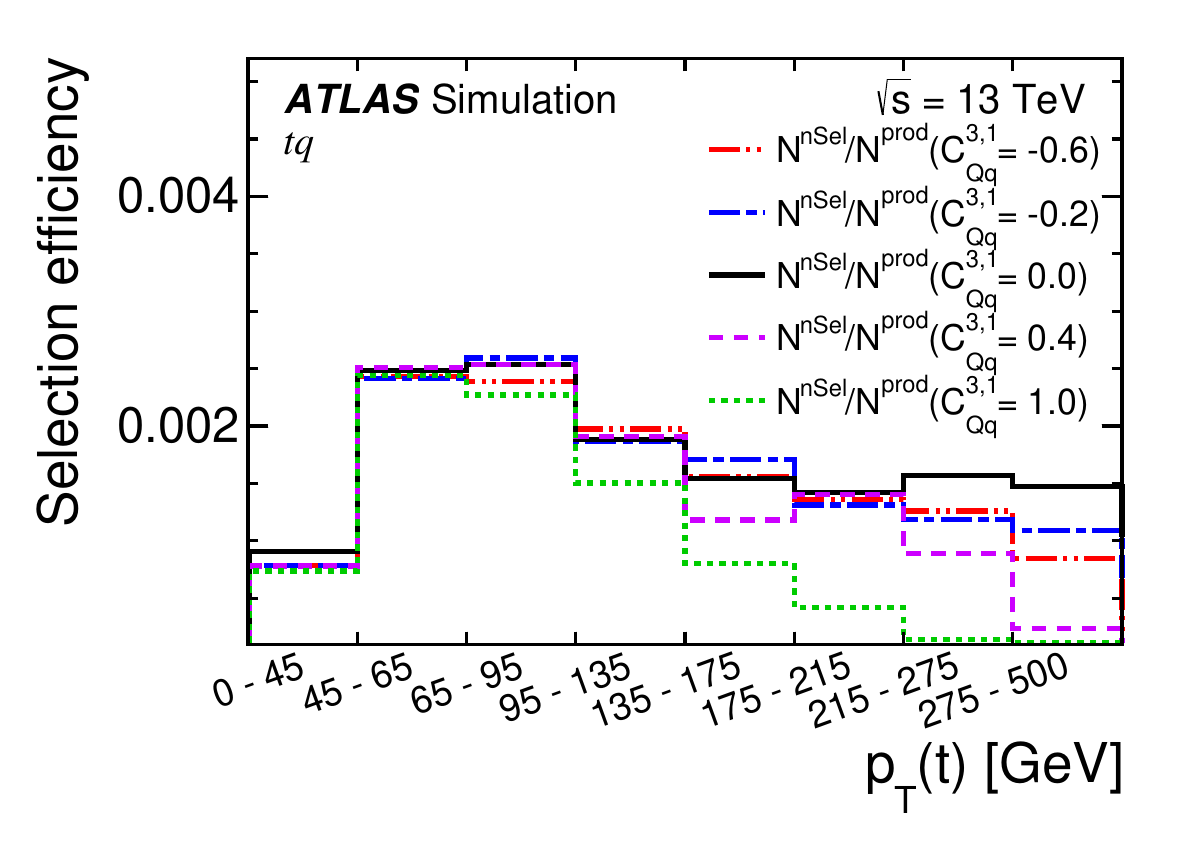}
\label{fig:EFTselectionEffSRp}
}
\subfigure[]{
\includegraphics[width=0.45\textwidth]{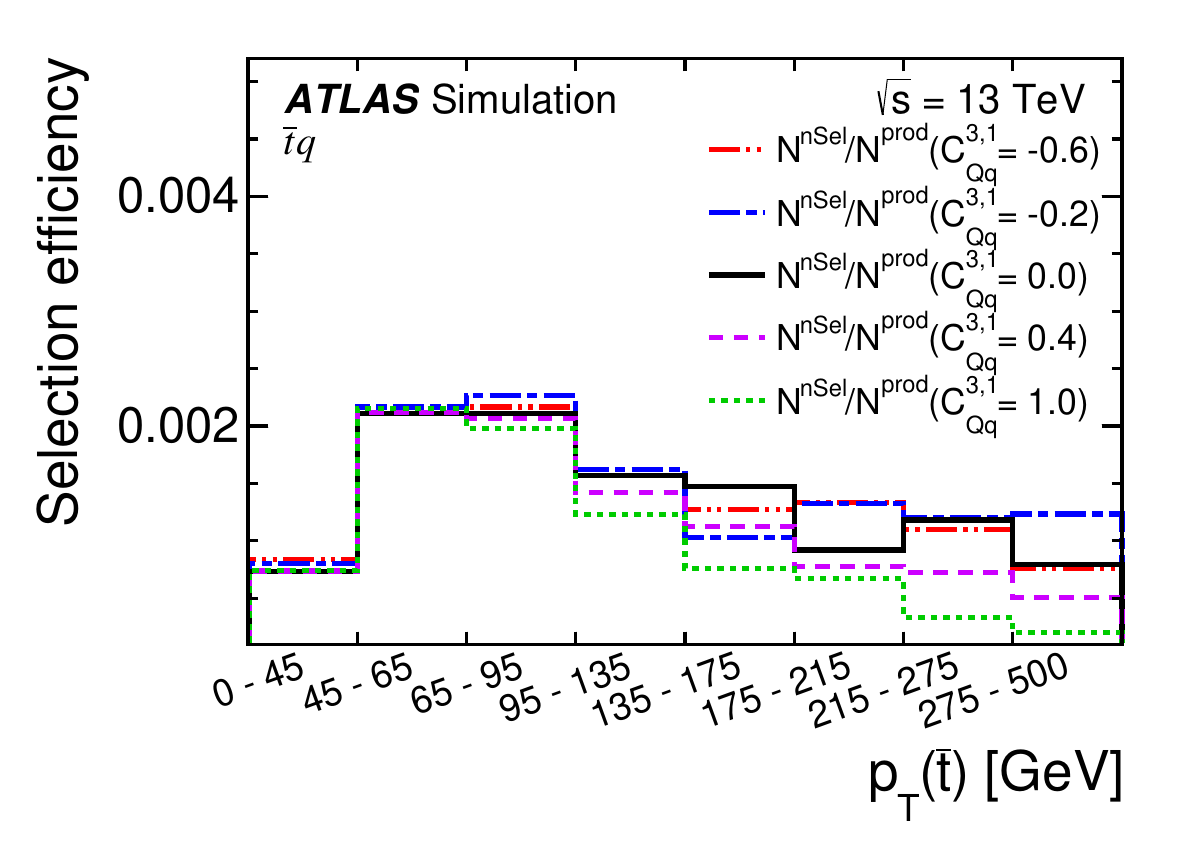}
\label{fig:EFTselectionEffSRn}
}
\caption{Selection efficiencies for the \protect\subref{fig:EFTselectionEffSRp} \pTt and \protect\subref{fig:EFTselectionEffSRn} \pTtbar spectra calculated for \tqtbarq production events generated with different values of \Cqq, with $\Lambda$ set to $\SI{1}{\TeV}$.
}
\label{fig:EFTselectionEff}
\end{figure}

\begin{figure}[!htbp]
\centering
\subfigure[]{
\includegraphics[width=0.7\textwidth]{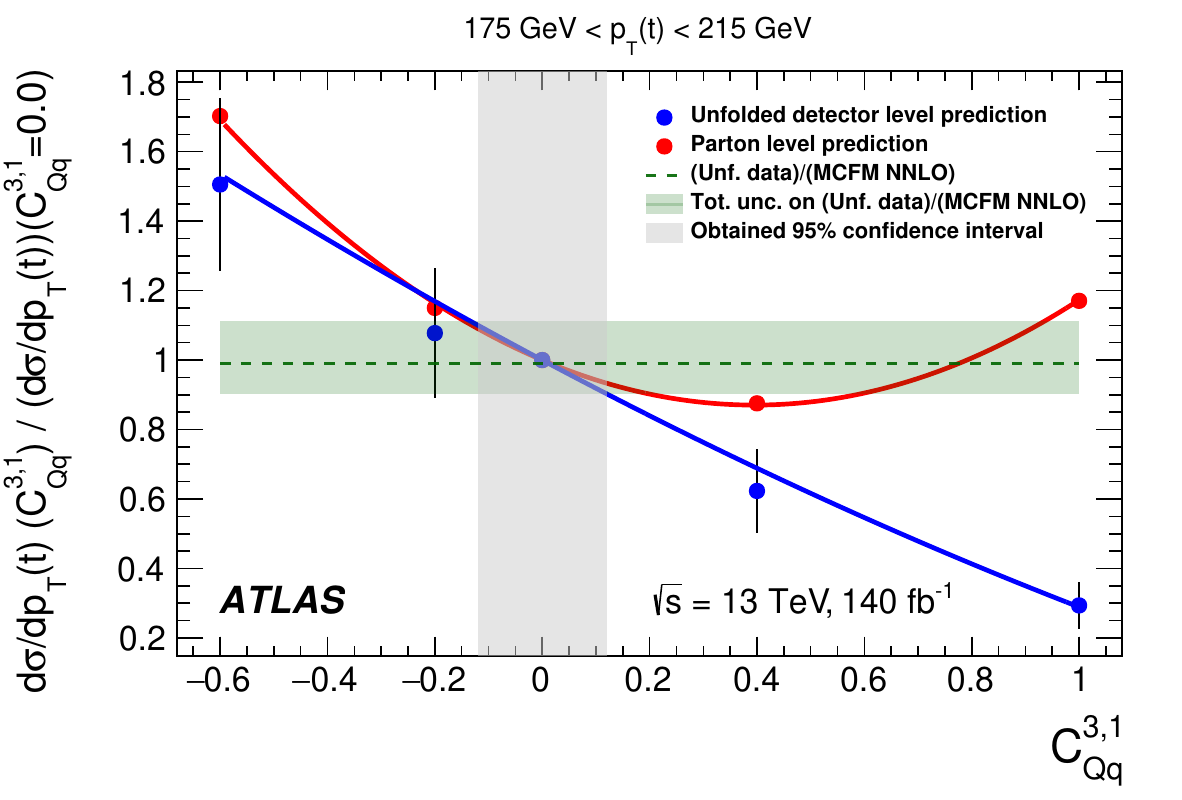}
}
\caption{ Comparison of the parameterisation for the expected relative change of the differential $tq$ production cross-section as a function of \Cqq, with $\Lambda$ set to $\SI{1}{\TeV}$, obtained
from parton level predictions and by unfolding the detector level predictions with the nominal unfolding corrections.
The uncertainties in the predictions are MC statistical uncertainties only.
The dashed line indicates the measurement result over the theoretical prediction calculated with MCFM at NNLO (cf. \Eqn{\ref{eq:Cqqdependency}}).
The uncertainty band includes the total uncertainty in the measurement result and the uncertainty in the theory prediction, which considers the uncertainty in the renormalisation and factorisation scale choice and the value of \alphas.
The vertical shaded area shows the 95\% confidence interval obtained in the EFT interpretation of the measurement.
}
\label{fig:EFTparameterisation}
\end{figure}

The parameters $a_1$ and $a_2$ in \Eqn{\ref{eq:Cqqdependency}} are determined via least-squares fits performed for each bin of the unfolded distributions.
The uncertainties in the measured result, the SM prediction and the statistical uncertainty due to the limited size
of the EFT MC samples are taken into account for the least-squares fit.
The uncertainty due to the limited statistical power is evaluated by unfolding pseudo-experiments which are produced by fluctuating each bin of the detector level distributions according to a Gaussian distribution whose spread corresponds to the expected uncertainty due to the number of MC events in that bin.
The standard deviation of all unfolded pseudo-experiments per bin is taken as statistical uncertainty.
The statistical uncertainties of the different EFT MC samples are treated as uncorrelated.
The correlation of the total uncertainty in the measurement result between bins is not considered for the least-squares fit.
The parameterisations for \dxsecEFT are given in \Tab{\ref{tab:EFTFitFuncs}}.

The EFTfitter tool~\cite{EFTfitter} is used to constrain \Cqq in a Bayesian statistical framework.
The total uncertainty in the measured results is taken into account and symmetrised by taking the larger of the two variations.
Correlations between the single bins of the distributions are accounted for via the correlation matrices of the differential measurements.
Uncertainties in the SM prediction are taken into account as well and comprise the uncertainty in the renormalisation and factorisation scale and the uncertainty in the value of \alphas.
The 95\% confidence level (CL) interval for \Cqq is determined to be
\begin{equation*}
\SI{-0.12}{\TeV^{-2}} < \Cqq < \SI{0.12}{\TeV^{-2}}.
\end{equation*}
The constraints improve upon the limits set by the ATLAS measurement of the total \tqtbarq cross-sections \Refn{\cite{TOPQ-2019-02}}, where a confidence interval $\Cqq\in[-0.37,0.06]\,\TeV^{-2}$ at 95\% CL was obtained.

%
%
%


%

%
%

%
%
\FloatBarrier
\section{Conclusion}

The production of single top quarks and top antiquarks in \tqtbarq final states proceeds in the SM via the \tchan exchange of a virtual $W$ boson.
Differential \tqtbarq production cross-sections are measured in proton--proton collisions at a centre-of-mass energy of \SI{13}{\TeV} using the full Run 2 data sample recorded by the
ATLAS detector at the LHC in the years 2015--2018, corresponding to an integrated luminosity of \SI{140}{\per\fb}.
Events are required to contain exactly one isolated electron or muon, high missing transverse momentum and exactly two jets, one of which must be identified as originating from a \bquark.
A neural network is employed to separate signal from background events and a requirement of a high neural-network output score is imposed to achieve a high purity in signal events.

The absolute and normalised \tqtbarq production cross-sections are measured differentially at parton level as a function
of the transverse momentum \pTt or \pTtbar and the absolute value of the rapidity \Rapt or \Raptbar of the top quark and top antiquark.
Differential \tqtbarq production cross-sections are measured separately and their ratio is measured as well.
The measured cross-sections are compared with various predictions obtained from fixed order calculations up to NNLO in QCD and from MC event generators interfaced to parton shower programs at NLO.
Overall, a good agreement between the measurement results and the theory predictions is observed within the measurement uncertainties.
The measurement is limited by systematic uncertainties, where the largest contributions are experimental uncertainties and
uncertainties related to the modelling of the signal processes.
Data statistical uncertainties are significant for the normalised differential cross-sections and the cross-section ratio.

This analysis provides the first separate measurement of differential \tqtbarq production at \SI{13}{\TeV} and the first measurement of the ratio of the differential $tq/\bar{t}q$ cross-sections.
Compared with the measurements of the differential cross-sections at \SI{8}{\TeV}, a significant reduction of the uncertainties is achieved and measurement range for
the differential cross-sections as a function of \pTt or \pTtbar is extended to \SI{500}{\gev}.

The differential cross-sections as a function of \pTt or \pTtbar are interpreted in an EFT approach, setting limits at the 95\% confidence level on contributions from the
four-quark operator \Oqq. The impact of \Oqq on the acceptance, caused by altered event kinematics, is taken into account.
The 95\% confidence interval $\SI{-0.12}{\TeV^{-2}} < \Cqq < \SI{0.12}{\TeV^{-2}}$ is obtained, improving the constraints set by an interpretation of the ATLAS measurement of the total \tqtbarq cross-sections.


%
%
%
%
%
\section*{Acknowledgements}
%

%

%
%

%
%

We thank CERN for the very successful operation of the LHC and its injectors, as well as the support staff at
CERN and at our institutions worldwide without whom ATLAS could not be operated efficiently.

The crucial computing support from all WLCG partners is acknowledged gratefully, in particular from CERN, the ATLAS Tier-1 facilities at TRIUMF/SFU (Canada), NDGF (Denmark, Norway, Sweden), CC-IN2P3 (France), KIT/GridKA (Germany), INFN-CNAF (Italy), NL-T1 (Netherlands), PIC (Spain), RAL (UK) and BNL (USA), the Tier-2 facilities worldwide and large non-WLCG resource providers. Major contributors of computing resources are listed in Ref.~\cite{ATL-SOFT-PUB-2026-001}.

We gratefully acknowledge the support of ANPCyT, Argentina; YerPhI, Armenia; ARC, Australia; BMWFW and FWF, Austria; ANAS, Azerbaijan; CNPq and FAPESP, Brazil; NSERC, NRC and CFI, Canada; CERN; ANID, Chile; CAS, MOST and NSFC, China; Minciencias, Colombia; MEYS CR, Czech Republic; DNRF and DNSRC, Denmark; IN2P3-CNRS and CEA-DRF/IRFU, France; SRNSFG, Georgia; BMFTR, HGF and MPG, Germany; GSRI, Greece; RGC and Hong Kong SAR, China; ICHEP and Academy of Sciences and Humanities, Israel; INFN, Italy; MEXT and JSPS, Japan; CNRST, Morocco; NWO, Netherlands; RCN, Norway; MNiSW, Poland; FCT, Portugal; MNE/IFA, Romania; MSTDI, Serbia; MSSR, Slovakia; ARIS and MVZI, Slovenia; DSI/NRF, South Africa; MICIU/AEI, Spain; SRC and Wallenberg Foundation, Sweden; SERI, SNSF and Cantons of Bern and Geneva, Switzerland; NSTC, Taipei; TENMAK, T\"urkiye; STFC/UKRI, United Kingdom; DOE and NSF, United States of America.

Individual groups and members have received support from BCKDF, CANARIE, CRC and DRAC, Canada; CERN-CZ, FORTE and PRIMUS, Czech Republic; COST, ERC, ERDF, Horizon 2020, ICSC-NextGenerationEU and Marie Sk{\l}odowska-Curie Actions, European Union; Investissements d'Avenir Labex, Investissements d'Avenir Idex and ANR, France; DFG and AvH Foundation, Germany; Herakleitos, Thales and Aristeia programmes co-financed by EU-ESF and the Greek NSRF, Greece; BSF-NSF and MINERVA, Israel; NCN and NAWA, Poland; La Caixa Banking Foundation, CERCA Programme Generalitat de Catalunya and PROMETEO and GenT Programmes Generalitat Valenciana, Spain; G\"{o}ran Gustafssons Stiftelse, Sweden; The Royal Society and Leverhulme Trust, United Kingdom; Eric and Wendy Schmidt Fund for Strategic Innovation, United States of America.

In addition, individual members wish to acknowledge support from CERN: European Organization for Nuclear Research (CERN DOCT); Chile: Agencia Nacional de Investigaci\'on y Desarrollo (ANID FONDECYT reg. 1230987, FONDECYT 1230812, FONDECYT 1240864, Fondecyt 3240661, Fondecyt Regular 1240721); China: Chinese Ministry of Science and Technology (MOST-2023YFA1605700, MOST-2023YFA1609300), National Natural Science Foundation of China (NSFC - 12175119, NSFC 12275265); Czech Republic: Czech Science Foundation (GACR - 24-11373S), Ministry of Education Youth and Sports (ERC-CZ-LL2327, FORTE CZ.02.01.01/00/22\_008/0004632), PRIMUS Research Programme (PRIMUS/21/SCI/017); EU: H2020 European Research Council (ERC - 101002463); European Union: European Research Council (BARD No. 101116429, ERC - 948254, ERC 101089007), European Regional Development Fund (HE COFUND GA No.101081355, ERDF), European Union, Future Artificial Intelligence Research (FAIR-NextGenerationEU PE00000013), Italian Center for High Performance Computing, Big Data and Quantum Computing (ICSC, NextGenerationEU), Marie Sklodowska-Curie Actions (GAP-101168829); France: Agence Nationale de la Recherche (ANR-21-CE31-0013, ANR-21-CE31-0022, ANR-22-EDIR-0002, ANR-24-CE31-0504-01); Germany: Deutsche Forschungsgemeinschaft (DFG - 469666862, DFG - CR 312/5-2); China: Research Grants Council (GRF); Italy: Istituto Nazionale di Fisica Nucleare (ICSC, NextGenerationEU), Ministero dell'Università e della Ricerca (NextGenEU 153D23001490006 M4C2.1.1, NextGenEU I53D23000820006 M4C2.1.1, NextGenEU I53D23001490006 M4C2.1.1, SOE2024\_0000023); Japan: Japan Society for the Promotion of Science (JSPS KAKENHI JP22H01227, JSPS KAKENHI JP22H04944, JSPS KAKENHI JP22KK0227, JSPS KAKENHI JP24K23939, JSPS KAKENHI JP24KK0251, JSPS KAKENHI JP25H00650, JSPS KAKENHI JP25H01291, JSPS KAKENHI JP25K01023); Norway: Research Council of Norway (RCN-314472); Poland: Ministry of Science and Higher Education (IDUB AGH, POB8, D4 no 9722), Polish National Science Centre (NCN 2021/42/E/ST2/00350, NCN OPUS 2023/51/B/ST2/02507, NCN OPUS nr 2022/47/B/ST2/03059, NCN UMO-2019/34/E/ST2/00393, UMO-2022/47/O/ST2/00148, UMO-2023/49/B/ST2/04085, UMO-2023/51/B/ST2/00920, UMO-2024/53/N/ST2/00869); Portugal: Foundation for Science and Technology (FCT); Spain: Agencia de Gestión de Ayudas Universitarias y de Investigación (AGAUR - 2023 BP 00141), Generalitat Valenciana (ASFAE/2022/008), Ministry of Science and Innovation (MCIN \& NextGenEU PCI2022-135018-2, MICIN \& FEDER PID2021-125273NB, RYC2019-028510-I, RYC2020-030254-I, RYC2021-031273-I, RYC2022-038164-I), Ministerio de Ciencia, Innovación y Universidades/Agencia Estatal de Investigaci\'on (PID2022-142604OB-C22); Sweden: Carl Trygger Foundation (Carl Trygger Foundation CTS 22:2312), Swedish Research Council (Swedish Research Council 2023-04654, VR 2021-03651, VR 2022-03845, VR 2022-04683, VR 2023-03403, VR 2024-05451), Knut and Alice Wallenberg Foundation (KAW 2018.0458, KAW 2022.0358, KAW 2023.0366); Switzerland: Swiss National Science Foundation (SNSF - PCEFP2\_194658); United Kingdom: Royal Society (NIF-R1-231091); United States of America: U.S. Department of Energy (ECA DE-AC02-76SF00515), Neubauer Family Foundation.

%
%


%
\clearpage
\appendix
\part*{Appendix}
\addcontentsline{toc}{part}{Appendix}

%
\subsection*{Migration matrices}
The migration matrices are optimised according to the criteria explained in \Sect{\ref{sec:inputs}}.
The obtained migration matrices are presented in \Fig{\ref{fig:MMs}}.
\begin{figure}[htbp]
\centering
\subfigure[]{
\includegraphics[width=0.45\textwidth]{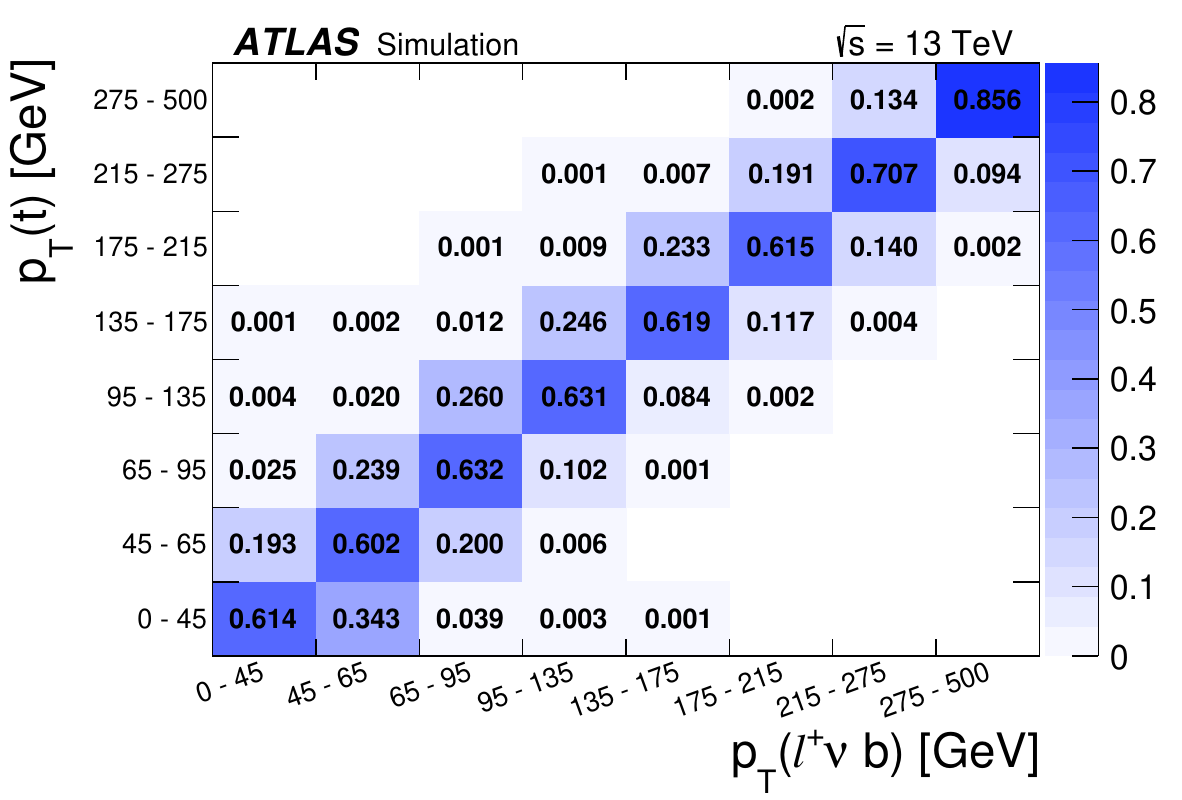}
\label{fig:MM_pt_SRp}
}
\subfigure[]{
\includegraphics[width=0.45\textwidth]{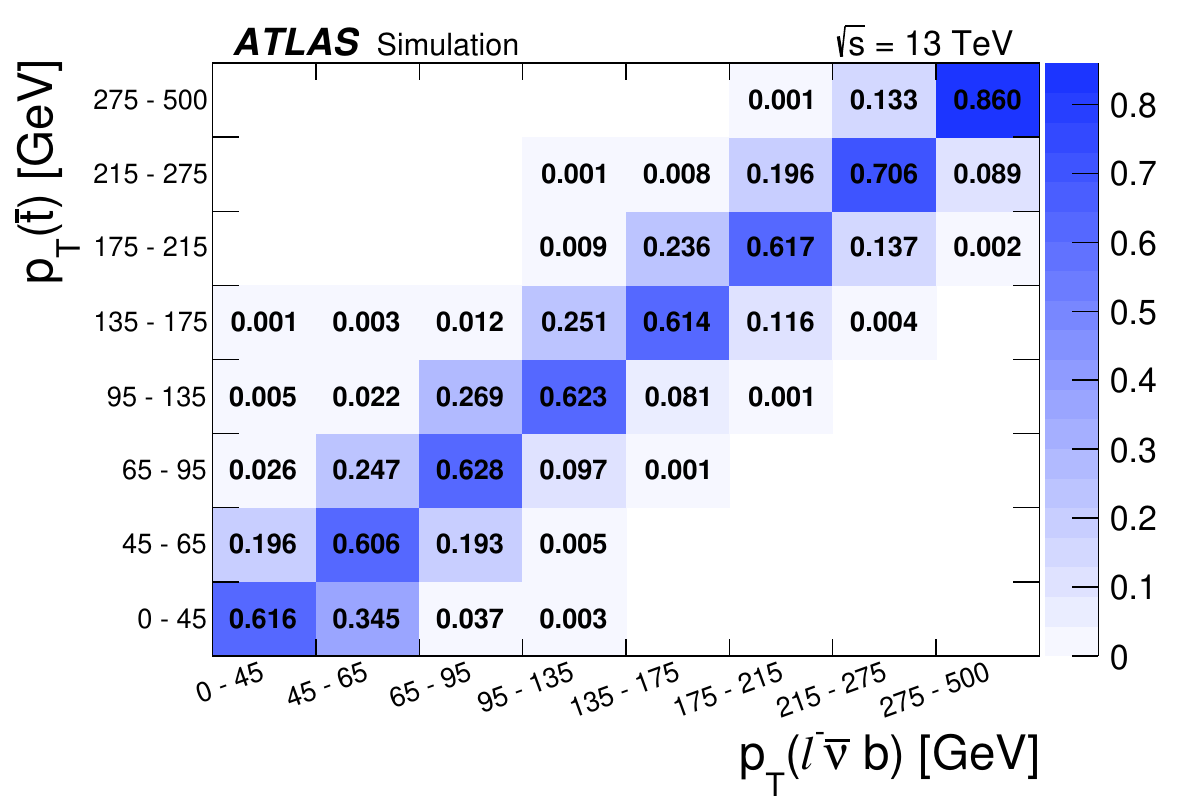}
\label{fig:MM_pt_SRn}
}

\subfigure[]{
\includegraphics[width=0.45\textwidth]{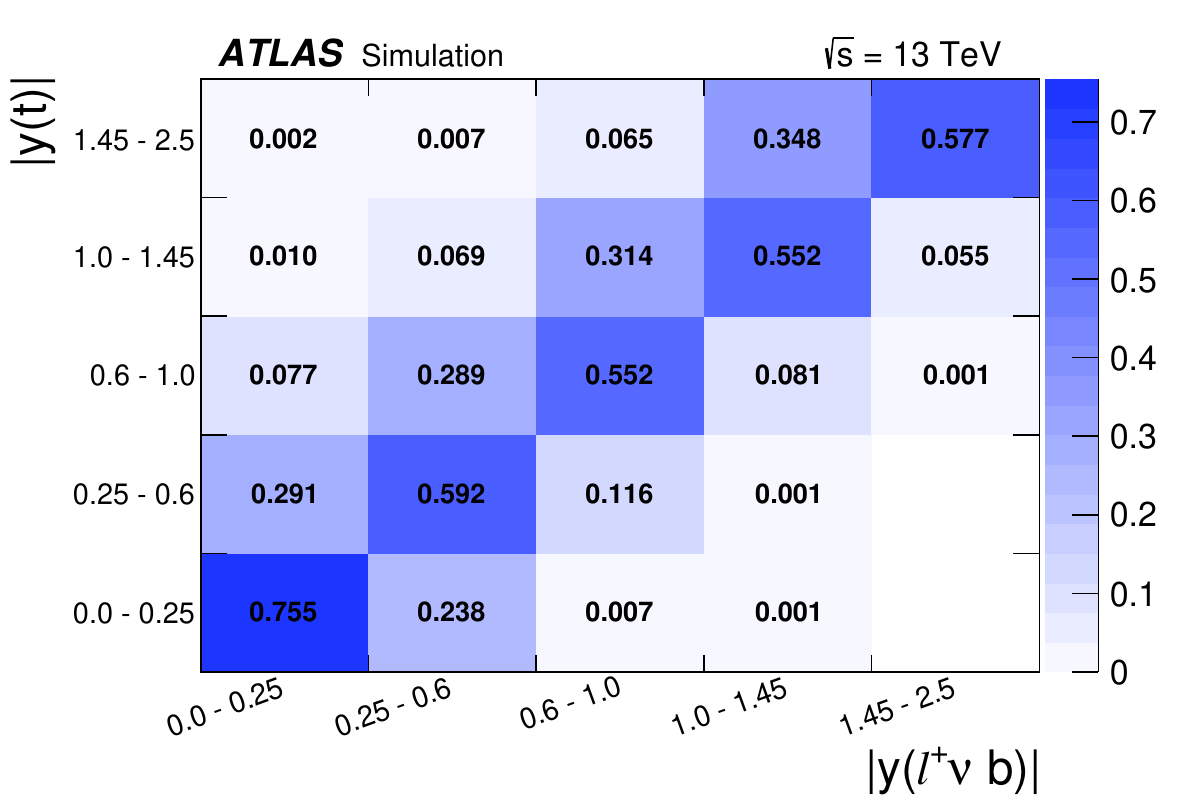}
\label{fig:MM_rap_SRp}
}
\subfigure[]{
\includegraphics[width=0.45\textwidth]{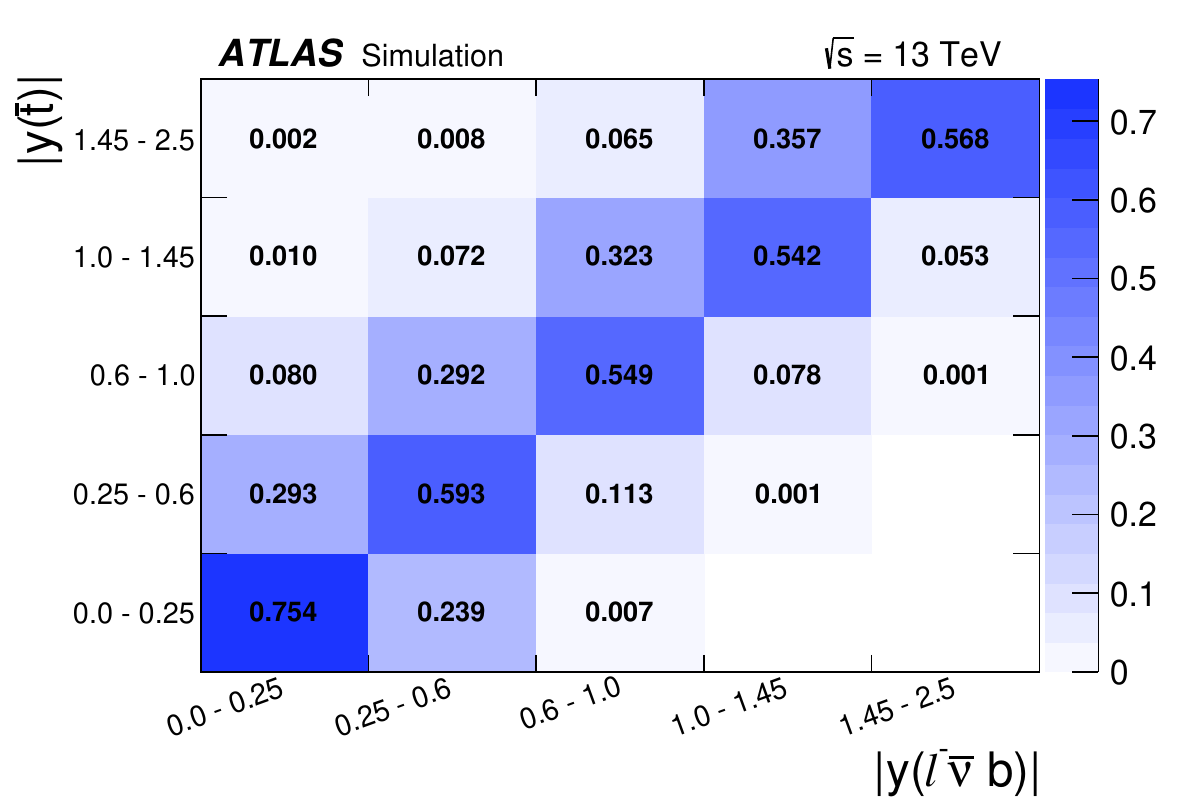}
\label{fig:MM_rap_SRn}
}
\caption{Migration matrices for the \protect\subref{fig:MM_pt_SRp},\protect\subref{fig:MM_pt_SRn} \pTtreco and \protect\subref{fig:MM_rap_SRp},\protect\subref{fig:MM_rap_SRn} \Raptreco distributions in the \protect\subref{fig:MM_pt_SRp},\protect\subref{fig:MM_rap_SRp} $\ell^+$ SR and \protect\subref{fig:MM_pt_SRn},\protect\subref{fig:MM_rap_SRn} $\ell^-$ SR\@.
The variables on parton-level are shown on the $y$-axis and the reconstructed variables are shown on the $x$-axis.
The rows are normalised to unity.}
\label{fig:MMs}
\end{figure}

\FloatBarrier
\subsection*{Systematic and statistical uncertainties}
The statistical and systematic uncertainties in the measured results are evaluated according to the description in \Sect{\ref{sec:uncertainties}}.
The impact of the different sources of uncertainties are given in \Tabrange{\ref{tab:CommonSubSyspTSRp}}{\ref{tab:CommonSubSysRatiorap}}.
Experimental uncertainties are further separated into components related to the JES, JER, (f)JVT, $b$--tagging, pile-up, \met and the charged lepton.

\begin{table}[!htbp]
\caption{\label{tab:CommonSubSyspTSRp}
Summary of the uncertainties in the absolute differential $tq$ production cross-sections measured as a function of \pTt.
Similar sources of uncertainties are grouped using their quadratic sum.
The total uncertainty is defined as the quadratic sum of all sources of uncertainties.
}
\begin{center}
\resizebox{\textwidth}{!}{
\begin{tabular}{l|c|c|c|c|c|c|c|c}
\toprule
Systematic & \multicolumn{8}{c}{Systematic uncertainty in [\%] on the absolute $tq$ cross-section in the given $p_\text{T}(t)$ interval}  \\ \midrule
& [0, 45] \GeV & [45, 65] \GeV & [65, 95] \GeV & [95, 135] \GeV & [135, 175] \GeV & [175, 215] \GeV & [215, 275] \GeV & [275, 500] \GeV   \\ \midrule
JER & 1.5/ --4.2 & 2.4/ --3.5 & 2.7/ --1.0 & 1.0/ --1.9 & 2.9/ --0.6 & 0.8/ --3.6 & 3.0/ --0.4 & 1.7/ --2.6\\
JES & 5.7/ --3.2 & 1.6/ --1.0 & 2.3/ --1.7 & 4.8/ --4.8 & 5.6/ --4.9 & 6.3/ --4.5 & 3.4/ --5.8 & 5.5/ --3.9\\
(f)JVT & 1.8/ --1.8 & 1.3/ --1.3 & 0.8/ --0.8 & 0.3/ --0.3 & 0.6/ --0.6 & 0.5/ --0.5 & 0.5/ --0.5 & 0.5/ --0.5\\
Lepton & 1.4/ --1.2 & 1.1/ --1.1 & 1.0/ --1.0 & 1.0/ --1.0 & 1.0/ --1.0 & 1.0/ --1.0 & 1.1/ --1.0 & 1.0/ --1.0\\
$E_\text{T}^{miss}$ & 2.1/ --0.0 & 0.7/ --0.3 & 0.8/ --0.3 & 1.0/ --0.2 & 1.0/ --0.1 & 1.8/ --0.0 & 0.7/ --0.8 & 2.5/ --0.0\\
$b$--tagging & 2.2/ --2.2 & 1.7/ --1.7 & 1.5/ --1.5 & 1.4/ --1.4 & 1.3/ --1.3 & 1.3/ --1.3 & 1.2/ --1.2 & 1.3/ --1.3\\
Pile-up & 1.4/ --1.4 & 1.3/ --1.2 & 1.5/ --1.3 & 1.4/ --1.3 & 1.3/ --1.2 & 1.1/ --1.1 & 1.2/ --1.1 & 1.3/ --1.3\\
Luminosity & 0.8/ --0.8 & 0.8/ --0.8 & 0.8/ --0.8 & 0.8/ --0.8 & 0.8/ --0.8 & 0.8/ --0.8 & 0.8/ --0.8 & 0.8/ --0.8\\
Signal modelling & 5.2/ --4.9 & 6.2/ --6.0 & 5.8/ --5.4 & 6.1/ --5.5 & 7.0/ --5.7 & 6.9/ --6.7 & 12/ --12 & 16/ --16\\
Background modelling & 1.5/ --2.0 & 0.9/ --0.6 & 1.1/ --0.7 & 0.7/ --0.7 & 1.0/ --1.0 & 1.1/ --1.6 & 2.4/ --2.4 & 2.5/ --2.2\\
Background rate  & 2.9/ --2.9 & 1.4/ --1.4 & 0.7/ --0.7 & 0.5/ --0.5 & 0.5/ --0.5 & 0.4/ --0.4 & 0.4/ --0.4 & 0.6/ --0.6\\ \hline
Total systematic & 9.5/ --8.7 & 7.6/ --7.7 & 7.4/ --6.4 & 8.3/ --7.9 & 9.8/ --8.0 & 9.9/ --9.3 & 13/ --13 & 18/ --17\\
Statistical  & 2.0/ --2.0 & 1.3/ --1.3 & 1.3/ --1.3 & 2.0/ --2.0 & 3.5/ --3.5 & 5.3/ --5.3 & 6.6/ --6.6 & 9.1/ --9.1\\ \hline
Total uncertainty & 9.7/ --8.9 & 7.7/ --7.9 & 7.5/ --6.5 & 8.5/ --8.1 & 10/ --9 & 11/ --10 & 15/ --15 & 20/ --19\\
\bottomrule
\end{tabular}}
\end{center}
\end{table}

\begin{table}[!htbp]
\caption{
Summary of the uncertainties in the absolute differential $\bar{t}q$ production cross-sections measured as a function of \pTtbar.
Similar sources of uncertainties are grouped using their quadratic sum.
The total uncertainty is defined as the quadratic sum of all sources of uncertainties.
}
\begin{center}
\resizebox{\textwidth}{!}{
\begin{tabular}{l|c|c|c|c|c|c|c|c}
\toprule
Systematic & \multicolumn{8}{c}{Systematic uncertainty in [\%] on the absolute $\bar{t}q$ cross-section in the given $p_\text{T}(\bar{t})$ interval}  \\ \midrule
& [0, 45] \GeV & [45, 65] \GeV & [65, 95] \GeV & [95, 135] \GeV & [135, 175] \GeV & [175, 215] \GeV & [215, 275] \GeV & [275, 500] \GeV   \\ \midrule
JER & 7.2/ --3.0 & 1.6/ --1.2 & 3.1/ --3.7 & 1.0/ --2.6 & 2.2/ --2.3 & 3.9/ --3.9 & 3.3/ --2.4 & 1.8/ --5.4\\
JES & 6.0/ --8.1 & 2.1/ --1.9 & 3.3/ --2.7 & 5.1/ --4.7 & 6.4/ --5.9 & 5.8/ --5.5 & 6.3/ --4.9 & 6.5/ --4.3\\
(f)JVT & 1.6/ --1.6 & 1.3/ --1.3 & 0.6/ --0.6 & 0.2/ --0.2 & 0.7/ --0.8 & 0.6/ --0.7 & 0.5/ --0.5 & 0.7/ --0.7\\
Lepton & 1.3/ --1.4 & 1.2/ --1.2 & 1.1/ --1.2 & 1.0/ --1.1 & 1.1/ --1.1 & 1.1/ --1.1 & 1.1/ --1.2 & 1.2/ --1.6\\
$E_\text{T}^{miss}$ & 1.8/ --0.0 & 1.5/ --0.1 & 1.2/ --0.7 & 0.9/ --0.3 & 2.9/ --0.6 & 1.4/ --0.3 & 0.1/ --1.2 & 1.6/ --1.6\\
$b$--tagging & 2.4/ --2.4 & 1.9/ --1.9 & 1.6/ --1.6 & 1.5/ --1.5 & 1.4/ --1.4 & 1.4/ --1.4 & 1.4/ --1.4 & 1.5/ --1.5\\
Pile-up & 0.9/ --1.2 & 1.3/ --1.2 & 1.6/ --1.5 & 1.5/ --1.3 & 1.4/ --1.2 & 1.6/ --1.5 & 1.0/ --0.7 & 1.5/ --1.3\\
Luminosity & 0.8/ --0.8 & 0.8/ --0.8 & 0.8/ --0.8 & 0.8/ --0.8 & 0.8/ --0.8 & 0.8/ --0.8 & 0.8/ --0.8 & 0.8/ --0.8\\
Signal modelling & 5.1/ --4.1 & 6.0/ --5.5 & 5.2/ --5.7 & 4.9/ --5.2 & 12/ --11 & 11/ --9 & 13/ --12 & 24/ --21\\
Background modelling & 1.7/ --3.1 & 0.6/ --1.4 & 1.4/ --1.2 & 1.0/ --1.2 & 1.9/ --2.0 & 2.6/ --2.9 & 4.9/ --4.9 & 3.9/ --2.2\\
Background rate  & 3.0/ --3.0 & 1.8/ --1.8 & 1.1/ --1.1 & 0.7/ --0.7 & 0.9/ --0.9 & 0.6/ --0.6 & 0.7/ --0.7 & 0.9/ --0.9\\ \hline
Total systematic & 11/ --11 & 7.6/ --7.1 & 7.8/ --8.0 & 7.7/ --8.0 & 14/ --13 & 13/ --11 & 16/ --15 & 25/ --22\\
Statistical  & 2.5/ --2.5 & 1.8/ --1.8 & 2.0/ --2.0 & 2.9/ --2.9 & 5.3/ --5.3 & 7.6/ --7.6 & 10/ --10 & 13/ --13\\ \hline
Total uncertainty & 12/ --11 & 7.8/ --7.3 & 8.0/ --8.2 & 8.2/ --8.5 & 15/ --14 & 15/ --13 & 19/ --18 & 28/ --26\\
\bottomrule
\end{tabular}}
\end{center}
\end{table}

\begin{table}[!htbp]
\caption{
Summary of the uncertainties in the absolute differential $tq$ production cross-sections measured as a function of \Rapt.
Similar sources of uncertainties are grouped using their quadratic sum.
The total uncertainty is defined as the quadratic sum of all sources of uncertainties.
}
\begin{center}
\resizebox{0.8\textwidth}{!}{
\begin{tabular}{l|c|c|c|c|c}
\toprule
\multirow{2}{*}{Systematic} & \multicolumn{5}{c}{Systematic uncertainty in [\%] on the absolute $tq$ cross-section}  \\
& \multicolumn{5}{c}{in the given $|y(t)|$ interval}  \\ \midrule
& [0.0, 0.25]& [0.25, 0.6]& [0.6, 1.0]& [1.0, 1.45]& [1.45, 2.5]  \\ \midrule
JER & 0.8/ --2.5 & 3.5/ --1.5 & 2.1/ --2.5 & 0.5/ --2.3 & 4.5/ --5.7\\
JES & 2.3/ --1.1 & 2.1/ --1.7 & 1.9/ --1.6 & 2.9/ --2.8 & 6.4/ --3.4\\
(f)JVT & 0.9/ --0.9 & 0.9/ --0.9 & 0.9/ --0.9 & 0.8/ --0.8 & 0.8/ --0.8\\
Lepton & 1.1/ --1.1 & 1.1/ --1.1 & 1.0/ --1.0 & 1.0/ --1.0 & 1.3/ --1.1\\
$E_\text{T}^{miss}$ & 0.0/ --0.5 & 1.0/ --0.1 & 0.9/ --0.1 & 1.3/ --0.0 & 3.4/ --0.5\\
$b$--tagging & 1.6/ --1.6 & 1.6/ --1.6 & 1.6/ --1.6 & 1.5/ --1.5 & 1.6/ --1.6\\
Pile-up & 1.3/ --1.2 & 1.3/ --1.2 & 1.4/ --1.3 & 1.3/ --1.3 & 2.0/ --1.7\\
Luminosity & 0.8/ --0.8 & 0.8/ --0.8 & 0.8/ --0.8 & 0.8/ --0.8 & 0.8/ --0.8\\
Signal modelling & 5.5/ --5.3 & 5.7/ --5.3 & 6.0/ --5.2 & 5.1/ --4.9 & 6.5/ --6.0\\
Background modelling & 1.0/ --0.6 & 0.8/ --0.5 & 1.1/ --1.0 & 1.0/ --0.9 & 0.8/ --1.8\\
Background rate  & 1.2/ --1.2 & 1.2/ --1.2 & 1.2/ --1.2 & 1.0/ --1.0 & 1.5/ --1.5\\ \hline
Total systematic & 6.7/ --6.7 & 7.7/ --6.5 & 7.3/ --6.7 & 6.7/ --6.7 & 11/ --10\\
Statistical  & 1.7/ --1.7 & 1.4/ --1.4 & 1.6/ --1.6 & 1.9/ --1.9 & 2.6/ --2.6\\ \hline
Total uncertainty & 6.9/ --6.9 & 7.8/ --6.6 & 7.5/ --6.9 & 6.9/ --7.0 & 11/ --10\\
\bottomrule
\end{tabular}}
\end{center}
\end{table}

\begin{table}[!htbp]
\caption{
Summary of the uncertainties in the absolute differential $\bar{t}q$ production cross-sections measured as a function of \Raptbar.
Similar sources of uncertainties are grouped using their quadratic sum.
The total uncertainty is defined as the quadratic sum of all sources of uncertainties.
}
\begin{center}
\resizebox{0.8\textwidth}{!}{
\begin{tabular}{l|c|c|c|c|c}
\toprule
\multirow{2}{*}{Systematic} & \multicolumn{5}{c}{Systematic uncertainty in [\%] on the absolute $\bar{t}q$ cross-section}  \\
& \multicolumn{5}{c}{in the given $|y(\bar{t})|$ interval}  \\ \midrule
& [0.0, 0.25]& [0.25, 0.6]& [0.6, 1.0]& [1.0, 1.45]& [1.45, 2.5]  \\ \midrule
JER & 3.5/ --5.4 & 2.4/ --1.4 & 3.7/ --1.1 & 2.4/ --1.4 & 3.6/ --3.0\\
JES & 3.9/ --2.6 & 2.5/ --3.6 & 3.5/ --2.2 & 4.0/ --2.5 & 4.0/ --6.5\\
(f)JVT & 0.8/ --0.8 & 0.9/ --0.9 & 0.8/ --0.8 & 0.8/ --0.8 & 0.7/ --0.7\\
Lepton & 1.3/ --1.2 & 1.2/ --1.3 & 1.3/ --1.2 & 1.1/ --1.4 & 1.2/ --1.2\\
$E_\text{T}^{miss}$ & 1.7/ --0.0 & 0.5/ --0.5 & 1.6/ --0.3 & 2.1/ --0.0 & 1.8/ --1.0\\
$b$--tagging & 1.8/ --1.8 & 1.8/ --1.8 & 1.7/ --1.7 & 1.8/ --1.8 & 1.8/ --1.8\\
Pile-up & 1.4/ --1.4 & 1.4/ --1.3 & 1.2/ --1.2 & 1.2/ --1.2 & 1.5/ --1.5\\
Luminosity & 0.8/ --0.8 & 0.8/ --0.8 & 0.8/ --0.8 & 0.8/ --0.8 & 0.8/ --0.8\\
Signal modelling & 7.4/ --7.3 & 5.8/ --5.4 & 4.1/ --4.0 & 4.1/ --4.6 & 8.6/ --7.8\\
Background modelling & 1.9/ --2.2 & 1.6/ --1.2 & 1.5/ --1.9 & 1.5/ --1.5 & 1.8/ --4.0\\
Background rate  & 1.5/ --1.5 & 1.6/ --1.6 & 1.5/ --1.5 & 1.5/ --1.5 & 1.9/ --1.9\\ \hline
Total systematic & 10/ --10 & 7.7/ --7.5 & 7.6/ --6.0 & 7.4/ --6.5 & 11/ --11\\
Statistical  & 2.4/ --2.4 & 1.9/ --1.9 & 2.1/ --2.1 & 2.7/ --2.7 & 3.7/ --3.7\\ \hline
Total uncertainty & 10/ --10 & 7.9/ --7.7 & 7.9/ --6.3 & 7.9/ --7.0 & 11/ --12\\
\bottomrule
\end{tabular}}
\end{center}
\end{table}

\begin{table}[!htbp]
\caption{
Summary of the uncertainties in the normalised differential $tq$ production cross-sections measured as a function of \pTt.
Similar sources of uncertainties are grouped using their quadratic sum.
The total uncertainty is defined as the quadratic sum of all sources of uncertainties.
}
\begin{center}
\resizebox{\textwidth}{!}{
\begin{tabular}{l|c|c|c|c|c|c|c|c}
\toprule
Systematic & \multicolumn{8}{c}{Systematic uncertainty in [\%] on the normalised $tq$ cross-section in the given $p_\text{T}(t)$ interval}  \\ \midrule
& [0, 45] \GeV & [45, 65] \GeV & [65, 95] \GeV & [95, 135] \GeV & [135, 175] \GeV & [175, 215] \GeV & [215, 275] \GeV & [275, 500] \GeV   \\ \midrule
JER & 0.5/ --2.5 & 1.5/ --1.4 & 2.4/ --0.3 & 1.2/ --1.4 & 3.5/ --0.2 & 1.5/ --3.7 & 4.3/ --0.8 & 2.2/ --2.0\\
JES & 3.8/ --2.4 & 1.7/ --2.2 & 1.8/ --2.6 & 3.1/ --3.6 & 3.5/ --3.6 & 4.1/ --3.8 & 2.7/ --5.6 & 4.0/ --4.2\\
(f)JVT & 0.8/ --0.8 & 0.4/ --0.4 & 0.2/ --0.2 & 0.7/ --0.7 & 1.6/ --1.6 & 1.4/ --1.5 & 1.4/ --1.4 & 1.5/ --1.5\\
Lepton & 0.6/ --0.2 & 0.0/ --0.3 & 0.1/ --0.4 & 0.2/ --0.3 & 0.2/ --0.3 & 0.2/ --0.3 & 0.4/ --0.3 & 0.3/ --0.4\\
$E_\text{T}^{miss}$ & 1.1/ --0.0 & 0.0/ --1.0 & 0.0/ --0.6 & 0.0/ --0.3 & 0.1/ --0.6 & 0.6/ --0.0 & 0.2/ --1.1 & 1.4/ --0.0\\
$b$--tagging & 0.8/ --0.8 & 0.2/ --0.2 & 0.5/ --0.5 & 0.6/ --0.6 & 0.6/ --0.7 & 0.7/ --0.7 & 0.9/ --0.9 & 1.1/ --1.1\\
Pile-up & 0.1/ --0.1 & 0.1/ --0.1 & 0.1/ --0.0 & 0.0/ --0.0 & 0.1/ --0.1 & 0.2/ --0.3 & 0.2/ --0.2 & 0.0/ --0.1\\
Signal modelling & 2.5/ --2.5 & 1.6/ --1.8 & 1.3/ --1.3 & 2.7/ --2.6 & 3.6/ --2.8 & 3.2/ --2.8 & 8.2/ --8.0 & 13/ --12\\
Background modelling & 0.9/ --1.3 & 0.5/ --0.4 & 1.1/ --0.9 & 0.6/ --0.7 & 1.1/ --1.0 & 1.2/ --1.7 & 2.6/ --2.7 & 2.7/ --2.5\\
Background rate  & 1.5/ --1.5 & 0.2/ --0.2 & 0.9/ --1.0 & 1.0/ --1.0 & 1.0/ --1.1 & 1.1/ --1.1 & 1.1/ --1.2 & 1.2/ --1.2\\ \hline
Total systematic & 5.1/ --4.8 & 2.9/ --3.3 & 3.7/ --3.4 & 4.5/ --4.9 & 6.5/ --5.1 & 5.9/ --6.6 & 10/ --10 & 14/ --13\\
Statistical  & 1.4/ --1.4 & 1.3/ --1.3 & 1.5/ --1.5 & 1.9/ --1.9 & 3.4/ --3.4 & 5.2/ --5.2 & 6.6/ --6.6 & 9.0/ --9.0\\ \hline
Total uncertainty & 5.3/ --5.0 & 3.2/ --3.5 & 3.9/ --3.7 & 4.9/ --5.3 & 7.3/ --6.2 & 7.9/ --8.4 & 12/ --12 & 17/ --16\\
\bottomrule
\end{tabular}}
\end{center}
\end{table}

\begin{table}[!htbp]
\caption{
Summary of the uncertainties in the normalised differential $\bar{t}q$ production cross-sections measured as a function of \pTtbar.
Similar sources of uncertainties are grouped using their quadratic sum.
The total uncertainty is defined as the quadratic sum of all sources of uncertainties.
}
\begin{center}
\resizebox{\textwidth}{!}{
\begin{tabular}{l|c|c|c|c|c|c|c|c}
\toprule
Systematic & \multicolumn{8}{c}{Systematic uncertainty in [\%] on the normalised $\bar{t}q$ cross-section in the given $p_\text{T}(\bar{t})$ interval}  \\ \midrule
& [0, 45] \GeV & [45, 65] \GeV & [65, 95] \GeV & [95, 135] \GeV & [135, 175] \GeV & [175, 215] \GeV & [215, 275] \GeV & [275, 500] \GeV   \\ \midrule
JER & 5.3/ --2.5 & 1.4/ --2.3 & 3.0/ --4.7 & 0.9/ --3.5 & 2.3/ --3.7 & 3.6/ --5.4 & 3.1/ --3.4 & 1.7/ --6.7\\
JES & 3.3/ --4.9 & 2.8/ --2.0 & 3.9/ --3.2 & 3.3/ --3.1 & 4.1/ --3.7 & 4.4/ --4.5 & 5.5/ --4.5 & 5.9/ --4.3\\
(f)JVT & 0.7/ --0.7 & 0.4/ --0.4 & 0.3/ --0.3 & 0.7/ --0.7 & 1.6/ --1.7 & 1.5/ --1.5 & 1.4/ --1.4 & 1.6/ --1.6\\
Lepton & 0.3/ --0.4 & 0.3/ --0.0 & 0.4/ --0.5 & 0.1/ --0.3 & 0.4/ --0.4 & 0.4/ --0.4 & 0.2/ --0.6 & 0.4/ --1.2\\
$E_\text{T}^{miss}$ & 0.8/ --0.1 & 0.2/ --0.1 & 0.1/ --0.9 & 0.2/ --0.8 & 1.6/ --0.6 & 0.2/ --0.5 & 0.0/ --2.3 & 0.6/ --1.7\\
$b$--tagging & 0.8/ --0.8 & 0.2/ --0.2 & 0.6/ --0.6 & 0.7/ --0.7 & 0.8/ --0.8 & 0.8/ --0.8 & 0.9/ --0.9 & 1.3/ --1.3\\
Pile-up & 0.1/ --0.4 & 0.0/ --0.0 & 0.4/ --0.2 & 0.2/ --0.1 & 0.1/ --0.0 & 0.4/ --0.2 & 0.5/ --0.3 & 0.2/ --0.1\\
Signal modelling & 2.7/ --2.6 & 1.0/ --1.0 & 0.5/ --1.3 & 1.1/ --2.0 & 8.2/ --8.0 & 7.0/ --4.8 & 9.2/ --8.4 & 19/ --17\\
Background modelling & 1.1/ --1.8 & 0.7/ --1.0 & 1.7/ --1.4 & 1.2/ --0.8 & 1.9/ --1.9 & 3.0/ --3.0 & 4.7/ --4.7 & 4.3/ --2.3\\
Background rate  & 1.3/ --1.3 & 0.1/ --0.1 & 0.8/ --0.9 & 1.2/ --1.2 & 1.2/ --1.2 & 1.4/ --1.4 & 1.4/ --1.4 & 1.4/ --1.4\\ \hline
Total systematic & 7.2/ --6.6 & 3.4/ --3.4 & 5.3/ --6.2 & 4.1/ --5.4 & 10/ --10 & 9.8/ --9.3 & 12/ --11 & 21/ --19\\
Statistical  & 1.9/ --1.9 & 1.8/ --1.8 & 2.1/ --2.1 & 2.8/ --2.8 & 5.1/ --5.1 & 7.6/ --7.6 & 10/ --10 & 13/ --13\\ \hline
Total uncertainty & 7.4/ --6.8 & 3.9/ --3.9 & 5.7/ --6.6 & 5.0/ --6.1 & 11/ --11 & 12/ --12 & 16/ --15 & 25/ --23\\
\bottomrule
\end{tabular}}
\end{center}
\end{table}

\begin{table}[!htbp]
\caption{
Summary of the uncertainties in the normalised differential $tq$ production cross-sections measured as a function of \Rapt.
Similar sources of uncertainties are grouped using their quadratic sum.
The total uncertainty is defined as the quadratic sum of all sources of uncertainties.
}
\begin{center}
\resizebox{0.8\textwidth}{!}{
\begin{tabular}{l|c|c|c|c|c}
\toprule
\multirow{2}{*}{Systematic} & \multicolumn{5}{c}{Systematic uncertainty in [\%] on the normalised $tq$ cross-section}  \\
& \multicolumn{5}{c}{in the given $|y(t)|$ interval}  \\ \midrule
& [0.0, 0.25]& [0.25, 0.6]& [0.6, 1.0]& [1.0, 1.45]& [1.45, 2.5]  \\ \midrule
JER & 1.3/ --2.2 & 3.1/ --0.3 & 1.4/ --1.5 & 2.0/ --3.2 & 2.5/ --2.8\\
JES & 1.3/ --1.6 & 0.7/ --1.7 & 1.0/ --1.9 & 0.5/ --1.6 & 3.0/ --1.2\\
(f)JVT & 0.0/ --0.0 & 0.1/ --0.1 & 0.0/ --0.0 & 0.0/ --0.0 & 0.0/ --0.0\\
Lepton & 0.2/ --0.4 & 0.2/ --0.2 & 0.2/ --0.3 & 0.3/ --0.3 & 0.5/ --0.2\\
$E_\text{T}^{miss}$ & 0.1/ --2.1 & 0.1/ --0.6 & 0.1/ --0.7 & 0.3/ --0.4 & 1.8/ --0.3\\
$b$--tagging & 0.1/ --0.1 & 0.1/ --0.1 & 0.1/ --0.1 & 0.2/ --0.2 & 0.1/ --0.1\\
Pile-up & 0.2/ --0.2 & 0.2/ --0.3 & 0.1/ --0.2 & 0.1/ --0.2 & 0.4/ --0.3\\
Signal modelling & 0.7/ --0.9 & 0.9/ --0.8 & 0.7/ --0.4 & 0.9/ --1.0 & 0.9/ --0.9\\
Background modelling & 0.8/ --0.4 & 0.7/ --0.4 & 0.7/ --0.6 & 1.2/ --0.8 & 0.7/ --1.3\\
Background rate  & 0.2/ --0.2 & 0.2/ --0.2 & 0.2/ --0.2 & 0.3/ --0.3 & 0.3/ --0.3\\ \hline
Total systematic & 2.1/ --3.6 & 3.3/ --2.0 & 2.0/ --2.7 & 2.6/ --3.9 & 4.5/ --3.5\\
Statistical  & 1.9/ --1.9 & 1.6/ --1.6 & 1.8/ --1.8 & 2.1/ --2.1 & 2.1/ --2.1\\ \hline
Total uncertainty & 2.9/ --4.1 & 3.7/ --2.6 & 2.7/ --3.2 & 3.3/ --4.4 & 5.0/ --4.1\\
\bottomrule
\end{tabular}}
\end{center}
\end{table}

\begin{table}[!htbp]
\caption{
Summary of the uncertainties in the normalised differential $\bar{t}q$ production cross-sections measured as a function of \Raptbar.
Similar sources of uncertainties are grouped using their quadratic sum.
The total uncertainty is defined as the quadratic sum of all sources of uncertainties.
}
\begin{center}
\resizebox{0.8\textwidth}{!}{
\begin{tabular}{l|c|c|c|c|c}
\toprule
\multirow{2}{*}{Systematic} & \multicolumn{5}{c}{Systematic uncertainty in [\%] on the normalised $\bar{t}q$ cross-section}  \\
& \multicolumn{5}{c}{in the given $|y(\bar{t})|$ interval}  \\ \midrule
& [0.0, 0.25]& [0.25, 0.6]& [0.6, 1.0]& [1.0, 1.45]& [1.45, 2.5]  \\ \midrule
JER & 2.6/ --5.5 & 2.5/ --1.8 & 3.3/ --1.5 & 2.2/ --1.7 & 2.5/ --2.5\\
JES & 1.7/ --0.6 & 0.9/ --1.6 & 1.9/ --1.3 & 2.0/ --0.7 & 1.4/ --3.2\\
(f)JVT & 0.0/ --0.0 & 0.1/ --0.1 & 0.1/ --0.0 & 0.1/ --0.1 & 0.1/ --0.1\\
Lepton & 0.5/ --0.3 & 0.2/ --0.5 & 0.6/ --0.3 & 0.4/ --0.7 & 0.3/ --0.3\\
$E_\text{T}^{miss}$ & 0.7/ --0.1 & 0.0/ --1.1 & 0.4/ --0.1 & 1.0/ --0.3 & 0.5/ --0.7\\
$b$--tagging & 0.1/ --0.1 & 0.1/ --0.1 & 0.1/ --0.1 & 0.1/ --0.1 & 0.2/ --0.2\\
Pile-up & 0.0/ --0.1 & 0.1/ --0.0 & 0.1/ --0.2 & 0.2/ --0.2 & 0.2/ --0.2\\
Signal modelling & 1.6/ --1.8 & 0.5/ --0.4 & 2.3/ --2.4 & 2.1/ --2.4 & 2.7/ --2.4\\
Background modelling & 2.1/ --2.1 & 1.3/ --0.5 & 1.6/ --1.4 & 2.0/ --1.4 & 1.7/ --2.9\\
Background rate  & 0.4/ --0.4 & 0.3/ --0.3 & 0.2/ --0.2 & 0.3/ --0.3 & 0.5/ --0.5\\ \hline
Total systematic & 4.2/ --6.2 & 3.0/ --2.8 & 4.8/ --3.4 & 4.3/ --3.4 & 4.3/ --5.6\\
Statistical  & 2.5/ --2.5 & 2.0/ --2.0 & 2.3/ --2.3 & 2.8/ --2.8 & 2.9/ --2.9\\ \hline
Total uncertainty & 4.9/ --6.7 & 3.7/ --3.5 & 5.3/ --4.1 & 5.1/ --4.4 & 5.2/ --6.3\\
\bottomrule
\end{tabular}}
\end{center}
\end{table}

\begin{table}[!htbp]
\caption{
Summary of the uncertainties in the ratio of the differential production cross-sections measured as a function of $p_\text{T}(\tortbar)$.
Similar sources of uncertainties are grouped using their quadratic sum.
The total uncertainty is defined as the quadratic sum of all sources of uncertainties.
}
\begin{center}
\resizebox{\textwidth}{!}{
\begin{tabular}{l|c|c|c|c|c|c|c|c}
\toprule
Systematic & \multicolumn{8}{c}{Systematic uncertainty in [\%] on the cross-section ratio in the given $p_\text{T}(\tortbar)$ interval}  \\ \midrule
& [0, 45] \GeV & [45, 65] \GeV & [65, 95] \GeV & [95, 135] \GeV & [135, 175] \GeV & [175, 215] \GeV & [215, 275] \GeV & [275, 500] \GeV   \\ \midrule
JER & 2.0/ --9.6 & 2.2/ --3.8 & 3.2/ --1.9 & 1.4/ --1.0 & 3.4/ --1.3 & 1.8/ --5.5 & 4.1/ --2.4 & 4.9/ --2.1\\
JES & 6.7/ --2.5 & 1.6/ --1.3 & 1.9/ --1.5 & 0.8/ --1.8 & 1.5/ --1.5 & 2.9/ --1.9 & 2.6/ --3.5 & 2.0/ --2.5\\
(f)JVT & 0.2/ --0.2 & 0.0/ --0.0 & 0.2/ --0.2 & 0.0/ --0.0 & 0.1/ --0.1 & 0.2/ --0.2 & 0.1/ --0.1 & 0.2/ --0.2\\
Lepton & 1.2/ --0.3 & 0.1/ --0.4 & 0.5/ --0.5 & 0.4/ --0.1 & 0.3/ --0.4 & 0.3/ --0.3 & 0.9/ --0.1 & 1.1/ --0.2\\
$E_\text{T}^{miss}$ & 0.5/ --0.0 & 0.0/ --1.2 & 0.4/ --0.3 & 0.3/ --0.3 & 0.5/ --2.3 & 0.7/ --0.1 & 1.8/ --0.6 & 2.4/ --0.0\\
$b$--tagging & 0.3/ --0.3 & 0.3/ --0.3 & 0.2/ --0.2 & 0.2/ --0.2 & 0.2/ --0.2 & 0.2/ --0.2 & 0.3/ --0.3 & 0.3/ --0.3\\
Pile-up & 0.5/ --0.2 & 0.0/ --0.0 & 0.2/ --0.2 & 0.1/ --0.1 & 0.0/ --0.1 & 0.4/ --0.6 & 0.3/ --0.4 & 0.0/ --0.2\\
Signal modelling & 1.4/ --1.7 & 1.4/ --1.5 & 1.4/ --0.6 & 2.6/ --1.5 & 8.1/ --6.3 & 3.4/ --5.3 & 4.1/ --4.1 & 8/ --10\\
Background modelling & 3.0/ --1.9 & 1.2/ --0.6 & 1.2/ --0.6 & 1.1/ --0.9 & 1.7/ --1.4 & 2.7/ --2.6 & 7.2/ --6.7 & 3.1/ --3.6\\
Background rate  & 3.9/ --3.9 & 2.0/ --2.0 & 0.9/ --0.9 & 0.7/ --0.7 & 0.8/ --0.8 & 0.7/ --0.7 & 0.7/ --0.7 & 0.6/ --0.6\\ \hline
Total systematic & 9/ --10 & 3.9/ --5.0 & 4.3/ --2.8 & 3.4/ --2.8 & 9.1/ --7.2 & 5.7/ --8.4 & 9.8/ --9.0 & 10/ --11\\
Statistical  & 3.1/ --3.1 & 2.2/ --2.2 & 2.3/ --2.3 & 3.4/ --3.4 & 6.2/ --6.2 & 9.0/ --9.0 & 12/ --12 & 16/ --16\\ \hline
Total uncertainty & 9/ --11 & 4.4/ --5.4 & 4.9/ --3.7 & 4.8/ --4.4 & 10/ --10 & 10/ --12 & 15/ --15 & 19/ --19\\
\bottomrule
\end{tabular}}
\end{center}
\end{table}

\begin{table}[!htbp]
\caption{\label{tab:CommonSubSysRatiorap}
Summary of the uncertainties in the ratio of the differential production cross-sections measured as a function of $|y(\tortbar)|$.
Similar sources of uncertainties are grouped using their quadratic sum.
The total uncertainty is defined as the quadratic sum of all sources of uncertainties.
}
\begin{center}
\resizebox{0.8\textwidth}{!}{
\begin{tabular}{l|c|c|c|c|c}
\toprule
\multirow{2}{*}{Systematic} & \multicolumn{5}{c}{Systematic uncertainty in [\%] on the cross-section ratio }  \\
& \multicolumn{5}{c}{in the given $|y(\tortbar)|$ interval}  \\ \midrule
& [0.0, 0.25]& [0.25, 0.6]& [0.6, 1.0]& [1.0, 1.45]& [1.45, 2.5]  \\ \midrule
JER & 3.7/ --3.5 & 4.3/ --3.1 & 2.1/ --5.1 & 0.7/ --3.7 & 4.5/ --6.6\\
JES & 1.7/ --1.9 & 2.4/ --0.9 & 1.4/ --2.0 & 0.7/ --2.0 & 5.0/ --2.3\\
(f)JVT & 0.1/ --0.1 & 0.0/ --0.1 & 0.1/ --0.1 & 0.0/ --0.0 & 0.2/ --0.2\\
Lepton & 0.1/ --0.6 & 0.7/ --0.2 & 0.2/ --0.6 & 0.8/ --0.2 & 1.2/ --0.6\\
$E_\text{T}^{miss}$ & 0.0/ --2.1 & 0.7/ --0.0 & 0.2/ --0.7 & 0.5/ --1.0 & 1.8/ --0.0\\
$b$--tagging & 0.2/ --0.2 & 0.3/ --0.3 & 0.3/ --0.3 & 0.4/ --0.4 & 0.3/ --0.3\\
Pile-up & 0.3/ --0.1 & 0.1/ --0.2 & 0.2/ --0.0 & 0.1/ --0.1 & 0.4/ --0.2\\
Signal modelling & 2.5/ --2.1 & 0.8/ --0.9 & 2.4/ --2.2 & 1.5/ --1.3 & 2.9/ --3.0\\
Background modelling & 2.5/ --2.0 & 1.0/ --0.9 & 1.5/ --1.0 & 1.5/ --1.5 & 4.7/ --2.9\\
Background rate  & 1.6/ --1.6 & 1.7/ --1.7 & 1.7/ --1.7 & 1.6/ --1.6 & 2.0/ --2.0\\ \hline
Total systematic & 5.7/ --5.6 & 5.5/ --3.9 & 4.1/ --6.3 & 3.1/ --5.0 & 9.2/ --8.4\\
Statistical  & 2.8/ --2.8 & 2.3/ --2.3 & 2.5/ --2.5 & 3.1/ --3.1 & 4.4/ --4.4\\ \hline
Total uncertainty & 6.3/ --6.3 & 5.9/ --4.5 & 4.9/ --6.8 & 4.4/ --5.9 & 10/ --10\\
\bottomrule
\end{tabular}}
\end{center}
\end{table}

\FloatBarrier

\subsection*{Measurement results}
The measured results for the parton-level \tqtbarq production cross-sections are shown graphically in \Sect{\ref{sec:results}}.
The numerical values are given in \Tabrange{\ref{tab:ResultsptSRp}}{\ref{tab:ResultsrapRatio}}.

\begin{table}
\caption{\label{tab:ResultsptSRp} Absolute and normalised differential $tq$ production cross-sections as a function of \pTt.}
\begin{center}
\begin{tabular}{c|ccc|ccc}
\toprule
$p_\text{T}(t)$ [\GeV] & \multicolumn{3}{c}{$\diff{\sigma_{tq}}{p_\text{T}(t)}$ [fb \GeV$^{-1}$]} & \multicolumn{3}{c}{$\frac{1}{\sigma}\diff{\sigma_{tq}}{p_\text{T}(t)}$ [10$^{-3}$ \GeV$^{-1}$]}  \\ \midrule
&  & stat. & syst. &   & stat. & syst.  \\ \midrule
0--45 & 966 & $\pm$ 18 & 90/--83 & 7.44 & $\pm$ 0.15 & 0.37/--0.34\\
45--65 & 1274 & $\pm$ 16 & 95/--98 & 9.81 & $\pm$ 0.13 & 0.28/--0.32\\
65--95 & 914 & $\pm$ 12 & 67/--58 & 7.04 & $\pm$ 0.09 & 0.26/--0.24\\
95--135 & 456 & $\pm$ 9 & 37/--35 & 3.51 & $\pm$ 0.07 & 0.16/--0.17\\
135--175 & 191 & $\pm$ 7 & 18/--15 & 1.470 & $\pm$ 0.052 & 0.095/--0.075\\
175--215 & 93.6 & $\pm$ 4.9 & 9.3/--8.7 & 0.721 & $\pm$ 0.038 & 0.043/--0.047\\
215--275 & 40.4 & $\pm$ 2.7 & 5.6/--5.6 & 0.311 & $\pm$ 0.021 & 0.032/--0.032\\
275--500 & 6.1 & $\pm$ 0.6 & 1.1/--1.1 & 0.0470 & $\pm$ 0.0043 & 0.0068/--0.0065\\
\bottomrule
\end{tabular}
\end{center}
\end{table}

\begin{table}
\caption{Absolute and normalised differential $\bar{t}q$ production cross-sections as a function of \pTtbar.}
\begin{center}
\begin{tabular}{c|ccc|ccc}
\toprule
$p_\text{T}(\bar{t})$ [\GeV] & \multicolumn{3}{c}{$\diff{\sigma_{\bar{t}q}}{p_\text{T}(\bar{t})}$ [fb \GeV$^{-1}$]} & \multicolumn{3}{c}{$\frac{1}{\sigma}\diff{\sigma_{\bar{t}q}}{p_\text{T}(\bar{t})}$ [10$^{-3}$ \GeV$^{-1}$]}  \\ \midrule
&  & stat. & syst. &   & stat. & syst.  \\ \midrule
0--45 & 649 & $\pm$ 16 & 76/--71 & 8.05 & $\pm$ 0.20 & 0.56/--0.51\\
45--65 & 815 & $\pm$ 14 & 60/--57 & 10.11 & $\pm$ 0.18 & 0.34/--0.34\\
65--95 & 526 & $\pm$ 10 & 40/--41 & 6.52 & $\pm$ 0.13 & 0.35/--0.41\\
95--135 & 269 & $\pm$ 8 & 20/--21 & 3.33 & $\pm$ 0.10 & 0.13/--0.18\\
135--175 & 112 & $\pm$ 6 & 16/--15 & 1.38 & $\pm$ 0.07 & 0.14/--0.14\\
175--215 & 50.1 & $\pm$ 3.8 & 6.9/--5.8 & 0.621 & $\pm$ 0.047 & 0.061/--0.058\\
215--275 & 20.6 & $\pm$ 2.3 & 3.4/--3.1 & 0.256 & $\pm$ 0.028 & 0.031/--0.030\\
275--500 & 3.42 & $\pm$ 0.47 & 0.87/--0.78 & 0.0425 & $\pm$ 0.0058 & 0.0091/--0.0082\\
\bottomrule
\end{tabular}
\end{center}
\end{table}

\begin{table}
\caption{Absolute and normalised differential $tq$ production cross-sections as a function of \Rapt.}
\begin{center}
\begin{tabular}{c|ccc|ccc}
\toprule
$|y(t)|$ & \multicolumn{3}{c}{$\diff{\sigma_{tq}}{|y(t)|}$ [pb]} & \multicolumn{3}{c}{$\frac{1}{\sigma}\diff{\sigma_{tq}}{|y(t)|}$}  \\ \midrule
&  & stat. & syst. &   & stat. & syst.  \\ \midrule
0.0--0.25 & 69.7 & $\pm$ 1.2 & 4.6/--4.6 & 0.534 & $\pm$ 0.009 & 0.012/--0.020\\
0.25--0.6 & 67.7 & $\pm$ 0.9 & 5.2/--4.4 & 0.519 & $\pm$ 0.007 & 0.018/--0.011\\
0.6--1.0 & 62.0 & $\pm$ 1.0 & 4.5/--4.2 & 0.475 & $\pm$ 0.007 & 0.010/--0.013\\
1.0--1.45 & 54.1 & $\pm$ 1.0 & 3.6/--3.6 & 0.414 & $\pm$ 0.008 & 0.011/--0.016\\
1.45--2.5 & 31.6 & $\pm$ 0.8 & 3.5/--3.1 & 0.2420 & $\pm$ 0.0064 & 0.0100/--0.0075\\
\bottomrule
\end{tabular}
\end{center}
\end{table}

\begin{table}
\caption{Absolute and normalised differential $\bar{t}q$ production cross-sections as a function of \Raptbar.}
\begin{center}
\begin{tabular}{c|ccc|ccc}
\toprule
$|y(\bar{t})|$ & \multicolumn{3}{c}{$\diff{\sigma_{\bar{t}q}}{|y(\bar{t})|}$ [pb]} & \multicolumn{3}{c}{$\frac{1}{\sigma}\diff{\sigma_{\bar{t}q}}{|y(\bar{t})|}$}  \\ \midrule
&  & stat. & syst. &   & stat. & syst.  \\ \midrule
0.0--0.25 & 48.4 & $\pm$ 1.1 & 4.8/--4.9 & 0.592 & $\pm$ 0.014 & 0.025/--0.037\\
0.25--0.6 & 43.9 & $\pm$ 0.8 & 3.4/--3.3 & 0.538 & $\pm$ 0.010 & 0.017/--0.016\\
0.6--1.0 & 41.3 & $\pm$ 0.9 & 3.1/--2.4 & 0.506 & $\pm$ 0.011 & 0.025/--0.018\\
1.0--1.45 & 33.5 & $\pm$ 0.9 & 2.5/--2.2 & 0.410 & $\pm$ 0.011 & 0.018/--0.014\\
1.45--2.5 & 18.5 & $\pm$ 0.7 & 2.0/--2.2 & 0.2268 & $\pm$ 0.0084 & 0.0082/--0.0120\\
\bottomrule
\end{tabular}
\end{center}
\end{table}

\begin{table}
\caption{Ratio of the absolute differential $tq$ and $\bar{t}q$ production cross-sections as a fuction of $p_\text{T}(\tortbar)$.}
\begin{center}
\begin{tabular}{c|ccc}
\toprule
$p_\text{T}(\tortbar)$ [\GeV] & \multicolumn{3}{c}{$(\diff{\sigma(tq)}{p_\text{T}(t)}) / (\diff{\sigma(\bar{t}q)}{p_\text{T}(\bar{t})})$} \\ \midrule
&  & stat. & syst.  \\ \midrule
0--45 & 1.573 & $\pm$ 0.048 & 0.137/--0.172\\
45--65 & 1.634 & $\pm$ 0.035 & 0.063/--0.081\\
65--95 & 1.683 & $\pm$ 0.039 & 0.072/--0.047\\
95--135 & 1.766 & $\pm$ 0.060 & 0.059/--0.049\\
135--175 & 1.86 & $\pm$ 0.11 & 0.17/--0.13\\
175--215 & 1.96 & $\pm$ 0.18 & 0.11/--0.16\\
215--275 & 2.06 & $\pm$ 0.26 & 0.20/--0.18\\
275--500 & 2.26 & $\pm$ 0.36 & 0.23/--0.25\\
\bottomrule
\end{tabular}
\end{center}
\end{table}

\begin{table}
\caption{\label{tab:ResultsrapRatio} Ratio of the absolute differential $tq$ and $\bar{t}q$ production cross-sections as a fuction of $|y(\tortbar)|$.}
\begin{center}
\begin{tabular}{c|ccc}
\toprule
$|y(\tortbar)|$ & \multicolumn{3}{c}{$(\diff{\sigma(tq)}{|y(t)|}) / (\diff{\sigma(\bar{t}q)}{|y(\bar{t})|})$ } \\ \midrule
&  & stat. & syst.  \\ \midrule
0.0--0.25 & 1.507 & $\pm$ 0.043 & 0.085/--0.085\\
0.25--0.6 & 1.525 & $\pm$ 0.034 & 0.083/--0.060\\
0.6--1.0 & 1.562 & $\pm$ 0.040 & 0.065/--0.099\\
1.0--1.45 & 1.642 & $\pm$ 0.052 & 0.050/--0.083\\
1.45--2.5 & 1.90 & $\pm$ 0.08 & 0.17/--0.16\\
\bottomrule
\end{tabular}
\end{center}
\end{table}

\FloatBarrier
\subsection*{EFT interpretation}

\begin{table}
\caption{\label{tab:EFTFitFuncs} Best fit results for the expected relative change in the differential $tq$ and $\bar{t}q$ cross-sections as a function of \Cqq.
The predictions for the expected relative change are obtained by unfolding the distributions of the EFT MC samples with the nominal migration matrix and efficiency.
The parameterisations are determined from least-square fits to the predictions normalised to the prediction from the EFT MC sample generated with \Cqq set to zero.
According to \Eqn{\ref{eq:Cqqdependency}}, the parameterisations take the form $\dxsecEFT = \left(1 + a_1 \cdot \Cqq + a_2 \cdot (\Cqq)^2\right)$}
\begin{center}
\begin{tabular}{c|c|c|c|c}
\toprule
\multirow{2}{*}{\pTt or \pTtbar range} & \multicolumn{2}{c|}{$\Delta\tilde{\sigma}_{tq}\left(\Cqq\right)$} & \multicolumn{2}{c}{$\Delta\tilde{\sigma}_{\bar{t}q}\left(\Cqq\right)$} \\
& $a_1$ & $a_2$ & $a_1$ & $a_2$ \\
\midrule
\SI{0}{\GeV} $\leq$ \pTt or \pTtbar $\leq$ \SI{45}{\GeV}  & --0.175 & 0.0  & --0.260 & +0.142 \\
\SI{45}{\GeV} $\leq$ \pTt or \pTtbar $\leq$ \SI{65}{\GeV}  & --0.145 & 0.0  & --0.205 & +0.187 \\
\SI{65}{\GeV} $\leq$ \pTt or \pTtbar $\leq$ \SI{95}{\GeV}  & --0.263 & +0.050  & --0.216 & +0.014 \\
\SI{95}{\GeV} $\leq$ \pTt or \pTtbar $\leq$ \SI{135}{\GeV}  & --0.389 & 0.0  & --0.429 & +0.063 \\
\SI{135}{\GeV} $\leq$ \pTt or \pTtbar $\leq$ \SI{175}{\GeV}  & --0.735 & +0.143  & --0.593 & +0.231 \\
\SI{175}{\GeV} $\leq$ \pTt or \pTtbar $\leq$ \SI{215}{\GeV}  & --0.813 & +0.099  & --1.107 & +0.695 \\
\SI{215}{\GeV} $\leq$ \pTt or \pTtbar $\leq$ \SI{275}{\GeV}  & --0.897 & +0.043  & --1.184 & +0.640 \\
\SI{275}{\GeV} $\leq$ \pTt or \pTtbar $\leq$ \SI{500}{\GeV}  & --2.452 & +1.694  & --2.576 & +3.426 \\
\bottomrule
\end{tabular}
\end{center}
\end{table}


\FloatBarrier
\printbibliography
\clearpage
 
\begin{flushleft}
\hypersetup{urlcolor=black}
{\Large The ATLAS Collaboration}

\bigskip

\AtlasOrcid[0000-0002-6665-4934]{G.~Aad}$^\textrm{\scriptsize 103}$,
\AtlasOrcid[0000-0001-7616-1554]{E.~Aakvaag}$^\textrm{\scriptsize 17}$,
\AtlasOrcid[0000-0002-5888-2734]{B.~Abbott}$^\textrm{\scriptsize 122}$,
\AtlasOrcid[0000-0002-0287-5869]{S.~Abdelhameed}$^\textrm{\scriptsize 118a}$,
\AtlasOrcid[0000-0002-1002-1652]{K.~Abeling}$^\textrm{\scriptsize 55}$,
\AtlasOrcid[0000-0001-5763-2760]{N.J.~Abicht}$^\textrm{\scriptsize 49}$,
\AtlasOrcid[0000-0002-8496-9294]{S.H.~Abidi}$^\textrm{\scriptsize 30}$,
\AtlasOrcid[0009-0003-6578-220X]{M.~Aboelela}$^\textrm{\scriptsize 45}$,
\AtlasOrcid[0000-0002-9987-2292]{A.~Aboulhorma}$^\textrm{\scriptsize 36e}$,
\AtlasOrcid[0000-0001-5329-6640]{H.~Abramowicz}$^\textrm{\scriptsize 156}$
% ...
\bigskip
\\

% ...
$^{17}$Department for Physics and Technology, University of Bergen, Bergen; Norway.\\
% ...
$^{30}$Physics Department, Brookhaven National Laboratory, Upton NY; United States of America.\\
% ...
$^{36}$$^{(a)}$Facult\'e des Sciences Ain Chock, Universit\'e Hassan II de Casablanca;$^{(b)}$Facult\'{e} des Sciences, Universit\'{e} Ibn-Tofail, K\'{e}nitra;$^{(c)}$Facult\'e des Sciences Semlalia, Universit\'e Cadi Ayyad, LPHEA-Marrakech;$^{(d)}$LPMR, Facult\'e des Sciences, Universit\'e Mohamed Premier, Oujda;$^{(e)}$Facult\'e des sciences, Universit\'e Mohammed V, Rabat;$^{(f)}$Institute of Applied Physics, Mohammed VI Polytechnic University, Ben Guerir; Morocco.\\
% ...
$^{45}$Physics Department, Southern Methodist University, Dallas TX; United States of America.\\
% ...
$^{49}$Fakult\"{a}t Physik, Technische Universit{\"a}t Dortmund, Dortmund; Germany.\\
% ...
$^{55}$II. Physikalisches Institut, Georg-August-Universit\"{a}t G\"ottingen, G\"ottingen; Germany.\\
% ...
$^{103}$CPPM, Aix-Marseille Universit\'e, CNRS/IN2P3, Marseille; France.\\
% ...
$^{118}$$^{(a)}$New York University Abu Dhabi, Abu Dhabi;$^{(b)}$United Arab Emirates University, Al Ain; United Arab Emirates.\\
% ...
$^{122}$Homer L. Dodge Department of Physics and Astronomy, University of Oklahoma, Norman OK; United States of America.\\
% ...
$^{156}$Raymond and Beverly Sackler School of Physics and Astronomy, Tel Aviv University, Tel Aviv; Israel.\\
% ...
$^{a}$ Also at Affiliated with an institute formerly covered by a cooperation agreement with CERN.\\
$^{b}$ Also at An-Najah National University, Nablus; Palestine.\\
$^{c}$ Also at Borough of Manhattan Community College, City University of New York, New York NY; United States of America.\\
$^{d}$ Also at Center for Interdisciplinary Research and Innovation (CIRI-AUTH), Thessaloniki; Greece.\\
$^{e}$ Also at Centre of Physics of the Universities of Minho and Porto (CF-UM-UP); Portugal.\\
$^{f}$ Also at CERN, Geneva; Switzerland.\\
% ...

\end{flushleft}


%

%
%

%

%
%
%
%

\end{document}